\documentclass[floatfix,citeautoscript,preprint,11pt,superscriptaddress]{revtex4-1}

\usepackage{graphicx}
\usepackage{comment}

\usepackage{subfigure}
\usepackage{setspace}[1.5] 
\usepackage{amsmath, amsthm, amssymb}
\usepackage{xfrac}
\usepackage{dcolumn}
\usepackage{gensymb}
\usepackage{threeparttable}
\usepackage[version=3]{mhchem} 
\usepackage{verbatim}

\usepackage{amssymb,amsfonts,amsmath}

\newcommand{\Elatt}{E_{\text{latt}}}
\newcommand{\DHsub}{ \Delta_\textrm{sub} H }
\newcommand{\DHcor}{ \Delta_\textrm{T\&QN} }

\begin{document}


\title{Fast and accurate quantum Monte Carlo for molecular crystals} 

\author{Andrea Zen}
\affiliation{Department of Physics and Astronomy, University College  London, Gower Street, London, WC1E 6BT, U.K.}
\affiliation{Thomas Young Centre and London Centre for Nanotechnology, 17--19 Gordon Street, London, WC1H 0AH, U.K.}
\author{Jan Gerit Brandenburg}
\affiliation{Department of Physics and Astronomy, University College  London, Gower Street, London, WC1E 6BT, U.K.}
\affiliation{Thomas Young Centre and London Centre for Nanotechnology, 17--19 Gordon Street, London, WC1H 0AH, U.K.}
\author{Ji\v{r}\'{i} Klime\v{s}}
\affiliation{
J. Heyrovsk\'{y} Institute of Physical Chemistry, Academy of Sciences of the Czech Republic, Dolej\v{s}kova 3,
CZ-18223 Prague 8, Czech Republic and Department of Chemical Physics and Optics, Faculty of Mathematics
and Physics, Charles University, Ke Karlovu 3, CZ-12116 Prague 2, Czech Republic
}
\author{Alexandre Tkatchenko}
\affiliation{Physics and Materials Science Research Unit, University of Luxembourg, L-1511 Luxembourg, Luxembourg}
\author{Dario Alf\`{e}}
\affiliation{Thomas Young Centre and London Centre for Nanotechnology, 17--19 Gordon Street, London, WC1H 0AH, U.K.}
\affiliation{Department of Earth Sciences, University College London,  Gower Street, London WC1E 6BT, U.K.}
\author{Angelos Michaelides}
\email{angelos.michaelides\@ucl.ac.uk}
\affiliation{Department of Physics and Astronomy, University College  London, Gower Street, London, WC1E 6BT, U.K.}
\affiliation{Thomas Young Centre and London Centre for Nanotechnology, 17--19 Gordon Street, London, WC1H 0AH, U.K.}

\begin{abstract}
Computer simulation plays a central role in modern day materials science. The utility of a given computational approach depends largely on the balance it provides between accuracy and computational cost. Molecular crystals are a class of materials of great technological importance which are challenging for even the most sophisticated \emph{ab initio} electronic structure theories to accurately describe. This is partly because they are held together by a balance of weak intermolecular forces but also because the primitive cells of molecular crystals are often substantially larger than those of atomic solids. Here, we demonstrate that diffusion quantum Monte Carlo (DMC) delivers sub-chemical accuracy for a diverse set of molecular crystals at a surprisingly moderate computational cost. As such, we anticipate that DMC can play an important role in understanding and predicting the properties of a large number of molecular crystals, including those built from relatively large molecules which are far beyond reach of other high accuracy methods.
\end{abstract}


\maketitle


\maketitle

\section*{ Significance }
Computational approaches based on the fundamental laws of quantum mechanics
are now integral to almost all materials design initiatives in academia and industry.
If computational materials science
is genuinely going to deliver on its
promises then an electronic
structure method with consistently high accuracy is urgently needed.
We show that, thanks to recent algorithmic advances and the strategy developed in our
manuscript, quantum Monte-Carlo (QMC) yields extremely accurate predictions for the lattice
energies of materials at a surprisingly modest computational cost. It is thus no longer a
technique that requires a world-leading computational facility in order to obtain meaningful
results. 
While we focus on molecular crystals the significance of our findings extends to all classes of materials.



\section*{Introduction}

Computer simulations, in particular those based on the fundamental laws of quantum mechanics, play a key role in modern materials science research. Such electronic structure approaches serve as an indispensable complement to experiment by helping to understand and predict the properties of materials as well as guiding the development of new ones. Density functional theory (DFT) is the leading electronic structure technique in materials science, thanks to its favorable balance between accuracy and computational cost as well as efficient implementation in modern codes. Indeed, DFT has been described as one of the great success stories of modern science~\cite{Burke2012:jcp_rev}, 
with widespread use in materials science and cognate disciplines~\cite{Curtarolo:natmat2013, Marzari:natmat2016,scan_natchem}. However, DFT as generally applied with standard exchange-correlation functionals, suffers from a number of well-known deficiencies~\cite{Cohen:2008fg,peverati2014,dft_dens,dft_dens_comment}.
One notable shortcoming is in the description of weak interactions such as London dispersion forces. Although the last decade has seen impressive developments with the incorporation of dispersion forces in the DFT framework~\cite{grimme_chemrev, vdw_perspective, beran_chemrev, ChemRev2017:Tkatchenko},
and new density functionals with improved accuracy have been developed~\cite{scan,wb97xv,revm06l}, such approaches are not systematically improvable and their accuracy for condensed phases is open to question. For example, DFT cannot be relied upon to routinely deliver an accuracy below 4 kJ/mol for the lattice energy of molecular crystals. This is the accuracy (so called chemical accuracy) that is often needed to discriminate between different polymorphs of a given material, 
and in some pharmaceutical molecules even more stringent accuracies (down to 1~kJ/mol) are needed~\cite{polymophism, CSP2016}.

Traditionally, highly accurate computations of interaction energies have been based on quantum chemistry techniques, in particular, coupled cluster with single, double and perturbative triple excitations (CCSD(T)). 
However, CCSD(T) calculations on solids have been notoriously challenging. Great strides forward have recently been made and
CCSD(T) can now be used to study solids through calculations in periodic boundary conditions \cite{FCIQMC:Nat2013}, or by employing 
approaches based on embedding and fragment decomposition \cite{Wen:2012cu, Bygrave:2012eo}.  The benzene crystal, for example, was recently considered in a \emph{tour de force} study using fragment decomposition \cite{GarnetChan:Sci2014}.
The recent introduction of local approaches~\cite{Werner:2011jv} promises to extend the range of applicability of CCSD(T) methods.
However, the cost of CCSD(T) calculations for large systems 
will remain high for the foreseeable future and their large-scale application is additionally hindered by enormous memory requirements.
The random phase approximation (RPA) is emerging as a promising approach for materials, in particular if singles corrections are introduced~\cite{RPA_Schimka:NatMatt2010, Ren:2011ht, Jiri:2016}. Although it is less accurate than CCSD(T) it is considerably more affordable and currently offers a very good balance between accuracy and computational cost. Quantum Monte Carlo, in particular within the fixed node diffusion Monte Carlo scheme \cite{foulkes01}, is an established method for reference quality calculations of molecular systems and condensed phases. Systematic studies in cases of non-covalent bonding have shown that DMC has an accuracy comparable to CCSD(T) \cite{noncov:chemrev2016}. 
An advantage of DMC over traditional quantum-chemical methods like CCSD(T) is that it is essentially unaffected by basis set issues, thanks to the deployment of an efficient ground state projection scheme and the use of B-splines.
However, the enormous computational cost of this approach means that DMC studies generally require a world-leading computational facility and even then it remains highly challenging to obtain demonstrably converged results. 
As a result, DMC studies are often restricted to one-off benchmark calculations of specific systems.

Building upon a recent algorithmic development~\cite{sizeconsDMC} which significantly reduces the computational cost of DMC, we have developed a scheme  (described in the Methods section) 
for obtaining converged DMC energies of molecular crystals in a fully periodic treatment, without the need for fragment decompositions.
The approach is effective for both small and large molecules and yields lattice energies of periodic molecular crystals at a computational cost comparable to RPA but with the accuracy of CCSD(T). For example, a chemically accurate lattice energy of a benzene crystal can be computed in as few as $10^4$ CPU hours. 
In addition, the scheme proposed comes with relatively low manpower costs and can be seen as a step towards the automated DMC treatment of molecular crystals. This opens up new horizons for DMC as a tool for accurately and rapidly predicting the properties of molecular crystals, including those built from pharmaceutical molecules. More broadly the insights obtained from this study should also prove beneficial to DMC studies of other crystalline materials and their surfaces.


\section*{Results}

The main quantity to consider in order to assess the stability of a crystal is its  
lattice energy $\Elatt$, which is the energy per molecule gained upon assuming the crystal form
with respect to the gas state.
It can be computed as
\begin{equation}\label{Elatt}
E_\textrm{latt} = { E_\textrm{crys} } - E_\textrm{gas}
\end{equation}
with $E_\textrm{crys}$
the energy per molecule in the crystal state 
and $E_\textrm{gas}$ the energy of the isolated molecule.
While the accurate computation of $E_\textrm{gas}$ is straightforward, $E_\textrm{crys}$ involves an ideally infinite system. 
Since any simulated system will be necessarily finite, the evaluation of $E_\textrm{crys}$ will be affected by finite size errors (FSE).
The most common approach to tackle crystals is to exploit their periodicity and to simulate a cell within periodic boundary conditions. 
In DFT the energy of the infinite system is recovered by simulating the primitive cell and by sampling the first Brillouin zone. 
In DMC 
the wave function is sampled by considering a number of electron configurations, including configurations with local dipoles that in a small simulation cell (such as the primitive cell)
couple with the periodic images, yielding substantial FSE. 
However, established procedures developed for DMC to reduce or eliminate these errors are available, such as the model periodic Coulomb (MPC) interaction~\cite{MPC:Fraser1996}; the approach used in this study.
MPC treats the explicit two-body interactions within the minimum-image convention in the simulation cell, while outside the cell the interaction is with a pre-calculated charge density normally obtained from a DFT calculation. 
In principle, MPC can miss some of the correlation energy. 
Thus, in DMC it is good practice to simulate supercells of increasing size and to extrapolate the energy of the infinite system. 
On the other hand, molecular crystals have primitive cells that are already large (and related to the size of the molecule) 
meaning that calculations in the larger cells needed for the extrapolations can be computationally prohibitive. 
However, thanks to recent developments~\cite{sizeconsDMC} it is now possible 
to simulate large supercells up to cases where FSE are negligible.
This, in turn, allows us to establish that accurate results from DMC can in practice be obtained in small primitive cells if appropriate corrections for FSE are taken into account.

Typically the computation of $\Elatt$ is performed at zero temperature and considering only the electronic contribution, {\em i.e.} quantum nuclear effects are neglected
~\cite{beran_accounts}. 
The lattice energy is not directly assessable experimentally, but it can be indirectly obtained from experimental measures of the sublimation enthalpy $\DHsub(T)$ at a given temperature $T$ by including a (theoretically evaluated) energy contribution $\DHcor(T)$ accounting for contributions from thermal and quantum nuclear effects:
\begin{equation}\label{DeltaH}
\DHsub(T) = -\Elatt + \DHcor(T).
\end{equation}
The evaluation of $\DHcor(T)$ can be challenging, especially for large molecules where anharmonic contributions are important~\cite{Elatt_ExpRT}. 
In the SI Appendix we provide details about the theoretical evaluations of $\DHcor(T)$ and the uncertainty associated with experimental evaluations of $\DHsub$. Since both $\DHsub(T)$ and $\DHcor(T)$ are affected by errors, accurate theoretical evaluations of $\Elatt$ are of help for comparison.
To assess the accuracy of a method one needs to use a diverse test set of systems. 
\begin{figure}[htb]
\centerline{\includegraphics[width=1.0\linewidth]{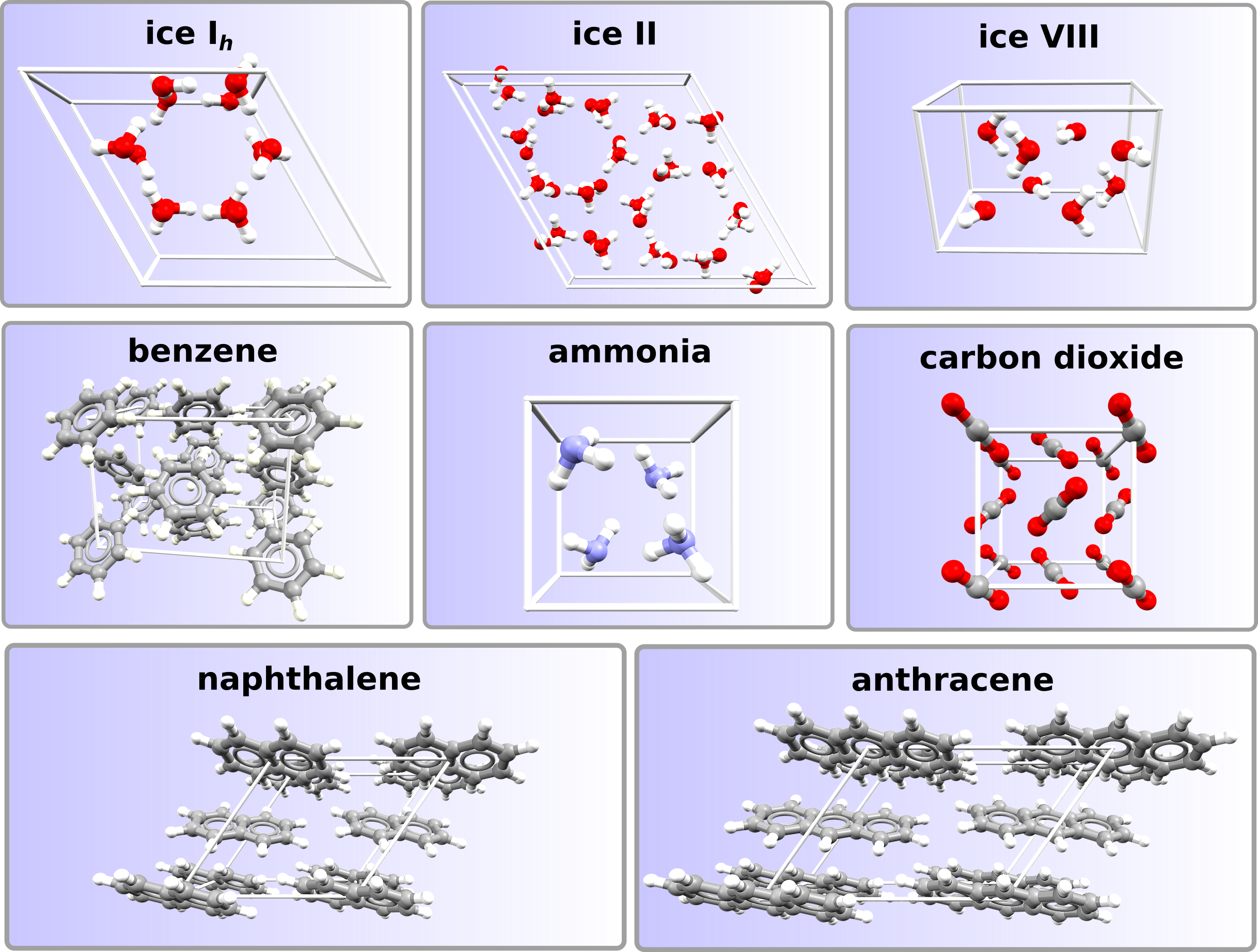}}
\caption{\label{fig:crys}
Molecular crystals considered in this work with DMC. Only the primitive cell is shown in each case.
The systems treated are of considerable size and contain up to 144 molecules (ice VIII) 
or 1728 electrons (CO$_2$). 
}
\end{figure}
\begin{table}[bht!]
\caption{\label{tab:crys}
Lattice energy [kJ/mol] for the molecular crystals under consideration in this work, computed with DMC 
compared to values from experimental measures of sublimation enthalpy.}
\begin{tabular}{   l   r r r }
\hline\\[-0.2cm]
& {\bf DMC(lc)}$^{\text{[a]}}$ & {\bf DMC(sc)}$^{\text{[b]}}$ & {\bf Experiment}$^{\text{[c]}}$ \\[0.1cm]
\cline{2-4}\\[-0.2cm]
Ice Ih           	&  -59.3$\pm$0.5  	&  -59.2$\pm$0.2	&  -58.8 	\\
Ice II           	&  -59.1$\pm$0.6  	&  -59.0$\pm$0.3	&  -58.8 	\\
Ice VIII         	&  -57.3$\pm$0.6  	&  -57.4$\pm$0.1	&  -57.4 	\\
Carbon dioxide		&  -28.2$\pm$1.3 	&  -28.5$\pm$0.4	&  -28.4 	\\
Ammonia         	&  -37.1$\pm$0.4  	&  -37.5$\pm$0.1	&  -37.2 	\\
Benzene          	&  -52.1$\pm$0.4  	&  -51.2$\pm$0.2	&  -50.6 	\\
Naphtalene      	&  -78.8$\pm$0.8  	&  -78.0$\pm$0.6	&  -79.2 	\\
Anthracene      	& -105.5$\pm$1.7  	& -103.9$\pm$1.0	& -105.8 	\\
\hline
\multicolumn{4}{p{9cm}}{ $^{\text{[a]}}$DMC using a large supercell. 
$^{\text{[b]}}$DMC using a small supercell, additional values in SI Appendix.
$^{\text{[c]}}$See SI Appendix, Sec.~S12 for details. 
}
\end{tabular}
\end{table}
In this study we considered eight molecular crystals (see Fig.~\ref{fig:crys}): 
carbon dioxide (CO$_2$), ammonia (NH$_3$), benzene (C$_6$H$_6$), naphthalene (C$_{10}$H$_8$) and anthracene (C$_{14}$H$_{10}$) crystals 
from the C21 test set of Otero-de-la-Roza and Johnson \cite{Elatt_ExpOJ},
plus three polymorphs of ice: the hexagonal ice I$_h$, ice II and a high-pressure phase ice VIII.
This set of molecular crystals comprises a diversity in intermolecular interactions involving strong hydrogen bonds and London dispersion of saturated and unsaturated molecules.
A range of interactions such as this is a tough test for any electronic structure method. It is a test that must be passed through if a method is to be truly predictive and applicable to complex molecular crystals, including those of industrial interest~\cite{csp_for_drugs, csp_nature}.
DMC values for $\Elatt$ of the eight molecular crystals are summarized in Table~\ref{tab:crys}. 
Two sets of DMC results are reported: DMC(lc) and DMC(sc).
The former is obtained using large supercells (containing around a thousand valence electrons);
the latter using smaller cells and relying on the FSE correction via the MPC interaction (see Methods and SI Appendix). 
We find that in all cases the DMC(sc) results are in excellent agreement with the DMC(lc) results, as a confirmation of the quality of the MPC approach for the correction of FSE.
MPC was introduced two decades ago~\cite{MPC:Fraser1996}, but here using the algorithm of Ref.~\cite{sizeconsDMC} we have been able to explicitly demonstrate how well MPC performs for complex systems such as molecular crystals.
Table~\ref{tab:crys} 
also reports lattice energies derived from experiments. 
The experimental values
are obviously not free from error and 
have an uncertainty coming  from both the actual measure of $\DHsub$ 
and the computed term $\DHcor$
on the order of the chemical accuracy, $\sim 4$~kJ/mol.
For naphthalene and anthracene which have the largest values for $\Elatt$ 
the experimental uncertainties are likely to be larger.
See the SI Appendix for a detailed discussion on the experimental values.
Upon comparing DMC to experiment we find 
that both DMC(lc) and DMC(sc) always fall within the accuracy of the experimental value.
This is remarkable if we consider that this accuracy is achieved over a large range of lattice energies, $\Elatt$ from 28 to more than 100~kJ/mol. 
DMC gets correct lattice energies for hydrogen bonded, dispersion bonded and mixed bonded crystals.
DMC also predicts the correct relative energies of the ice polymorphs, yielding slightly improved lattice energies over those reported in ref.~\cite{santra_hydrogen_2011}.


\section*{Discussion}
\begin{figure}[htb]
\centerline{\includegraphics[width=0.85\linewidth]{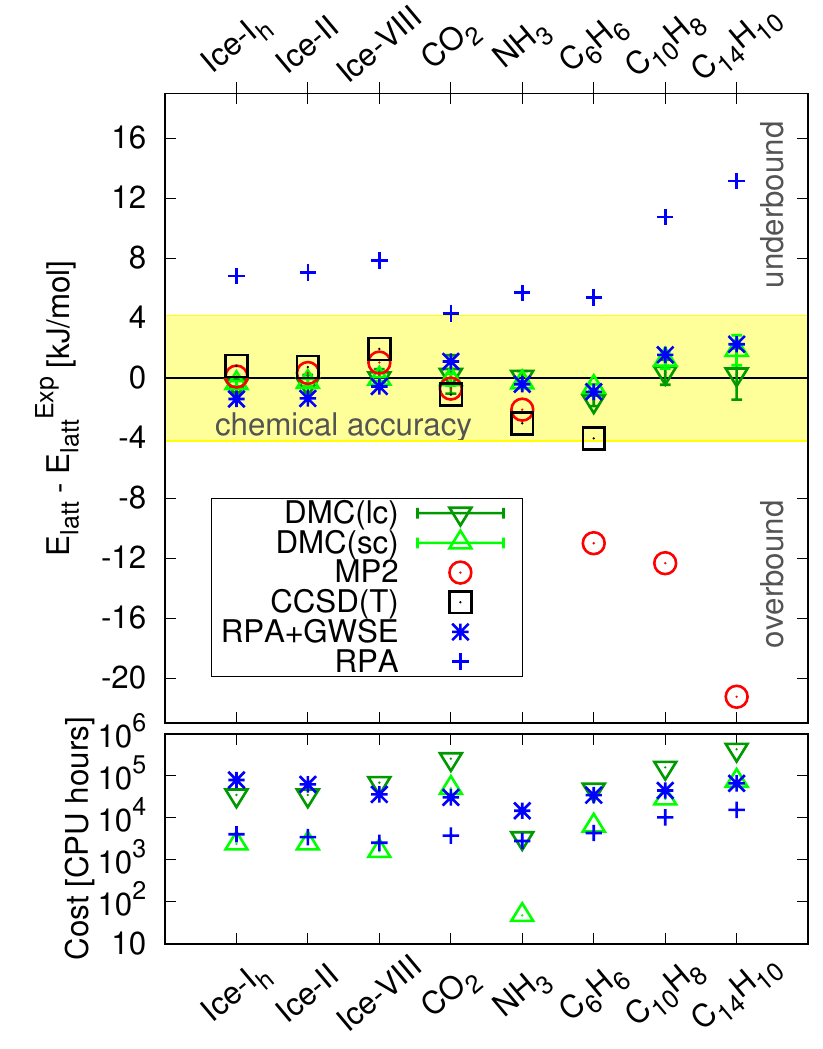}}
\caption{\label{fig:summary}
Accurate and fast DMC results for a range of molecular crystals.
(Top panel) Difference in the value of $\Elatt$ between the experimental value and several computational approaches often used as reference methods.
Here DMC(lc) and DMC(sc) indicates that large or small supercells have been employed, respectively.
RPA and RPA+GWSE values for ice have been computed in this work, other values are  from Ref.~\cite{Jiri:2016}.
MP2 and CCSD(T) values for ice are from Ref.~\cite{Gillan:enebench:jcp2013}, 
benzene from Ref.~\cite{GarnetChan:Sci2014}, 
MP2 values for naphthalene and anthracene from Ref.~\cite{Cutini:2016fh},
other values are from Ref.~\cite{Wen:2011gm}.
(Bottom panel) Approximate computational cost for DMC(sc), DMC(lc), RPA and RPA+GWSE  (see SI Appendix for details). The DMC cost is intended for a precision of 0.7~kJ/mol. Reported timings are intended only to provide an indication; differences in the codes and computation facilities can yield very different timings }
\end{figure}

A comparison of the results obtained from DMC to experiment and to other reference-quality computational approaches is shown in Fig.~\ref{fig:summary}.
This includes MP2 results for all systems~\cite{Gillan:enebench:jcp2013, Cutini:2016fh, Wen:2011gm},
and CCSD(T) for all molecules up to benzene~\cite{Gillan:enebench:jcp2013, GarnetChan:Sci2014, Wen:2011gm}.
RPA and RPA with singles corrections (RPA+GWSE) lattice energies for ice are calculated in this work, and the other values are from  Ref.~\cite{Jiri:2016}.
From this comparison 
we notice that CCSD(T) and RPA+GWSE perform well, whereas RPA systematically underbinds all systems and MP2 severely overbinds in systems with delocalized electrons such as benzene, naphthalene and anthracene.
Among the computational approaches reported in Fig.~\ref{fig:summary}, only CCSD(T) is acknowledged for an accuracy comparable to DMC, and indeed they show excellent agreement.
However, all CCSD(T) (and MP2) values reported come from fragment decomposition approaches, which involves the computation of many small contributions to the lattice energy, all of which must be converged to high accuracy and typically the correlation contribution from long-range fragments is computed at a lower level of theory. 
This can be a painstaking process. Also the range of values obtained from the widely studied benzene crystal (-50 to -56~kJ/mol)~\cite{GarnetChan:Sci2014, Wen:2011gm, Podeszwa:2008fq, Bludsky:2008bu, Ringer:2008kq},
suggests that the decisions made in carrying out the fragment decomposition can have a noticeable effect on the final result.
A big advantage of methods employing periodic boundary conditions, such as DMC, is that the $\Elatt$ is obtained from a single calculation (provided that FSE are corrected for), which makes such approaches more suitable for rapid screening. 
As an added bonus, methods such as DMC also yield information on the electronic structure and electron density on the full periodic system; information that can be used for the calculation of experimental observables and to obtain deeper understanding of the system under consideration.

Computational cost is of utmost importance when making comparisons of computational methods.
Whilst DMC(lc) and DMC(sc) produce almost equal values for $\Elatt$, each DMC(sc) is much cheaper than DMC(lc).
For example, as shown in the bottom panel of Fig.~\ref{fig:summary},
DMC(sc) is typically 1-2 orders of magnitude cheaper.
Indeed, most of the DMC(sc) results take around $10^4$ core hours, and can be obtained in around a day on a few hundred processors.
This is much more affordable than CCSD(T), 
which is also only feasible for relatively small molecules with the fragment decomposition approach or small crystals in periodic  boundary conditions.
RPA+GWSE was so far providing a good compromise between accuracy and computational cost.
Fig.~\ref{fig:summary} shows that the cost for DMC(sc), for a precision on $\Elatt$ of around 1~kJ/mol, is in general comparable to RPA.

The computational efficiency of the DMC simulations 
and the fact that we have periodic boundary conditions makes it relatively straightforward to investigate 
other properties beyond the lattice energy. For instance, we have obtained the equation of state (EOS) for both ammonia and benzene; these are crystals held together predominantly by hydrogen bonds and dispersion interactions, respectively. The results of these simulations, along with fits to the Murnaghan EOS, are reported in Fig.~\ref{fig:EOS}. From this we find that the equilibrium volumes ($V_0$) predicted by DMC agree very well with experiment, coming out $\sim$3\% smaller than experiment for both crystals. Slightly smaller DMC volumes are to be expected since our calculations do not take into account anharmonic thermal expansion and quantum nuclear effects present in experiment. 
The EOS calculations are also useful because they allow us to test the sensitivity of our computed $\Elatt$ to the volume used in our calculations. The DMC $\Elatt$ values listed in Table~\ref{tab:crys} have been obtained at experimentally measured densities. For the two crystals reported in Fig.~\ref{fig:EOS}, the bias on $\Elatt$ arising from the use of the experimental volume appears very small, less than 0.2 kJ/mol. For the other crystals reported in Table~\ref{tab:crys} we expect, on the basis of DFT tests~\cite{Jiri:2016}, a bias on $\Elatt$ due to the volume on the order of 1 kJ/mol or less. A second source of bias on the values of $\Elatt$ reported in Table~\ref{tab:crys} is due to the geometries used for the DMC calculations. Indeed, DMC, CCSD(T), MP2 and RPA are typically too expensive for a geometry optimization, which is often performed via DFT with a reliable functional. In the SI Appendix we report the EOS of ammonia and benzene obtained using the geometries from two different DFT functionals. The uncertainty on $\Elatt$ appears to be less than 1 kJ/mol both for ammonia and benzene.
\begin{figure}[htb]
\centerline{\includegraphics[width=0.99\linewidth]{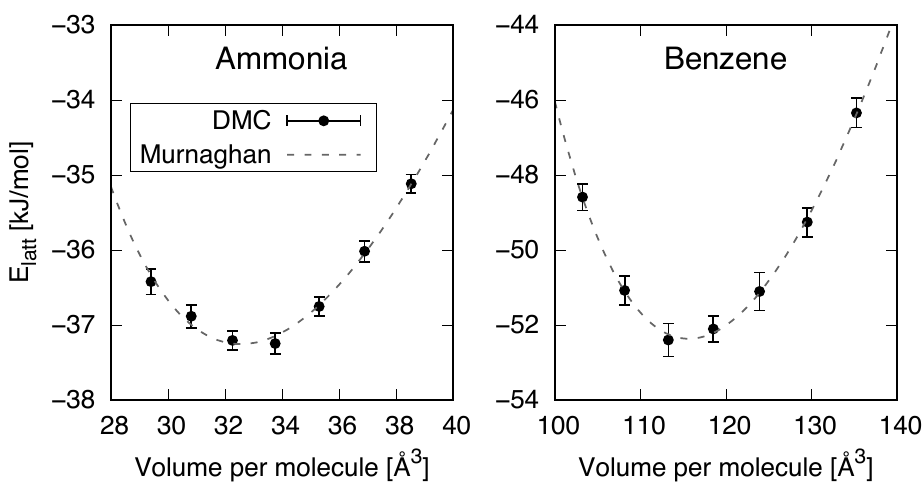}}
\caption{\label{fig:EOS}
Equation of state (EOS) for the ammonia and benzene crystals (with zero-point motion not accounted for). 
In both cases, the DMC $\Elatt$ values are calculated using the $2\times 2\times 2$ cell 
and FSE are corrected for with MPC. 
The dashed line is the Murnaghan EOS fitting the DMC values, which 
for ammonia yields
a minimum $E_0$ of $-37.25\pm 0.05$~kJ/mol 
at a volume $V_0$ of $32.6\pm 0.1$~\AA$^3$ per molecule 
and a bulk modulus $B_0$ of $8.5\pm 0.5$~GPa, 
and for benzene 
$E_0$ is $-52.37\pm 0.06$~kJ/mol,
$V_0$ is $115.7\pm 0.1$~\AA$^3$ per molecule
and $B_0$ is $7.7\pm 0.2$~GPa.
The value of $B_0'$ in the Murnaghan EOS is set to 4.
%
}
\end{figure}

To conclude, we have demonstrated that DMC provides a route towards the fast and accurate determination of the properties of molecular crystals. In essence, the scheme makes use of the size-consistent DMC algorithm introduced earlier~\cite{sizeconsDMC} and an accurate approach for correcting for finite size errors. We have applied this approach to a range of exemplar systems held together with a range of intermolecular interactions (hydrogen bonds to London dispersion). The calculations have confirmed previous results on water-ice polymorphs but with minimal computational cost and with much more control over the numerical accuracy of the results than before. Our results also include EOS calculations for benzene -- the ``fruitfly'' molecular crystal in computational materials science -- and anthracene, the largest molecule in the C21 dataset. The consistently high accuracy demonstrated by DMC along with its moderate computational cost suggests that DMC can play an increasingly important role in studies of molecular crystals. In particular, DMC could prove to be the method of choice in challenging polymorph prediction studies. Similarly, molecules of direct pharmaceutical interest could now be tackled with DMC; opening up their study with a high-level {\em ab initio} approach for the first time. To provide full phase diagrams for molecular crystals, our accurate lattice energies have to be combined with estimates of zero-point and thermal effects. This is traditionally computed at the DFT level, where our study will further provide an important benchmark to test and calibrate these approximate methods. Looking further to the future we note that several steps of the proposed methodology could be used in a full configuration interaction QMC approach~\cite{FCIQMC:Nat2013}, which would yield essentially exact solutions to the Schr{\"o}dinger equation for molecular crystals. Finally, we note that beyond molecular crystals the improved efficiencies and improved understanding of finite size errors obtained here will also be of direct relevance to DMC simulations on other classes of material, e.g. absorption in metal organic frameworks and binding to surfaces.


\section*{Materials and Methods}

Geometries for the C21 crystals are taken from \cite{Jiri:2016} 
(where the geometries for molecules and crystals are optimized via DFT using the optB88-vdW functional \cite{klimes-vdW-DF}, and crystals are in the experimental unit cell).
For ice phases we took the geometries used in ref.~\cite{santra_hydrogen_2011}.
DMC simulations were carried out with the {\sc casino} code~\cite{casino} in order to evaluate $E_\textrm{crys}$ and $E_\textrm{gas}$. 
We used Dirac-Fock pseudopotentials~\cite{trail05_NCHF, trail05_SRHF} with the locality approximation~\cite{mitas91}. 
The trial wavefunctions were of the Slater-Jastrow type with single Slater determinants and the single particle orbitals obtained from DFT-LDA plane-wave calculations performed with  {\sc pwscf}~\cite{pwscf}
and re-expanded in terms of B-splines~\cite{alfe04}.
The Jastrow factor included electron-electron, electron-nucleus and electron-electron-nucleus terms. 
Further details on the wave-function and the optimization are provided in the SI Appendix,
as well as some comparative tests with the recently introduced correlated electron pseudopotentials~\cite{CEPP:JCP2013}.

In the computation of $E_\textrm{crys}$, periodic boundary conditions are employed. 
Simulations with DMC in periodic boundary conditions can be subject to significant FSE, as previously discussed.
In order to assess the converged value of $\Elatt$, for any molecular crystal several simulation cells were considered as well as twist boundary conditions~\cite{Lin:qmctwistavg:pre2001} for the smallest cells. 
This has revealed that the $\Elatt$ obtained from the primitive cell can be overestimated by as much as 300\% due to FSE. However, correction schemes to reduce FSE are available in DMC, such as the model periodic Coulomb (MPC) interaction~\cite{MPC:Fraser1996,MPC:Will1997,MPC:Kent1999}, 
the correction proposed in ref.~\cite{Chiesa:size_effects:prl2006} and the one in ref.~\cite{KZK:prl2008}.
We have tested all of them and observed that MPC provides the best results, as shown in the SI Appendix. 
Here we report results obtained exclusively with MPC. 
A second and smaller source of FSE in DMC stems from the use of single particle orbitals obtained from a DFT calculation on a single point in the Brillouin zone (typically the $\Gamma$-point). 
This error, called the independent particle finite size error (IPFSE), can be easily estimated and corrected for by performing few additional DFT calculations.
Further computational details are reported in the SI Appendix, including the atomic coordinates for each molecular system studied.

\begin{figure}[htb!]
\centerline{\includegraphics[width=0.85\linewidth]{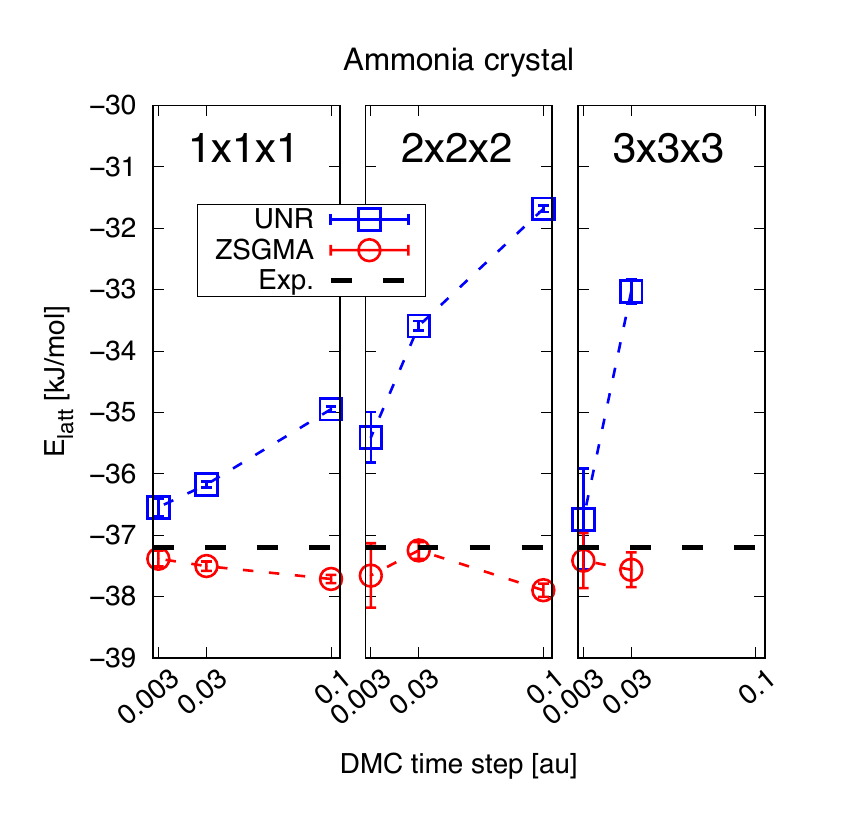}}
\caption{\label{fig:dmc_tau}
Converged lattice energies from DMC in small and large unit cells.
Lattice energies, $\Elatt$, for the ammonia crystal, as obtained from DMC (with the MPC interaction to reduce FSE) by using different time steps and cell sizes, from the primitive 1x1x1 cell (comprising 4 molecules)
to a 3x3x3 supercell (108 molecules).
Blue squares represent the results obtained by employing the algorithm by Umrigar, Nightingale and Runge (UNR)~\cite{umrigar93}; red circles correspond to the algorithm by \cite{sizeconsDMC} (ZSGMA), which consistently yields accurate results. 
The black dashed line is the value obtained from experimental sublimation enthalpies by
\cite{Elatt_ExpRT}.
}
\end{figure}

The time step, $\tau$, is a key issue affecting the accuracy of DMC calculations.
In DMC a propagation according to the imaginary time Schr\"{o}dinger equation is performed in order to project out the exact ground state from a trial wave function \cite{foulkes01}.
A time step $\tau$ must be chosen, keeping in mind that the efficiency of DMC is directly proportional to $\tau$ but the projection is exact only in the continuous limit $\tau\to 0$. 
Thus, $\tau$ has to be small enough to yield converged results,
but as large as possible to make DMC efficient.
The time step dependence is system dependent, so it has to be evaluated on a case by case basis. 
In periodic systems this can be computationally very expensive, because each supercell is possibly affected by the time step differently.
As noted, an improved DMC algorithm \cite{sizeconsDMC} was recently presented.
The new algorithm, denoted ZSGMA from the authors' initials, gives better convergence with respect to $\tau$ than the one proposed by Umrigar, Nightingale and Runge (UNR)~\cite{umrigar93} 
which is
implemented as standard in DMC codes.
In the evaluation of $\Elatt$ an important practical point is the influence of the simulation cell size on the time step error. 
This is shown in Fig. 4 for the example of the ammonia crystal. Specifically, in Fig.~\ref{fig:dmc_tau} the dependence of $\Elatt$ on $\tau$ is shown for a range of different unit cells. First, it can be seen that with ZSGMA, $\Elatt$ exhibits almost no dependence on $\tau$ for the range of $\tau$ reported. In contrast, values of $\Elatt$ from UNR show a pronounced and non-linear dependence on $\tau$. This means that for UNR simulations for small values of $\tau$ (say $\tau \le 0.001$~au) are required in order to generate a reliable $\tau\to 0$ extrapolation. Second, and shown here for the first time, we find that with ZSGMA the time step error on $\Elatt$ is independent of the size of the unit cell (c.f. the 1x1x1, the 2x2x2 and the 3x3x3 cells in Fig.~\ref{fig:dmc_tau}). In contrast with UNR the time step error increases significantly as the size of the simulation cell is increased. This general behavior can be rationalized by the fact that ZSGMA is (approximatively) size consistent up to relatively large values of $\tau$, while UNR is size consistent only in the limit $\tau\to 0$.

In this work we have verified the time step convergence with the ZSGMA algorithm for each molecular crystal, as reported in the SI Appendix. It is the larger time step that ZSGMA facilitates and the insensitivity of the time step error to the size of the cell that enables the converged DMC calculations on large crystals reported in this study. All results reported in the work are obtained with the ZSGMA algorithm and a time step that yields a bias of $< 1$~kJ/mol. With the UNR algorithm the same accuracy would have required difficult extrapolations and a computational cost around two orders of magnitude larger. 

We now describe the scheme employed here to compute accurate values of $\Elatt$ with DMC.
We recommend the use of the ZSGMA algorithm~\cite{sizeconsDMC} in all DMC calculations, and MPC for all DMC calculations in periodic systems.
The following 5 step procedure can be used to assess the lattice energy for a given molecular crystal.
\begin{itemize}
\item[\bf (i)] {\bf Geometries --} Obtain geometries for the molecular crystal and the isolated molecule. Since geometry optimizations of large systems are challenging with DMC, we recommend the use of DFT and an exchange-correlation functional that accounts for vdW dispersion forces.
If reliable experimental structures are available, the optimization should be performed at the experimental volume.
\item[\bf (ii)] {\bf Independent particle finite size error (IPFSE) --} 
Using the structure obtained in (i), 
converge the energy per molecule in the crystal, 
$E_{\text{crys}}^{\text{DFT},\infty}$, 
using the functional that is used to obtain the single particle orbitals for the DMC calculations (we generally use LDA). 
Convergence is reached by considering $l$x$m$x$n$ Monkhorst-Pack grids of increasing size.
The difference $E_{\text{crys}}^{\text{DFT},l\text{x}m\text{x}n}$ - $E_{\text{crys}}^{\text{DFT},\infty}$={IPFSE}$^{\text{DFT}}_{l\text{x}m\text{x}n}$
provides a good indication of the independent particle contribution to the FSE in DMC calculations for a $l$x$m$x$n$ supercell, see SI Appendix.
\item[\bf (iii)] {\bf Jastrow optimization --}
Take the smallest supercell that is compatible with the Jastrow factor 
(typically the Jastrow factor has cutoffs related to the size of the simulated cell; we suggest to use supercells with the maximum radius of a sphere inscribed within the Wigner-Seitz cell $>$5~\AA)
and optimize the Jastrow factor of the quantum Monte Carlo wave function by minimizing the variance (or, alternatively, the variational energy). 
An optional test of the reliability of the Jastrow can be performed by calculating the DMC binding energy in a molecular dimer extracted from the crystal, and comparing it with a reference value obtained from CCSD(T). 
\item[\bf (iv)] {\bf DMC time step --} 
Check the time step dependence either on the cell used in step (iii) or on the molecular dimer.  
\item[\bf (v)] {\bf Final DMC calculation of $\Elatt$ --} 
Take a supercell from step (ii) with the estimation
{IPFSE}$^{\text{DFT}}_{l\text{x}m\text{x}n}$
smaller that 10~kJ/mol. Perform the DMC simulation for this crystal using MPC, and the DMC calculation with open conditions for the molecule. Calculate $\Elatt$ and correct for the independent particle FSE using {IPFSE}$^{\text{DFT}}_{l\text{x}m\text{x}n}$. 
This yields the final DMC(sc) result. 
Optionally, consider larger supercells in order to reduce {IPFSE}$^{\text{DFT}}_{l\text{x}m\text{x}n}$ and the MPC correction. 
\end{itemize}

The threshold {IPFSE}$^{\text{DFT}}_{l\text{x}m\text{x}n}$ $<$ 10~kJ/mol (in step v) is motivated by the
target accuracy of $\sim$1~kJ/mol and a $\sim$10~\% reliability of the DFT based IPFSE correction.
A more accurate alternative to evaluate the IPFSE is possible (twist averaging) and discussed in the SI Appendix.

\subsection*{Supporting Information (SI)}
Supporting Information Appendix provides details of the setup for the DMC, RPA and RPA+GWSE calculations, a discussion of the finite size errors and additional DMC results. In addition, the computational cost of DMC is discussed, as well as of RPA and RPA+GWSE. An extended version of Table~\ref{tab:crys} is given, and the evaluation of lattice energies from experimental sublimation enthalpies is discussed. Geometries of the molecular crystals and reference molecules used for the DMC, RPA and RPA-GWSE calculations are given.

\begin{acknowledgments}
A.Z. and A.M. are supported by the European Research Council
under the European Union's Seventh Framework Program
(FP/2007-2013) / ERC Grant Agreement number 616121 (HeteroIce project).
A.Z. and A.M.'s work is also sponsored 
by the Air Force Office of Scientific Research, Air Force Material Command, USAF, under grant number FA8655-12-1-2099.
A.M. is also supported by the Royal Society through a Royal Society Wolfson Research Merit Award.
J.G.B acknowledges support by the Alexander von Humboldt foundation within the Feodor-Lynen program.
J.K. is supported by the European Union's Horizon 2020 research and innovation program under the Marie Sklodowska-Curie grant agreement No 658705.
We are also grateful for computational resources to ARCHER, UKCP consortium (EP/ F036884/1), 
the London Centre for Nanotechnology, UCL Research Computing, Oak Ridge Leadership Computing Facility (No. DE-AC05-00OR22725), and IT4Innovations Centre of Excellence (CZ.1.05/1.1.00/02.0070 and LM2015070).
\end{acknowledgments}

\setcounter{section}{0}
\renewcommand{\thesection}{S\arabic{section}}%
\setcounter{table}{0}
\renewcommand{\thetable}{S\arabic{table}}%
\setcounter{figure}{0}
\renewcommand{\thefigure}{S\arabic{figure}}%
\setcounter{equation}{0}

\begin{widetext}
\newpage
\section*{Supporting Information for ``Fast and accurate quantum Monte Carlo for molecular crystals''}

\noindent
As supporting information we provide details about the setup for the DMC calculations in Section~\ref{sec:DMCsetup}; 
a discussion of the finite size errors in DMC in Section~\ref{sec:FSE}; 
additional results on the DMC calculations performed for carbon dioxide in Section~\ref{sec:dmc_CO2}, 
ammonia in Section~\ref{sec:dmc_NH3}, 
benzene in Section~\ref{sec:dmc_C6H6}, 
naphthalene in Section~\ref{sec:dmc_C10H8}, 
anthracene in Section~\ref{sec:dmc_C14H10}, 
and the three ice polymorphs in Section~\ref{sec:ice_dmc}. 
A discussion on the DMC scaling and computational cost is reported in Section~\ref{sec:DMCcost}. 
Furthermore, details on the RPA and RPA+GWSE calculations are given in Section~\ref{sec:RPA}. 
An extended version of Table~I of the main paper is given in Section~\ref{sec:othermethods}. 
We describe in Section~\ref{sec:ElattEXP} how the lattice energy can be obtained from experimental evaluations of the sublimation enthalpy, and we discuss the corresponding uncertainty. 
Finally, we provide the geometries of the molecular crystals used for the DMC, RPA and RPA-GWSE calculations in Section~\ref{sec:geoCrys}, as well as the geometries of the reference molecules in Section~\ref{sec:geoMol}.
%


\section{ DMC setup }\label{sec:DMCsetup}

The geometries used for the molecular crystals and for the reference molecule are reported in Sec.~\ref{sec:geoCrys} and \ref{sec:geoMol}, respectively.
%
In all the reported DMC calculations we used a trial wavefunctions of the Slater-Jastrow type with a single Slater determinants, the 
Trail-Needs-Dirac-Fock pseudopotentials (NTDF)~\cite{trail05_NCHF, trail05_SRHF} and the locality approximation. 
Single particle orbitals were obtained using DFT with the LDA functional and a plane-wave cutoff of 600 Ry, 
re-expanded in terms of B-splines with the natural grid spacing $a=\pi/G_{\rm max}$, where $G_{\rm max}$ is the magnitude of the largest plane wave in the expansion.  
The Jastrow factor used in the trial wavefunction of the system included a two-body electron-electron (e-e) term; two-body electron-nucleus (e-n) terms and three-body electron-electron-nucleus (e-e-n) terms specific for any atom type. 
The variational parameters of the Jastrow have been optimized in order to minimize the variance in the smallest considered simulated cell for each molecular crystal, and the same Jastrow has been used for larger supercells and the isolated molecule. The size of the simulation cell impose some constraints of the Jastrow variational freedom, in the form of cut-offs in the e-n, e-e and e-e-n terms. For this reason, the primitive cell could be too small and a larger simulation cell should be considered in order to do not impose a suboptimal Jastrow. In particular, the smallest simulation cell that we considered is the primitive cell for the ice polymorphs, ammonia and benzene; the $2\times 2 \times 2$ for the carbon dioxide and the $1\times 2 \times 1$ for naphthalene and anthracene. The Jastrow parametrization that we used for ice is the same as already used previously in Ref.~\onlinecite{santra_hydrogen_2011}; for naphthalene and anthracene we used the Jastrow optimized for the benzene crystals in the primitive cell.
The properties (e.g., variance, the time step) used for the calculations reported in Table~I of the main paper are given in Table~\ref{tab:crysprop}, and an estimate of the computational cost for each of the molecular crystals with different supercells are reported in Table~\ref{tab:Supercells}.

The chosen setup can surely influence the absolute DMC energy value, but in this work we are interested in energy differences, so a big error cancellation is expected and slightly different choices should not yield drastic differences.
For instance, in Fig.~\ref{fig:pseudoNH3} we compare the values of the binding energy of the ammonia dimer, obtained using the locality approximation and the TNDF pseudopotentials ({\em i.e.}, the set up used for all the other systems) with those obtained using the  T-move~\cite{CasulaTmove, casula10} and the recently introduced correlated electron pseudopotentials (CEPP)~\cite{CEPP:JCP2013}.
The latter choice shows a slightly larger time step dependance, but for a timestep $\tau\le0.01$~au the results are in good agreement. 
%
\begin{figure}[tb!]
\includegraphics[width=3.5in]{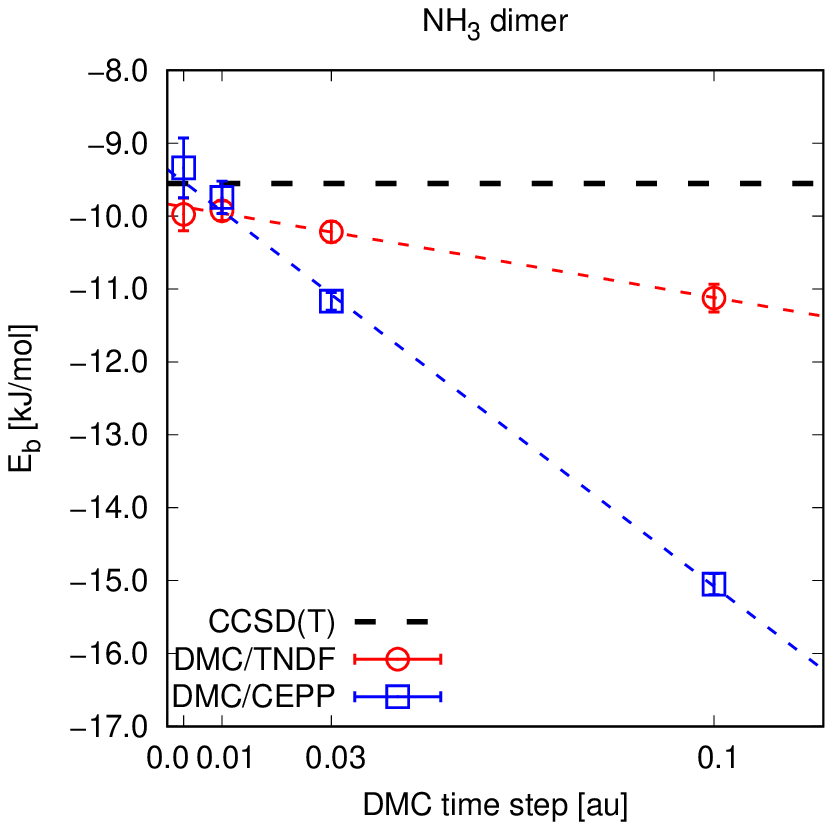}
\caption{\label{fig:pseudoNH3}
Binding energy for the ammonia dimer calculated via DMC as a function of the DMC time steps. 
Results are obtained using locality approximation and 
Trail-Needs-Dirac-Fock pseudopotentials (TNDF)~\cite{trail05_NCHF, trail05_SRHF} (red circles),
or T-move and  
correlated electron pseudopotentials (CEPP)~\cite{CEPP:JCP2013} (blue squares).
We report as a black dashed line the reference CCSD(T) value.
}
\end{figure}

The geometries used for the isolated molecule and the molecular crystal can also influence the value of $\Elatt$, as discussed in the main paper.
In Fig.~\ref{fig:EOS} we report the equation of state (EOS) for ammonia and benzene, as obtained with DMC calculations using DFT geometries from the optB88-vdW and the HSE-3c functionals.
The fitted parameters of the Murnaghan EOS for the two sets of geometries are in quite good agreement.

\begin{figure}[tb!]
\includegraphics[width=5.5in]{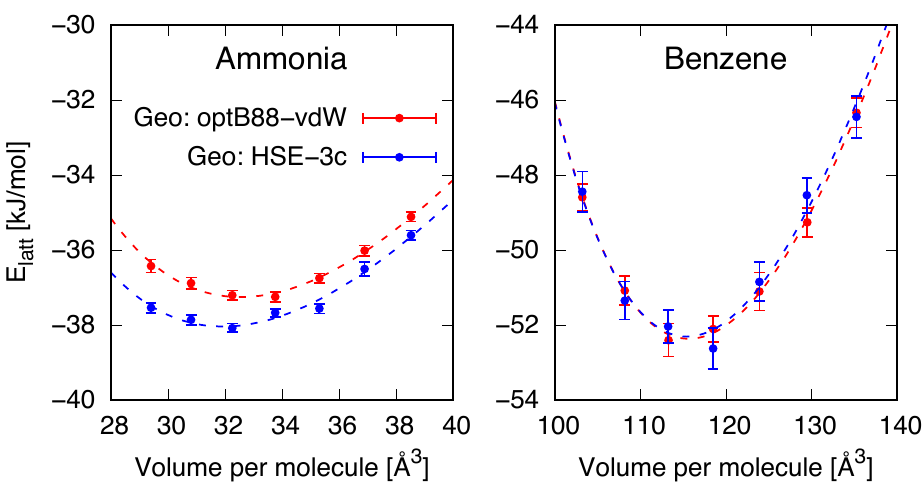}
\caption{\label{fig:EOS}
EOS for ammonia and benzene, evaluated using DMC calculations on the $2\times 2\times 2$ cell, FSE corrected via MPC, and geometries from DFT optimizations with the optB88-vdW and the HSE-3c functionals. 
Dashed lines are obtained fitting the DMC values with the Murnaghan EOS, where the  $B_0'$ term (the derivative of the bulk modulus with the pressure) is set to a value of 4.
}
\end{figure}


\begin{table}[bhp]
\caption{
Results from 
DMC and CCSD(T) at complete basis set (CBS), for molecular dimers. The geometry of the dimers is extracted from the molecular crystals. Zero-point effects are not included.
}\label{tab:dimers}
\begin{center}
\begin{tabular}{   l   c c c c c }
\hline \hline
	&	Carbon dioxide			&	Ammonia			&	Benzene			&	Naphthalene			&	Anthracene			\\
	&	CO$_2$			&	NH$_3$			&	C$_6$H$_6$			&	C$_{10}$H$_8$			&	C$_{14}$H$_{10}$			\\
\hline \hline																					
DMC($\tau$=0.1)	&	-14.6	$\pm$	0.2	&	-11.1	$\pm$	0.2	&				&				&				\\
DMC($\tau$=0.03)	&	-6.9	$\pm$	0.3	&	-10.2	$\pm$	0.1	&	-10.6	$\pm$	0.2	&	-19.5	$\pm$	0.6	&	-29.7	$\pm$	0.8	\\
DMC($\tau$=0.01)	&	-5.4	$\pm$	0.2	&	-9.9	$\pm$	0.1	&	-9.9	$\pm$	0.3	&	-18.6	$\pm$	0.5	&	-30.0	$\pm$	0.8	\\
DMC($\tau$=0.003)	&	-4.6	$\pm$	0.4	&	-10.0	$\pm$	0.2	&				&				&				\\
DMC($\tau\to 0$)	&	-4.2	$\pm$	0.2	\\
\hline
CCSD(T)$^a$	&	-4.1			&	-9.6			&	-9.9	$\pm$	0.3	&	-19.1	$\pm$	0.4	&	-29.8	$\pm$	0.6	\\
\hline \hline
\multicolumn{6}{p{5.5in}}{\footnotesize
$^a$ In all the systems, chemical core is kept fixed in the correlation treatment, no relativistic are included; 
\ce{CO2} is obtained at the CCSD(T)/CP/CBS level;
\ce{NH3} at CCSD(T)-F12/CBS(cc-pVTZ-F12,cc-pVQZ)/CP;
\ce{C6H6}, \ce{C10H8} and \ce{C14H10} at L-CCSD(T) interaction energy with TZ, QZ extrapolation.
}\\
\end{tabular}
\end{center}
\end{table}

In order to test the reliability of all the choices made, system by system, we extracted from each molecular crystal a dimer, and we compared the DMC results from our approach, and for different choices of the time step $\tau$, with the predictions of CCSD(T). 
Results are provided in Table~\ref{tab:dimers}.  
From the results in the dimer, we have an indication that in order to have a target accuracy of around 1~kJ/mol, we can use a time step of $\tau = 0.03$~au for ammonia, benzene, naphthalene and anthracene, while in carbon dioxide there is a strong time step dependance, thus we have performed calculations at different values of $\tau$, down to 0.003~au, and we have extrapolated for $\tau\to 0$.
Additional tests on the DMC time step have also been performed on these molecular crystals, as reported in Sec.~\ref{sec:dmc_CO2}--\ref{sec:dmc_C14H10}, confirming the reliability of the over mentioned choices.
In the three ice polymorphs we have used $\tau = 0.003$~au, as discussed in Sec.~\ref{sec:ice_dmc}.
However, we first discuss in Sec.~\ref{sec:FSE} details about the FSE corrections.

\section{ Finite size errors in DMC }\label{sec:FSE}

In the computation of $E_\textrm{crys}$, periodic boundary conditions (PBC) are employed and the value $E_\textrm{crys}^{N_\textrm{mol}}$ computed is obtained from a cell with only $N_\textrm{mol}$ molecules. The quantity that we need is the thermodynamic limit for $E_\textrm{crys}^{\infty}$, obtained for $N_\textrm{mol}\to \infty$, that is by using an infinitively large simulation cell.
The difference 
$ E_\textrm{crys}^{\infty} - E_\textrm{crys}^{N_\textrm{mol}} $
is the finite size error (FSE).
FSEs in DMC are quite tricky, but they have been studied in detail and are well characterized \cite{FSEqmc:PRB2008, FSE:Ceperley2016}.
They essentially arise from
an independent particle contribution (IPFSE) due to quantization of momentum, and 
a spurious Coulomb interaction (CIFSE) between periodic images.


Methods employing independent orbitals (even effectively, such as using the Kohn-Sham orbitals in DFT) often allows to calculate the energy (per electron, per atom or per primitive cell) by performing a reduction of the problem to the primitive cell.
Many-body approaches, such as QMC, do not allow this reduction, because correlations can have long ranges, thus there is often the need to use large simulation cells.
Assuming that we are simulating a supercells having $N$ electrons, with positions $\left\{ {\bf r}_1, \ldots, {\bf r}_N \right\}$, and that ${\bf R}_p$ are the primitive-cell lattice vectors, then the Hamiltonian has to satisfy the following translational symmetry:
\begin{equation}\label{eq:transsymp}
\hat H( {\bf r}_1+{\bf R}_p, \ldots, {\bf r}_i + {\bf R}_p, \ldots, {\bf r}_N + {\bf R}_p) =
\hat H( {\bf r}_1, \ldots, {\bf r}_i, \ldots, {\bf r}_N)  
\end{equation}
which leads to the many-body Bloch condition:
\begin{equation}
\Psi_{{\bf k}_p} ( {\bf r}_1, \ldots, {\bf r}_N) = 
W_{{\bf k}_p} ( {\bf r}_1, \ldots, {\bf r}_N) \;
{\rm e}^{i \, {\bf k}_p \cdot {1\over N} \sum_i {\bf r}_i }
\end{equation}
where $W$ has the same translational symmetry of the Hamiltonian in Eq.~\ref{eq:transsymp},
and the Bloch momentum ${{\bf k}_p}$ can be restricted to the Brillouin zone of the primitive cell.
On top of that, we need to impose periodic boundary conditions (PBC) across the simulations cell,
yielding the following additional translational symmetry for the Hamiltonian:
\begin{equation}\label{eq:transsyms}
\hat H( {\bf r}_1, \ldots, {\bf r}_i + {\bf R}_s, \ldots, {\bf r}_N) =
\hat H( {\bf r}_1, \ldots, {\bf r}_i, \ldots, {\bf r}_N)  
\quad \textrm{for each }i=1,\ldots,N
\end{equation}
where ${\bf R}_s$ are the simulation cell lattice vectors.
Notice that the symmetry in Eq.~\ref{eq:transsyms} is an artifact of the PBC, which leads to the many-body Bloch condition:
\begin{equation}\label{eq:psiBlock_s}
\Psi_{{\bf k}_s} ( {\bf r}_1, \ldots, {\bf r}_N) = 
U_{{\bf k}_s} ( {\bf r}_1, \ldots, {\bf r}_N) \;
{\rm e}^{i \, {\bf k}_s \cdot \sum_i {\bf r}_i }
\end{equation}
where $U$ has the periodicity of the simulation cell for each electron,
and the simulation cell Bloch momentum ${\bf k}_s$, also called {\em twist} vector, can be restricted to the Brillouin zone for the simulation-cell lattice.
The employment of ${\bf k}_s$ vectors other than zero is ofter referred as the application of twisted boundary conditions \cite{Lin:qmctwistavg:pre2001}, and when the expectation value is taken as the average over many vectors ${\bf k}_s$ in the Brillouin zone of a gives simulation cell, we talk about {\em twist averaging} boundary condition (TABC)~\footnote{This process can become tricky for metals, but in this work we are only interested to insulators, for which there are no expected difficulties.}.  
The fluctuations of the QMC energy due to the use of different twists is almost proportional to the corresponding fluctuations observed by performing Hartree-Fock or DFT calculations at the same ${\bf k}_s$ point.  For instance, this can be seen for the specific case of ammonia crystal in the lower panel of Fig.~\ref{fig:twist_NH3}.
Moreover, in insulators it is often possible to find a special point in the Brillouin zone \cite{Rajagopal:1994cc, Rajagopal:1995fk, Dagrada:2016}, such as the Baldereschi point \cite{Baldereschi},  which provides an energy very close to the one obtained by averaging over many ${\bf k}_s$ points in the Brillouin zone. For the ammonia crystal, this aspect is clear in the lower panel in Fig.~\ref{fig:twist_NH3}.

\begin{figure}[hbp]
\includegraphics[width=4.5in]{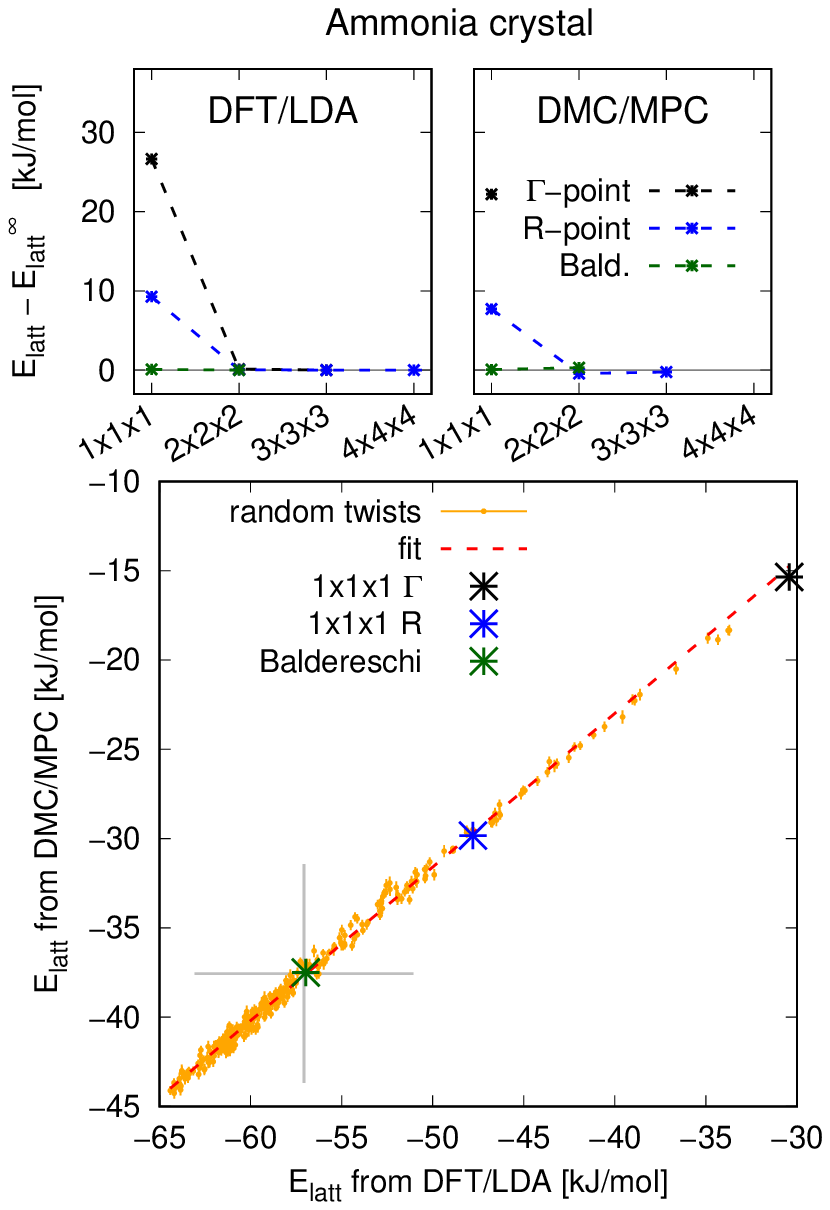}
\caption{
({\em Upper left panel}) Difference between the lattice energy $\Elatt$ obtained  with the a finite simulation cell, with respect to the extrapolated lattice energy $\Elatt^\infty$ for the infinite system, as obtained from a DFT calculation with LDA functional, plotted as a function of the simulation cell size (where $1\times 1\times 1$ indicated the primitive cell). We show the convergence for three special points in the Brillouin zone (notice that the system is a simple cubic with lattice $a\sim$5.13~\AA): the $\Gamma$-point corresponding to ${\bf k}_s = (0,0,0)$, the corner point R  corresponding to ${\bf k}_s = {\pi \over a}(1,1,1)$, and the Baldereschi point ${\bf k}_s = {\pi \over 2 a}(1,1,1)$.
({\em Upper right panel}) Same as the left panel, but for DMC calculations with MPC. Notice that the  $\Gamma$ and R points correspond to twists that make the QMC wave function real (see Eq.~\ref{eq:psiBlock_s}), so we have used the fixed-node approximation. On the other hand, the Baldereschi point makes the QMC wave function complex and we used the fixed-phase approximation.
({\em Lower panel}) scatter plot of the lattice energy obtained using DMC with MPC versus DFT with LDA, using only the primitive cell and 320 random twists. The special twists corresponding to the $\Gamma$, R and Baldereschi points are highlighted. All the twists (excluded the $\Gamma$ and R points) correspond to a complex wave function and thus employ the fixed-phase approximation. The big gray cross indicates $\Elatt^\infty$ for the two approaches.
The red fitting line has an angular coefficient of 0.860$\pm$0.003. 
}
\label{fig:twist_NH3}
\end{figure}

It has to be noticed that in general the QMC wave function of a periodic system is a complex function, according to Eq.~\ref{eq:psiBlock_s}.
Indeed, there are a finite set of points in the Brillouin zone which make $\Psi_{{\bf k}_s}$ a real wave function, such as the $\Gamma$-point ${\bf k}_s = (0,0,0)$, or the corner points, or the center of edge or of face points, but a general $\Psi_{{\bf k}_s}$ yield a complex wave function.
Thus, it has to be remembered that with complex wave functions the fixed-node (FN) approximation \cite{FNApp:Anderson, FNApp:Reynolds} cannot be used and has to be substituted by the fixed-phase (FP) approximation \cite{FPapp}. The two approximations are strictly related but not equivalent. 
%
%
Thus, whenever we use a complex wave function, in the calculation of the lattice energy 
$\Elatt = { E_\textrm{crys} } - E_\textrm{gas}$,
also $E_\textrm{gas}$ needs to be computed with the FP approximation.
In order to do so, we have employed a big simulation cell with the single molecule into the cell, such that the interaction between the molecule and the periodic images is negligible, and performed a DMC/MPC calculation with PBC and at a twist ${\bf k}_s = \epsilon (1,1,1)$, with $\epsilon\sim 10^{-6}$, in order to use the FP approximation also for the reference energy.
In this work we have employed TABC with complex wave functions only for the cases of water and ammonia crystals. In these systems the have that, if there is any difference in $E_\textrm{gas}$ due to the employment of FP versus FN approximation, it is smaller than the statistical accuracy of the results.
However, this cannot be assumed in general.


Earlier we mentioned that DMC can have large FSEs in periodic systems, and that the error can be split into a CIFSE and a IPFSE. 
At this point we can clarify what are the contributions of the CIFSE and IPFSE terms.
Given a finite supercell, the IPFSE is the error that we make by taking only one ${\bf k}_s$ point (for instance, the $\Gamma$-point) instead of performing TABC.
The remaining of the FSE of due to the CIFSE.
%
Fig.~\ref{fig:twist_NH3} gives 
the impression that the IPFSE is the largest part FSE. In fact, it is the other way around: CIFSE is much larger, one to two orders of magnitude, than the IPFSE. 
The feeling in Fig.~\ref{fig:twist_NH3}  
that there is no CIFSE is due to the fact that the DMC calculations are employing the model periodic Coulomb (MPC) model, which is a very effective method recover from the CIFSE. 
If we would have used the standard Ewald summation, the lattice energies for the primitive cell would have been severely overestimated, as it can be seen for the ammonia crystal in Fig.~\ref{fig:dmc_FSE}.
%

\begin{figure}[tb!]
\includegraphics[width=5in]{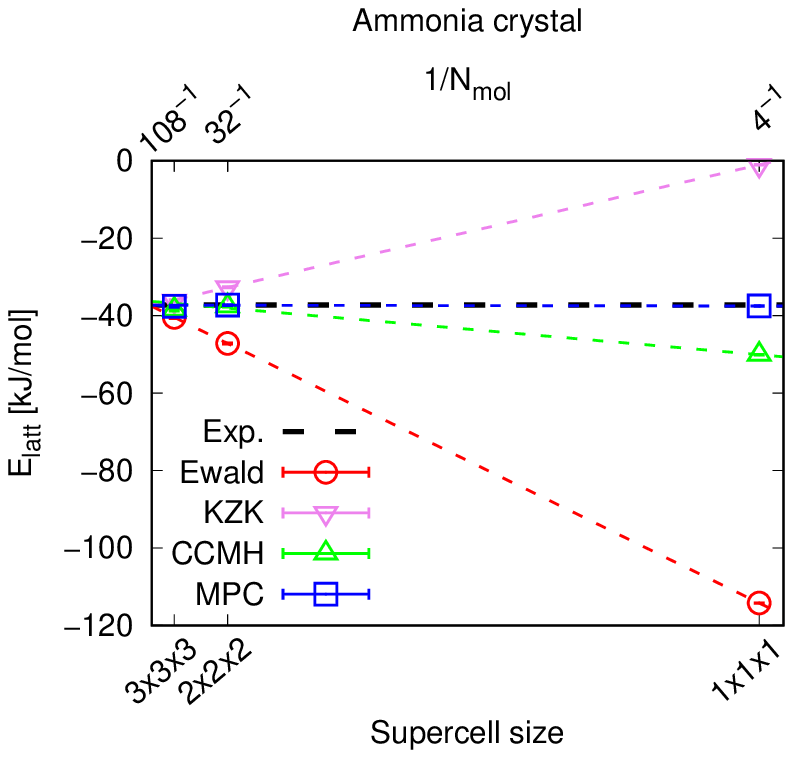}
\caption{\label{fig:dmc_FSE}
$E_\text{latt}$ for the ammonia crystal versus the inverse of the number $N_\textrm{mol}$ of molecules in the supercell.
The reported DMC values are obtained via bare Ewald interaction, KZK correction \cite{KZK:prl2008}, CCMH correction \cite{Chiesa:size_effects:prl2006}, and the MPC interaction \cite{MPC:Fraser1996,MPC:Will1997,MPC:Kent1999}.
Primitive and $2\times 2\times 2$ cells are at the Baldereschi point; $3\times 3\times 3$ cell is at the R-point.
Clearly the MPC interaction yields accurate results for all simulation cells considered.
}
\end{figure}

The source of CIFSE can be explained in different ways (see Refs~\cite{FSEqmc:PRB2008, FSE:Ceperley2016}). 
An intuitive explanation is that the sampling of configurations with electrons that are always in the same relative position in the periodic images of the simulation cell will induce a spurious Coulomb interaction between images (e.g., dipole-dipole).
Large supercells decorrelate the electron configurations and eliminate the issue, however this is not a practical solution because the convergence is slow and the calculation becomes very expensive.
There are at least three complementary ways available to deal with the CIFSE: 
the employment in the simulation of a model periodic Coulomb (MPC) potential
for the long-range interaction \cite{MPC:Fraser1996,MPC:Will1997,MPC:Kent1999} which eliminates the spurious Coulomb interaction between images;
or {\em a posteriori} corrections to standard DMC simulations employing Ewald summation, 
as proposed by \citet{Chiesa:size_effects:prl2006} (CCMH) or by \citet{KZK:prl2008} (KZK).
KZK is computationally the fastest but the least accurate; it employs the usage of a special DFT functional that mimics the FSEs of QMC. It is of usage for an initial estimate of the order of magnitude of the FSE in a specific system, before any DMC calculation.
MPC and CCMH are almost equivalent, as it is shown theoretically in Ref.~\onlinecite{FSEqmc:PRB2008},
and computationally they involve only a little overhead in the DMC simulation. 
In Fig.~\ref{fig:dmc_FSE} we show the behavior of the three correction schemes and of the standard (Ewald) simulation  as a function of the supercell size for the representative case of ammonia crystal.
In the primitive 
cell the Ewald estimate 
is three times as large as the converged value, and also in the $2\times 2\times 2$ the FSE is much larger than the chemical accuracy.
KZK overcorrects, by around 40\%, the FSE.
Both MPC and CCMH provide very effective corrections and the are within 1~kJ/mol already in the $2\times 2\times 2$, whilst in the primitive cell MPC is slightly more accurate than CCMH.
We have reported in the following paragraphs the results from the three corrections for any of the studied molecular crystals. In general, MPC appears as the most accurate method.

\newpage

\section{DMC results for Carbon Dioxide}\label{sec:dmc_CO2} 

Table~\ref{tab:dmc_CO2} reports the lattice energies values obtained using DMC with Ewald, KZK, CCMH or MPC approaches. 
The values show a strong dependence on $\tau$, thus we also performed the extrapolation for $\tau\to 0$.
FSEs for Ewald and KZK are quite large for the $2\times 2\times 2$ simulation cell, and are larger than 1~kJ/mol also for a $3\times 3\times 3$ simulation cell.
On the other hand, both CCMH and MPC appears already very close for a $2\times 2\times 2$ simulation cell.
Concerning the IPFSE, a DFT with LDA calculation shows that the expected error on $2\times 2\times 2$ simulation cell at the $\Gamma$-point is $<0.1$~kJ/mol, thus negligible, and it is even smaller for the $3\times 3\times 3$ simulation cell.

\begin{table}[hbtp]
\caption{
$\Elatt$ of Carbon Dioxide, in kJ/mol, from DMC results using Ewald, KZK, CCMH or MPC.
$\sigma$ indicated the associated stochastic error.
Values in boldface have been used for Table~I of the main paper.
}\label{tab:dmc_CO2}
\begin{center}
\begin{tabular}{   l   c c c c c }
\hline \hline
\multicolumn{6}{c}{\em $2\times 2\times 2$ simulation cell, $\Gamma$-point } \\								\hline			
$\tau$	&	Ewald	&	KZK	&	CCMH	&	MPC	&	$\sigma$	\\
\hline 
0.100	&	-50.5	&	-34.0	&	-39.4	&	-38.9	&	0.3	\\
0.030	&	-42.8	&	-26.3	&	-31.7	&	-31.5	&	0.3	\\
0.010	&	-40.6	&	-24.0	&	-29.5	&	-29.3	&	0.3	\\
0.003	&	-40.5	&	-24.0	&	-29.4	&	-29.2	&	0.4	\\
$\to 0$	&	-39.8	&	-23.3	&	-28.7	&	{\bf -28.5}	&	{\bf 0.4}	\\
\hline \hline
\multicolumn{6}{c}{\em $3\times 3\times 3$ simulation cell, $\Gamma$-point } \\								\hline			
$\tau$	&	Ewald	&	KZK	&	CCMH	&	MPC	&	$\sigma$	\\
\hline			
0.030	&	-34.9	&	-30.0	&	-31.5	&	-31.5	&	0.7	\\
0.003	&	-31.9	&	-27.0	&	-28.5	&	-28.5	&	1.2	\\
$\to 0$	&	-31.6	&	-26.6	&	-28.2	&	{\bf -28.2}	&	{\bf 1.3}	\\
\hline \hline
\end{tabular}
\end{center}
\end{table}

\section{DMC results for Ammonia}\label{sec:dmc_NH3} 

The results for the ammonia crystal are shown in Figs.~1 and 2 of the main paper, and in Figs.~\ref{fig:twist_NH3} and \ref{fig:dmc_FSE}.
The results reported in Table~I of the main paper are obtained with DMC/MPC and a $\tau=$0.03~au;
DMC(lc) is from a $3\times 3\times 3$ simulation cell at the R-point (for which the IPFSE is negligible);
DMC(sc) is from the primitive cell at the Baldereschi point.

\section{DMC results for Benzene}\label{sec:dmc_C6H6} 

The results for the benzene crystal are reported in Table~\ref{tab:dmc_C6H6}, using Ewald, KZK, CCMH or MPC.
The IPFSE associated to each simulation cell is also reported. It has been estimated via DFT/LDA calculation, as the difference
IPFSE$^{\textrm{cell}, {\bf k_s}} = E_\textrm{crys}^{\textrm{cell}, {\bf k_s}} - E_\textrm{crys}^\textrm{BZ}$
between the DMC/LDA energy per molecule $E_\textrm{crys}^{\textrm{cell}, {\bf k_s}}$ calculated at the ${\bf k_s}$-point of the Brillouin zone for the specified simulation cell 
minus the DMC/LDA energy per molecule $E_\textrm{crys}^\textrm{BZ}$ for the fully sampled Brillouin zone (which is equivalent to $E_\textrm{crys}^{\infty}$ in DFT). 
Thus, in order to correct both for the CIFSE and the IPFSE, we have to subtract from the MPC value the IPFSE contribution.

\begin{table}[hbtp]
\caption{
$\Elatt$ of benzene, in kJ/mol, from DMC with $\tau=$0.03~au and using Ewald, KZK, CCMH or MPC.
$\sigma$ indicated the associated stochastic error.
The IPFSE has been estimated via a DFT/LDA calculation, and it should be subtracted to any $\Elatt$ value in the row in order to correct for that, as reported in the last column.
Values in boldface have been used for Table~I of the main paper.
}\label{tab:dmc_C6H6}
\begin{center}
\begin{tabular}{   l   c c c c c c c }
\hline \hline
Cell						& Ewald	& KZK	& CCMH	& MPC	& $\sigma$	& IPFSE	& DMC-FSE	\\
\hline
$1\times 1\times 1 (\Gamma)$	& -143.1	& -18.7	& -52.1	& -52.4	& 0.2	 		& -1.2	& {\bf -51.2$\pm$0.2} \\
$1\times 1\times 1 $(R)		& -134.7	& -10.5	& -43.6	& -44.3	& 0.2	 		&  +7.0 	& { -51.3$\pm$0.2} \\
$2\times 2\times 2 (\Gamma)$	& -63.7	& -48.2	& -51.3	& -52.1	& 0.4	 		&  0.0	& {\bf -52.1$\pm$0.4} \\
$2\times 2\times 2 $(R)		& -64.1	& -48.6	& -51.6	& -52.5	& 0.6	 		&  0.0	& { -52.5$\pm$0.6} \\
$3\times 2\times 3 $(R)		& -58.5	& -51.6	& -53.7	& -54.0	& 4.4	 		&  0.0	& { -54.0$\pm$4.4} \\
\hline \hline
\end{tabular}
\end{center}
\end{table}

In addition to the test with the dimer, in Table~\ref{tab:dimers}, we also tested the size-consistency error.
In order to do that, we considered a benzene dimer, with the two benzene molecules far away, such that their residual interaction is negligible. The size-consistency error, defined as the residual energy per molecule ({\em i.e.}, a half of the energy of the far away dimer minus the the energy of the benzene molecule) is reported in Fig.~\ref{fig:tauErrorC6H6} as a function of the DMC time step, both for the UNR and the ZSGMA algorithms. It is clear that the ZSGMA algorithm is more accurate, and the error is smaller than 0.5~kJ/mol for $\tau<0.03$~au, while in the UNR algorithm a  $\tau<0.003$~au is still too large.
The results reported in Table~I of the main paper are obtained with DMC/MPC and a $\tau=$0.03~au;
DMC(lc) is from a $2\times 2\times 2$ simulation cell at the $\Gamma$-point;
DMC(sc) is from the primitive cell at the $\Gamma$-point.


\begin{figure}[htbp]
\includegraphics[width=4in]{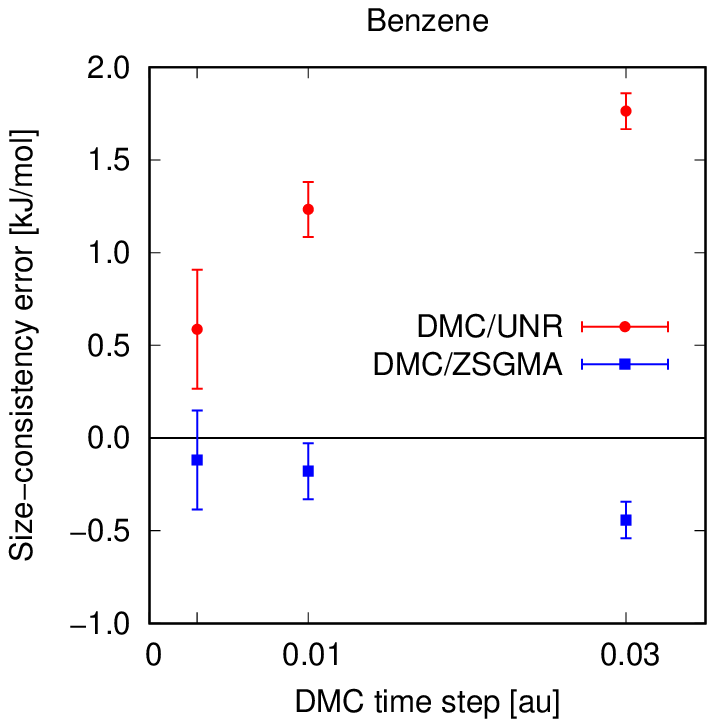}
\caption{
Estimation of DMC time step error in $\Elatt$  for benzene, obtained from benzene dimer 
with the two molecules
far away (distance $>12$~\AA).
}
\label{fig:tauErrorC6H6}
\end{figure}


\section{DMC results for Naphthalene}\label{sec:dmc_C10H8} 

Table~\ref{tab:dmc_C10H8} reports the lattice energies values obtained using DMC with Ewald, KZK, CCMH or MPC approaches. 
The results reported in Table~I of the main paper are obtained with DMC/MPC and a $\tau=$0.03~au;
DMC(lc) is from a $2\times 3\times 2$ simulation cell at the $\Gamma$-point (for which the IPFSE is negligible);
DMC(sc) is from the $1\times 2\times 1$ simulation cell at the $\Gamma$-point, corrected for the IPFSE according to DFT/LDA.

\begin{table}[hbtp]
\caption{
$\Elatt$ of naphthalene, in kJ/mol, from DMC with $\tau=$0.03~au and using Ewald, KZK, CCMH or MPC.
$\sigma$ indicated the associated stochastic error.
The IPFSE has been estimated via a DFT/LDA calculation, and it should be subtracted to any $\Elatt$ value in the row in order to correct for that.
Values in boldface have been used for Table~I of the main paper.
}\label{tab:dmc_C10H8}
\begin{center}
\begin{tabular}{   l   c c c c c c c }
\hline \hline
Cell	&	Ewald	&	KZK	&	CCMH	&	MPC	&	$\sigma$	& IPFSE	& DMC-FSE	\\
\hline
$1\times 2\times 1 (\Gamma)$	&	-170.5	&	-294.5	&	-68.5	&	-68.7	&	0.6	&	+9.3	& {\bf -78.0$\pm$0.6 } \\
$2\times 2\times 2 (\Gamma)$	&	-104.7	&	-137.8	&	na	&	-77.9	&	0.7	&	+0.3	& { -78.2$\pm$0.7 } \\
$2\times 3\times 2 (\Gamma)$	&	-96.6	 	&	-118.8	&	na	&	-78.7	&	0.8	&	0.0	& {\bf -78.8$\pm$0.8 } \\
\hline \hline
\end{tabular}
\end{center}
\end{table}

\section{DMC results for Anthracene}\label{sec:dmc_C14H10} 

Table~\ref{tab:dmc_C14H10} reports the lattice energies values obtained using DMC with Ewald, KZK, CCMH or MPC approaches. 
The results reported in Table~I of the main paper are obtained with DMC/MPC and a $\tau=$0.03~au;
DMC(lc) is from a $2\times 3\times 2$ simulation cell at the R-point, which has a negligible IPFSE;
DMC(sc) is from the $1\times 2\times 1$ simulation cell at the $\Gamma$-point, corrected for the IPFSE according to DFT/LDA.

\begin{table}[hbtp]
\caption{
$\Elatt$ of anthracene, in kJ/mol, from DMC with $\tau=$0.03~au and using Ewald, KZK, CCMH or MPC.
$\sigma$ indicated the associated stochastic error.
The IPFSE has been estimated via a DFT/LDA calculation, and it should be subtracted to any $\Elatt$ value in the row in order to correct for that.
}\label{tab:dmc_C14H10}
\begin{center}
\begin{tabular}{   l   c c c c c c c }
\hline \hline
Cell					&	Ewald	&	KZK	&	CCMH	&	MPC	&	$\sigma$	& IPFSE	& DMC-FSE	\\
\hline 
$1\times 2\times 1(\Gamma)$	& -213.0	& -77.7	&	-151.6	& -105.9	&	1.0	& -2.0	& {\bf -103.9$\pm$1.0 } \\
$2\times 2\times 1(\Gamma)$	& -163.0	& -95.3	&	-120.3	& -103.9	&	1.4	& +2.0	& { -105.9$\pm$1.4 } \\
$2\times 3\times 2$(R)		& -123.6	& -101.1	&	-104.3	& -105.5	&	1.7	& 0.0	 	& {\bf -105.5$\pm$1.7 } \\
\hline \hline
\end{tabular}
\end{center}
\end{table}

\section{ DMC results for the ice polymorphs }\label{sec:ice_dmc}
Fig.~\ref{fig:iceIh_dmc} shows the value of $\Elatt$ calculated via DMC with MPC, using either the primitive or the $2\times 2\times 2$ simulation cell, as a function of the time step.
It shows that the employment of MPC allows to obtain accurate results already with the primitive cell.
Moreover,  the time step dependance is quite small, smaller that 1~kJ/mol for $\tau<0.02$~au.
However, we have noticed that the time step dependence can be different for the different ice polymorphs, as shown in Fig.~\ref{fig:ice_dmc}.
In particular, the time step dependence is larger for ice VIII.
Thus, both the DMC(sc) and the DMC(lc) results reported in the main paper, Table~I, have been obtained for a time step of 0.003~au. 
The DMC(sc) results comes for the primitive cell of each ice polymorphs, and performing  a TABC of 4 twists (those corresponding to a $2\times 2\times 2$ Monkhorst-Pack grid, with grid displaced by half a grid step in each direction) for the I$_h$ and II polymorphs, and 9 twists (those corresponding to a $3\times 3\times 2$ Monkhorst-Pack grid, with grid displaced by half a grid step in each direction) for ice VIII.
The DMC(lc) results comes from a $2\times 2\times 2$ simulation cell at $\Gamma$-point for ice I$_h$, 
a $2\times 2\times 2$ simulation cell at R-point for ice II, and 
a $3\times 3\times 2$ simulation cell at $\Gamma$-point for ice VIII.
All the DMC results use MPC.
Considering the stochastic error, the results here reported are in quite good agreement with those in Ref.~\cite{santra_hydrogen_2011}, which were obtained with the size-inconsistent UNR algorithm but with a small time step.

\begin{figure}[hbp]
\includegraphics[width=4in]{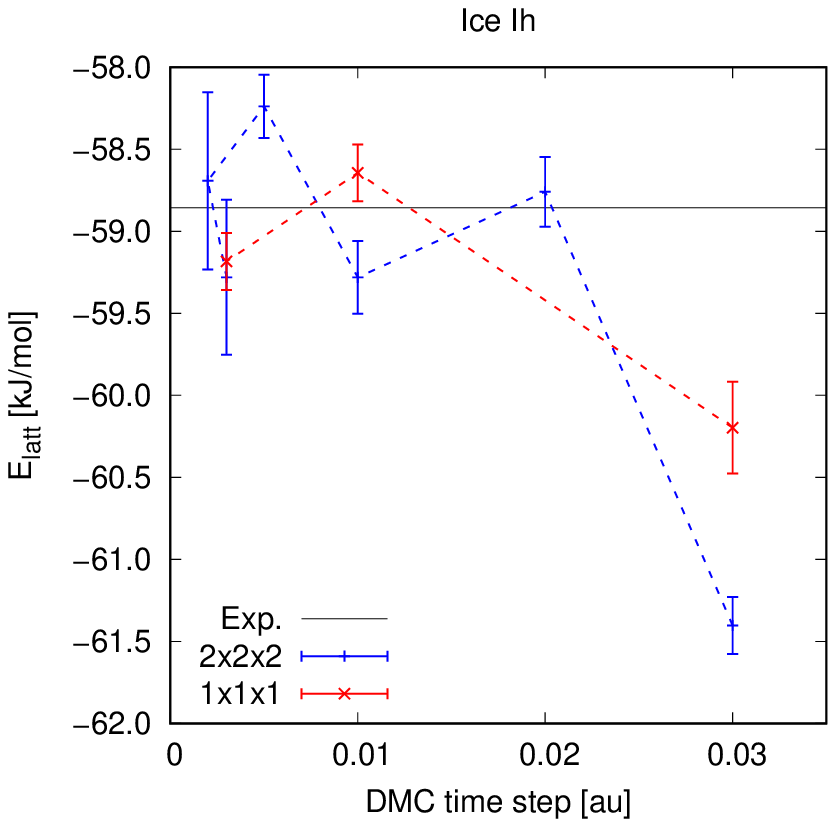}
\caption{
Lattice energy for ice I$_h$, by DMC with MPC, as a function of the DMC time step $\tau$.
We show results for a $2\times 2\times 2$ simulation cell at the $\Gamma$-point, with the results for the primitive cell ($1\times 1\times 1$) for which we performed a TABC of 4 twists (those corresponding to a $2\times 2\times 2$ Monkhorst-Pack grid).
}
\label{fig:iceIh_dmc}
\end{figure}

\begin{figure}[hbtp]
\includegraphics[width=4in]{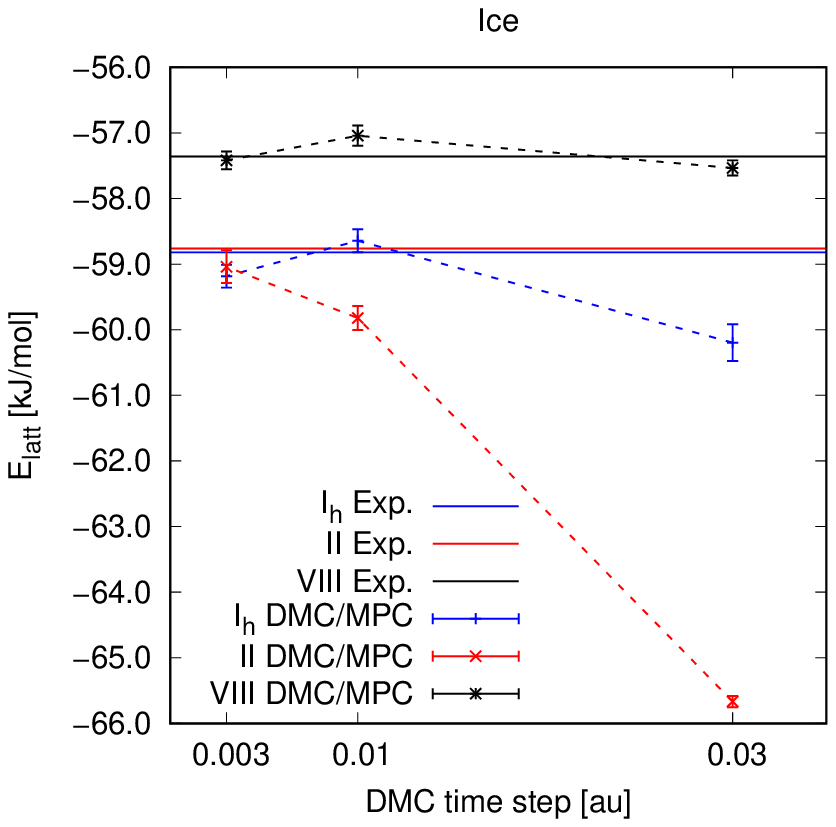}
\caption{
Lattice energy for ice I$_h$, II and VIII, by DMC with MPC, as a function of the DMC time step $\tau$.
We show results for the primitive cell for which we performed a TABC of 4 twists (those corresponding to a $2\times 2\times 2$ Monkhorst-Pack grid, with grid displaced by half a grid step in each direction) for the I$_h$ and II polymorphs, and 9 twists (those corresponding to a $3\times 3\times 2$ Monkhorst-Pack grid, with grid displaced by half a grid step in each direction) for ice VIII.
}
\label{fig:ice_dmc}
\end{figure}

\newpage

\section{ DMC scaling and computational cost }\label{sec:DMCcost}

\subsection{DMC scaling}

Standard DMC is usually claimed to scale as the third power of the system. 
At one level this is correct, but a more detailed analysis is needed if we want to understand the benefits from having reliable results from a small simulation cell.

The total CPU time $T_\textrm{sim}$ for a DMC simulation is the summation 
\begin{equation}\label{eq:TsimALL}
T_\textrm{sim} = T_\textrm{sim}^\textrm{eq} + T_\textrm{sim}^\textrm{samp}
\end{equation}
of the time $T_\textrm{sim}^\textrm{eq}$ spent for the equilibration and the time $T_\textrm{sim}^\textrm{samp}$ for the statistical sampling.
The equilibration time is can be estimated as follows:
\begin{equation}\label{eq:Tsimeq}
T_\textrm{sim}^\textrm{eq} = T_\textrm{step} N_\textrm{steps}^\textrm{eq} N_w \, ,
\end{equation}
where $T_\textrm{step}$ is the CPU time needed for a single DMC step, 
$N_\textrm{steps}^\textrm{eq}$ is the number of DMC equilibration steps to be performed, and
$N_w$ is the (average) number of walkers in the DMC branching process.
 It is clear that $N_\textrm{steps}^\textrm{eq}$ depends on the DMC time step $\tau$, indeed it can be spotted that 
$t^\textrm{eq} \equiv \tau \cdot N_\textrm{steps}^\textrm{eq}$, which is the imaginary time needed in the DMC propagation to project out the excited stated from the initial trial wave function, is a quantity that remains almost constant for different choices of $\tau$.
Both $T_\textrm{step}$ and $t^\textrm{eq}$ are system dependent. 
In particular, 
the value of $T_\textrm{step}$ depends strongly on the size of the simulated system, as we will discuss later, while
the value of $t^\textrm{eq}$ depends on the atomic species, bonds formed, and other details, but it is essentially unaffected by the number of molecules in the simulated system. 
For all the molecular crystals considered here we have that $t^\textrm{eq} \sim 6$~au.
The value of $N_w$ is more related to the available resources: if the simulation is executed using a large number of processors running in parallel, $N_w$ needs to be large enough to allow an efficient parallelization ({\em i.e.}, $N_w$ is roughly proportional to the number of processors used in parallel) and to produce negligible population bias (usually it is enough to use more than a few thousands walkers). In our simulations we have used values of $N_w$ of the order of $10^4$.
In many DMC implementations $N_w$ is not fixed but can fluctuate, however these fluctuations are relatively quite small if $N_w$ is large enough, thus for our purposes the average number of walkers is the value we need here.
The sampling time is instead given by:
\begin{equation}\label{eq:Tsimsamp}
T_\textrm{sim}^\textrm{samp} 
= T_\textrm{step} 
N_\textrm{steps}^\textrm{samp}
N_w 
\, , \qquad \textrm{with} \quad 
N_\textrm{steps}^\textrm{samp} =
N_\textrm{steps}^\textrm{ac}
{\sigma_\textrm{sys}^2 \over N_w \sigma_\textrm{target}^2 } 
\, ,
\end{equation}
the number of sampling steps, 
where 
$\sigma_\textrm{sys}^2$ is the variance of DMC local energies generated in the DMC process,  
$\sigma_\textrm{target}$ is our target stochastic error of the DMC energy evaluation,
and 
$N_\textrm{steps}^\textrm{ac}$ is the autocorrelation time, given in terms of number of steps, of the local energies generated in the DMC process.
As for $N_\textrm{steps}^\textrm{eq}$, also $N_\textrm{steps}^\textrm{ac}$ depends on $\tau$, and it can be seen that the autocorrelation time in atomic units 
$t^\textrm{ac} \equiv \tau \cdot N_\textrm{steps}^\textrm{ac}$
is almost unaffected by the choice of $\tau$.
Both $t^\textrm{ac}$ and $\sigma_\textrm{sys}^2$ are system dependent.
In particular, 
the variance $\sigma_\textrm{sys}^2$ depends strongly on the size of the system, while
the autocorrelation time $t^\textrm{ac}$, similarly to the equilibration time $t^\textrm{eq}$, depends on many details of the system but it is mostly unaffected by the number of molecules in the simulation.
In the molecular crystals considered in this work, $t^\textrm{ac}$ ranges from around 0.3~au to 1~au, depending of the molecules involved. 
By including Eqs.~\ref{eq:Tsimeq} and \ref{eq:Tsimsamp} in Eq.~\ref{eq:TsimALL} we obtain:
\begin{equation}\label{eq:TsimALL2}
T_\textrm{sim} = T_\textrm{step} { t^\textrm{ac} \over \tau}
\left(
{t^\textrm{eq} \over t^\textrm{ac}} \, N_w +
{\sigma_\textrm{sys}^2 \over \sigma_\textrm{target}^2 }
 \right) \, .
\end{equation}
Notice the dependance on the DMC time step: $T_\textrm{sim} \propto \tau^{-1}$.

In order to have the scaling of DMC with system size, we have to see how the right hand side of Eq.~\ref{eq:TsimALL2} depends on the number $N_e$ of electrons in the DMC simulation.
As anticipated, the only quantities that we need to consider are: the variance $\sigma_\textrm{sys}^2$ of the DMC local energy, and the time $T_\textrm{step}$ for a single DMC step.
It can be shown (see \citet{foulkes01}) that
$
T_\textrm{step} \propto {N_e}^2 + \epsilon {N_e}^3
$,
where $\epsilon\sim 10^{-4}$.
Moreover, for $N_e$ large enough (say, more than a few tens of electrons)
we have that
$
\sigma_\textrm{sys}^2 \sim N_e  \sigma_{1e}^2
$,
where $\sigma_{1e}^2$ is the variance per electron.
Including these relations into Eq.~\ref{eq:TsimALL2}, and using 
$t^\textrm{eq} \sim c \cdot t^\textrm{ac}$ ($c\sim$10 in our systems),
it yields:
\begin{equation}\label{eq:TsimALL3}
T_\textrm{sim} \propto 
({N_e}^2 + \epsilon {N_e}^3) \,
\left(
c \cdot N_w +
N_e { \sigma_{1e}^2 \over \sigma_\textrm{target}^2 }
 \right) \, .
\end{equation}
Into the parenthesis, the term $N_w$ is the contribution due to the equilibration time, and $N_e { \sigma_{1e}^2 / \sigma_\textrm{target}^2 }$ is due to the statistical sampling.

In order to simplify further we have to go into some concrete examples.
Let us consider the cases of benzene molecule ($N_e$=30), naphthalene molecule ($N_e$=48) and anthracene molecule ($N_e$=66).
With the setup used in this work, they all have $\sigma_{1e}^2 \sim 0.02$~Ha$^2$.
If we aim at a target precision of $\sigma_\textrm{target}=10^{-4}$~Ha, equivalent to 0.26~kJ/mol, then the ratio 
${ \sigma_{1e}^2 / \sigma_\textrm{target}^2 } \sim 2\cdot 10^6$.
So, it is clear that with $N_w \sim 10^{4}$ the impact of the equilibration time is negligible (it is around two order of magnitude smaller than the sampling time).
Moreover, as long as $N_e \ll \epsilon^{-1} \sim 10^4$,
the actual scaling is:
$$
T_\textrm{sim} \propto {N_e}^3 \, ,
$$
in agreement with that it is usually claimed for DMC.
For instance, this relation works well if we compare the computational cost for the molecules of benzene, naphthalene and anthracene, and also with their dimers or trimers.
However, when 
we simulate systems for a thousand of electrons or more (an unfeasible size until the very recent years), 
the dependance of $T_\textrm{step}$ on $\epsilon {N_e}^3$ cannot be neglected anymore and the scaling is $T_\textrm{sim} \propto {N_e}^4$.

\subsection{ Cost of DMC lattice energy evaluations with increasing supercell size}

What is the computational cost for DMC evaluations of the lattice energy for increasing supercells?
The important aspect to notice here is that we have a target precision on $\Elatt$, which is not the total energy of the system!
Indeed, as defined in the main paper, 
$
\Elatt =  { E_\textrm{crys}^\textrm{cell} / N_\textrm{mol}} - E_\textrm{gas}
$,
with $E_\textrm{crys}^\textrm{cell}$ being the total energy of the simulated cell having $N_\textrm{mol}$ molecules. 
What we want is a constant precision on our evaluation of $\Elatt$.
Of course both $E_\textrm{crys}^\textrm{cell} / N_\textrm{mol}$ and $E_\textrm{gas}$ contribute to the stochastic error of $\Elatt$, however, given the superlinear scaling of DMC, it is clear that the computationally expensive part of the calculation is $E_\textrm{crys}^\textrm{cell}$ (because $E_\textrm{gas}$ is the energy for a single molecule system, for which it is easier to reduce the stochastic error).
Thus, our aim is actually to have a constant stochastic error $\sigma_\textrm{target,1mol}$ on $E_\textrm{crys}^\textrm{cell} / N_\textrm{mol}$ even when we consider larger and larger supercells, that is, larger and larger $N_\textrm{mol}$. 
Keeping the notation introduced in the previous section, this implies that we have a target stochastic error $\sigma_\textrm{target}$ on the total energy $E_\textrm{crys}^\textrm{cell}$ of:
\begin{equation}
\sigma_\textrm{target} = N_\textrm{mol} \, \sigma_\textrm{target,1mol} \, .
\end{equation}
Moreover, if $M$ is the number of electrons per molecule, we have 
$N_e = M \cdot N_\textrm{mol}$,
and 
\begin{equation}
\sigma_\textrm{sys}^2 = N_\textrm{mol} \, \sigma_\textrm{sys,1mol}^2 \, ,
\end{equation}
where the variance per molecule $\sigma_\textrm{sys,1mol}^2 = M \cdot \sigma_{1e}^2$ is unaffected by $N_\textrm{mol}$.
Including these relations into Eq.~\ref{eq:TsimALL2} we obtain:
\begin{equation}\label{eq:TsimMOL}
T_\textrm{sim} = T_\textrm{step} { t^\textrm{ac} \over \tau}
\left(
{t^\textrm{eq} \over t^\textrm{ac}} \, N_w +
{ 1 \over N_\textrm{mol} } \cdot 
{\sigma_\textrm{sys,1mol}^2 \over \sigma_\textrm{target,1mol}^2 }
 \right) \, ,
\end{equation}
which clearly indicated that in large supercell the equilibration time becomes the computationally more expensive part of the calculation.
Indeed, if the go to see the scaling of $T_\textrm{sim}$ with $N_\textrm{mol}$,
using Eq.~\ref{eq:TsimALL3} and the above relations, we obtain:
\begin{equation}\label{eq:TsimMOL2}
T_\textrm{sim} \propto 
( N_\textrm{mol} + \tilde\epsilon  N_\textrm{mol}^2 ) \,
\left( c \, N_w\,  N_\textrm{mol} + {\sigma_\textrm{sys,1mol}^2 \over \sigma_\textrm{target,1mol}^2 } \right) \, ,
\end{equation}
where $\tilde\epsilon = M\cdot \epsilon$.
Optimistically, one would hope that the computational cost 
is linear with $N_\textrm{mol}$,  as would be if both the $\tilde\epsilon$  and the equilibration contributions would be negligible. Unfortunately, they are not!
When we try to reduce the FSE by considering large supercells
\footnote{For instance, the primitive cell of the ammonia crystal has 4 molecules, each having 8 electrons, for a total of 32, but if FSE is large we have to take the $2\times 2\times 2$ supercell, having $4\times 2^3 = 32$ molecules, 256 electrons; if FSE are still non negligible we need the $3\times 3\times 3$ supercell, having 108 molecules, 864 electrons, and so on. The situation is even worse with large molecules, where already a $2\times 2\times 2$ supercell may have over a thousand electrons.},
when $N_\textrm{mol}$ grows very quickly 
and those terms as to be considered pretty soon. 
For instance, if we simulate a $l\times m\times n$ supercell, the computational cost will roughly be 
proportional to $(l\cdot m\cdot n)^\alpha$, with $\alpha$ being almost 1 for small supercells and becoming 3 as the supercell increases.

The properties of the primitive cell for the molecular crystals considered in this work are given in Table~\ref{tab:crysprop}.
The computational cost for the different choices of the simulation cells are reported in Table~\ref{tab:Supercells}.


%
\begin{table*}[hp!]
\caption{
Properties of the {\em primitive} cell (${1\times 1\times 1}$) in the studied molecular crystals.
$M$ is the number of valence electrons per molecule, $N_\textrm{mol}^{1\times 1\times 1}$ the number of molecules in the primitive cell, $N_e^{1\times 1\times 1}$ is the total number of valence electrons in the primitive cell (clearly, $N_e^{1\times 1\times 1}=M\cdot N_\textrm{mol}^{1\times 1\times 1}$),  $V^{1\times 1\times 1}$ is the volume of the primitive cell, $\sigma_\textrm{sys,1mol}^2$ is the local energy variance per molecule, $\tau$ is the DMC time step,  $T_\textrm{step}$ (in seconds) is the computational time for a single DMC step on the machine used for these simulations, $N_\textrm{steps}^\textrm{eq}$ and $N_\textrm{steps}^\textrm{ac}$ are the number of steps of equilibration and of sampling for a target precision $\sigma_\textrm{target,1mol}$ of 0.5 and of 0.1~kJ/mol  on the total energy estimation, assuming $N_w=5120$.\footnote{ 
In the calculation of $N_\textrm{steps}^\textrm{eq}$ we have used $t^\textrm{eq}$=6~au for all the systems, and for $N_\textrm{steps}^\textrm{ac}$ we have used $t^\textrm{ac}$=1~au for benzene, naphthalene and anthracene, and 0.3~au for the remaining systems. } 
The volume, variance and $\tau$ are given in atomic units.
}\label{tab:crysprop}
\begin{center}
\begin{tabular}{   l   c c c  c c c  c c  | r | r | }
\cline{10-11}
&&&&&&&&&\multicolumn{2}{ c |}{ $N_\textrm{steps}^\textrm{ac}$ for $\sigma_\textrm{target,1mol}$ = } \\
\cline{10-11}
System & $M$ & $N_\textrm{mol}^{1\times 1\times 1}$ & $N_e^{1\times 1\times 1}$ & $V^{1\times 1\times 1}$ & $\sigma_\textrm{sys,1mol}^2$ & $\tau$ & $T_\textrm{step}$ [sec] & $N_\textrm{steps}^\textrm{eq}$ &  0.5~kJ/mol & 0.1~kJ/mol \\ 
\hline \hline
\ce{CO2} & 16 & 4 & 64 & 1200 & 0.62 & 0.003 & 0.041 & 2000 & 83476 & 2086891 \\ 
\ce{NH3} & 8 & 4 & 32 & 911 & 0.18 & 0.03 & 0.010 & 200 & 2423 & 60587 \\ 
\ce{C6H6} & 30 & 4 & 120 & 3199 & 0.54 & 0.03 & 0.146 & 200 & 24235 & 605872 \\ 
\ce{C10H8} & 48 & 2 & 96 & 2300 & 0.94 & 0.03 & 0.093 & 200 & 84373 & 2109331 \\ 
\ce{C14H10} & 66 & 2 & 132 & 3080 & 1.3 & 0.03 & 0.177 & 200 & 116686 & 2917160 \\ 
\ce{ice-Ih} & 8 & 12 & 96 & 2583 & 0.33 & 0.003 & 0.093 & 2000 & 14810 & 370255 \\ 
\ce{ice-II} & 8 & 12 & 96 & 2023 & 0.33 & 0.003 & 0.093 & 2000 & 14810 & 370255 \\ 
\ce{ice-VIII} & 8 & 8 & 64 & 1120 & 0.33 & 0.003 & 0.041 & 2000 & 22215 & 555382 \\ 
\cline{10-11}
\end{tabular}
\end{center}
\end{table*}
\begin{table*}[p]
\caption{
Evaluation of the computational cost $T_\textrm{sim}$, given in CPU-hours, for DMC energy evaluations for different simulation cells, having a target precision of  0.5 and 0.1~kJ/mol on the total energy and assuming to use 5,120 walkers.
The evaluation is based on the discussion in Sec.~\ref{sec:DMCcost} and on the values reported in  Table~\ref{tab:crysprop} and this table.
In particular, the time $T_\textrm{step}$ for each DMC time step, which is a quantity that can change by a factor two or even more depending on the used architecture\footnote{In the table we have calculated $T_\textrm{step}$ taking as reference the {\bf Rhea} machine available at the Oak Ridge Leadership Computing Facility, for which we have noticed that $T_\textrm{step} \sim 10^{-5} \cdot ({N_e}^2 + 10^{-4} {N_e}^3)$. The {\bf ARCHER} machine, that we have used for some calculations, has processors almost 2 times faster, yielding a $T_\textrm{step}$ reduced by 40~\%. } and additional setups in the DMC calculation.\footnote{For instance, in these calculations we are performing both the Ewald and MPC evaluations for each DMC step, and we are evaluating the structure factor. If we would be performing only Ewald or only MPC, without evaluating the structure factor, $T_\textrm{step}$ would be reduced by $\sim$50~\% or more. }
The fraction of time spent for the equilibration and for the sampling are proportional of the number of steps to be performed, reported in this table.
}\label{tab:Supercells}
\begin{center}
\begin{tabular}{  | l   c c c c c   | r r | r r | }
\hline 
&&&&&&
\multicolumn{2}{ c |}{ $\sigma_\textrm{target,1mol}$ = 0.5 kJ/mol } &
\multicolumn{2}{ c |}{ $\sigma_\textrm{target,1mol}$ = 0.1 kJ/mol } \\
System & Supercell & $N_e$ & $\tau$ & $T_\textrm{step}$ [sec] & $N_\textrm{steps}^\textrm{eq}$ & 
$N_\textrm{steps}^\textrm{ac}$ & $T_\textrm{sim}$ [CPUh] & 
$N_\textrm{steps}^\textrm{ac}$ & $T_\textrm{sim}$ [CPUh] \\ 
\hline \hline 
\ce{CO2} & mol & 16 & 0.003 & 0.003 & 2000 & 333903 & 1225 & 8347565 & 30448 \\ 
\ce{CO2} & 1$\times$1$\times$1 & 64 & 0.003 & 0.041 & 2000 & 83476 & 5011 & 2086891 & 122466 \\ 
\ce{CO2} & 2$\times$2$\times$2 & 512 & 0.003 & 2.76 & 2000 & 10434 & 48733 & 260861 & 1030195 \\ 
\ce{CO2} & 3$\times$3$\times$3 & 1728 & 0.003 & 35 & 2000 & 3092 & 253595 & 77292 & 3949206 \\ 
\hline 
\ce{NH3} & mol & 8 & 0.03 & 0.001 & 200 & 9155 & 9 & 228885 & 209 \\ 
\ce{NH3} & 1$\times$1$\times$1 & 32 & 0.03 & 0.01 & 200 & 2423 & 38 & 60587 & 888 \\ 
\ce{NH3} & 2$\times$2$\times$2 & 256 & 0.03 & 0.672 & 200 & 303 & 481 & 7573 & 7431 \\ 
\ce{NH3} & 3$\times$3$\times$3 & 864 & 0.03 & 8.11 & 200 & 90 & 3342 & 2244 & 28189 \\ 
\hline 
\ce{C6H6} & mol & 30 & 0.03 & 0.009 & 200 & 96939 & 1247 & 2423487 & 31116 \\ 
\ce{C6H6} & 1$\times$1$\times$1 & 120 & 0.03 & 0.146 & 200 & 24235 & 5064 & 605872 & 125613 \\ 
\ce{C6H6} & 2$\times$2$\times$2 & 960 & 0.03 & 10.1 & 200 & 3029 & 46391 & 75734 & 1090829 \\ 
\ce{C6H6} & 3$\times$2$\times$3 & 2160 & 0.03 & 56.7 & 200 & 1346 & 124774 & 33660 & 2732056 \\ 
\hline 
\ce{C10H8} & mol & 48 & 0.03 & 0.023 & 200 & 163361 & 5385 & 4084024 & 134474 \\ 
\ce{C10H8} & 1$\times$1$\times$1 & 96 & 0.03 & 0.093 & 200 & 84373 & 11192 & 2109331 & 279155 \\ 
\ce{C10H8} & 1$\times$2$\times$1 & 192 & 0.03 & 0.376 & 200 & 42187 & 22649 & 1054666 & 563672 \\ 
\ce{C10H8} & 1$\times$2$\times$2 & 384 & 0.03 & 1.53 & 200 & 21093 & 46370 & 527333 & 1148799 \\ 
\ce{C10H8} & 2$\times$2$\times$2 & 768 & 0.03 & 6.35 & 200 & 10547 & 97073 & 263666 & 2383466 \\ 
\ce{C10H8} & 2$\times$3$\times$2 & 1152 & 0.03 & 14.8 & 200 & 7031 & 152205 & 175778 & 3704099 \\ 
\hline 
\ce{C14H10} & mol & 66 & 0.03 & 0.044 & 200 & 224397 & 14006 & 5609923 & 349852 \\ 
\ce{C14H10} & 1$\times$1$\times$1 & 132 & 0.03 & 0.177 & 200 & 116686 & 29348 & 2917160 & 732488 \\ 
\ce{C14H10} & 1$\times$2$\times$1 & 264 & 0.03 & 0.715 & 200 & 58343 & 59562 & 1458580 & 1484164 \\ 
\ce{C14H10} & 2$\times$2$\times$1 & 528 & 0.03 & 2.94 & 200 & 29172 & 122605 & 729290 & 3045093 \\ 
\ce{C14H10} & 2$\times$2$\times$2 & 1056 & 0.03 & 12.3 & 200 & 14586 & 259262 & 364645 & 6397374 \\ 
\ce{C14H10} & 2$\times$3$\times$2 & 1584 & 0.03 & 29.1 & 200 & 9724 & 410220 & 243097 & 10057094 \\ 
\hline 
\ce{H2O} & mol & 8 & 0.003 & 0.001 & 2000 & 166951 & 154 & 4173783 & 3804 \\ 
\ce{ice-Ih} & 1$\times$1$\times$1 & 96 & 0.003 & 0.093 & 2000 & 14810 & 2224 & 370255 & 49261 \\ 
\ce{ice-Ih} & 2$\times$2$\times$2 & 768 & 0.003 & 6.35 & 2000 & 1851 & 34788 & 46282 & 436123 \\ 
\ce{ice-II} & 1$\times$1$\times$1 & 96 & 0.003 & 0.093 & 2000 & 14810 & 2224 & 370255 & 49261 \\ 
\ce{ice-II} & 2$\times$2$\times$2 & 768 & 0.003 & 6.35 & 2000 & 1851 & 34788 & 46282 & 436123 \\ 
\ce{ice-VIII} & 1$\times$1$\times$1 & 64 & 0.003 & 0.041 & 2000 & 22215 & 1420 & 555382 & 32678 \\ 
\ce{ice-VIII} & 2$\times$2$\times$2 & 512 & 0.003 & 2.76 & 2000 & 2777 & 18721 & 69423 & 279917 \\ 
\ce{ice-VIII} & 3$\times$3$\times$2 & 1152 & 0.003 & 14.8 & 2000 & 1234 & 68075 & 30855 & 691546 \\ 
\hline  
\end{tabular}
\end{center}
\end{table*}

\newpage

\section{RPA calculations}\label{sec:RPA}

To perform the RPA calculations we used the Vienna ab-initio simulation package
(VASP)~\cite{kresse_efficient_1996,kresse_ultrasoft_1999} and the recently developed
algorithm with quartic scaling~\cite{kaltak2014rpa2}.
The GWSE corrections were obtained as described in Ref.~\onlinecite{klimes2015}.
VASP uses plane-waves as a basis set and in all the calculations the plane-wave 
basis-set cut-off for the response function ({\tt ENCUTGW} tag in VASP)
was set to one half of the orbital basis-set cut-off ({\tt ENCUT} tag).
The PBE functional was used to provide input orbitals and energies for the RPA 
calculations~\cite{perdew_generalized_1996}
and the number of frequency and time points was set to 8~\cite{kaltak2014rpa1}.

To obtain the lattice energies we used the same approach as in Ref.~\onlinecite{Jiri:2016}.
The exact-exchange energies (EXX) are not computationally demanding and we obtained a k-point and 
volume converged energies, for the solid and molecule, respectively, using a basis-set cut-off of 1100~eV.
A cell of 20$\times$21$\times$22~\AA\ was used for the molecule
and a k-point sets of 4$\times$4$\times$4 for ice Ih and ice II and 5$\times$5$\times$5 for ice VIII. 
The Coulomb cut-off technique (tag {\tt} HFRCUT in VASP) was used to remove the slow convergence
of the Fock energy with the volume and we checked that the EXX energies agree with values
obtained by extrapolation to within 0.1~kJ/mol~\cite{rozzi2006}.

For the RPA and GWSE calculations, we obtained volume and k-point converged energies for 
a set of basis-set cut-offs starting at 600~eV and increasing in steps of 100~eV.
The largest cut-off was 800~eV for ice~Ih and ice~II, and 900~eV for ice~VIII and the isolated molecule.
In the case of the molecule, the RPA energies were obtained for a set of cells with increasing
volume and extrapolated to infinite volume assuming $1/V^2$ dependence of the energy on the volume $V$.
The GWSE energy was obtained in a similar way, but assuming a $1/V$ dependence of the energy.
The largest cell size that we used was 12$\times$13$\times$14~\AA\ for a plane-wave cut-off of 600~eV.
For solids, the RPA energy exhibits a fast convergence, it was sufficient to use a 3$\times$3$\times$3 k-point grid 
for ice Ih and ice II  and a 4$\times$4$\times$4 grid for ice~VIII.
For these k-point grids, the calculations were only feasible for the basis-set cut-off of 600~eV in the case
of ice Ih and ice II and for a cut-off of 700~eV for ice~VIII.
The RPA energies for larger cut-offs were approximated by adding a k-point correction
to an energy obtained with the desired cut-off using a sparse k-point grid.
The k-point correction is the difference between the RPA energies obtained using 
the dense and sparse k-point grids and such a basis-set for which both calculations are possible.
For the GWSE energies of solids we used extrapolations to obtain the k-point converged values. 
We assumed $1/N_k$ dependence of the energy on the number of k-points $N_k$.

For hydrogen bonded systems the internal basis-set extrapolation procedure of the RPA energies
developed and implemented~\cite{RPA:Kresse2008} in the VASP code works well.
For example, the binding energies of ice~VIII obtained with the RPA energies extrapolated 
by VASP change by less than 0.04~kJ/mol per water molecule 
when the basis-set cut-off is increased from 700~eV to 900~eV.
Therefore, the RPA energies produced by the internal extrapolation were used.
There is currently no extrapolation implemented for the GWSE energies.
Therefore, to obtain basis-set converged values, we extrapolated the energies assuming 
an $ENCUT^{-3/2}$ dependence on the basis-set size, in the same way as done in our previous work~\cite{Jiri:2016}.
 
Finally, we checked how the use of more accurate, ``hard," PAW data-sets affects the results.
To this end, we used a 2$\times$2$\times$2 k-point grid and a cell of dimensions 9$\times$10$\times$11~\AA\
to obtain converged EXX, RPA, and GWSE energies with normal and hard PAW data-sets.
We find that for water ice the use of the hard potentials affects the lattice energies only little.
As observed previously, the EXX and RPA corrections almost cancel each other, so that the
lattice energy is decreased only by about 0.8, 0.5, and 0.3~kJ/mol for ice Ih, II, and VIII, respectively.
The GWSE contribution gives more binding with the hard potentials.
For ice Ih the EXX, RPA, and GWSE corrections cancel so that the lattice energy is identical,
to within the specified precision, for the normal and hard potentials.
For ice II and VIII, the corrections for hard potentials increase the RPA+GWSE energies by 
approximately 0.4 and 0.7~kJ/mol in magnitude, respectively.

Table~\ref{tab:rpa_CPU} gives the computer time required to obtain the lattice energies, 
separately for RPA and for the RPA+GWSE scheme.
The data contain contributions from calculations at different basis set cut-offs, different
k-point sets or different cell volumes, as well as the time required to obtain the correction
with `hard' PAW potentials, see~\cite{Jiri:2016} for details. 
As one can see, for the systems presented here, the time required to obtain the energy of the
isolated molecule is similar to the time needed to obtain the energy of the solid.
This is caused by the fact that the current implementation treats vacuum in the same was as 
regions with electron density.
Both RPA and the GWSE correction scale mostly linearly with the number of k-points and we observe
this behaviour.
For example, the CPU time required to obtain RPA energies for basis set cut-offs 500 to 1000~eV
is 140~CPUhours for 2$\times$2$\times$2 k-points and 466~CPUhours for 3$\times$3$\times$3 k-points.
That is, the latter calculations require 3.3 times more CPU time, in agreement with the 
3.4-fold increase in the number of k-points.
The ratio is 2.7 for the GWSE correction.
The dominant parts of the implementations of the RPA and GWSE methods scale as $O(N^2)$ or $O(N^3)$.
For example, parts with $O(N^2)$ scaling are the construction of the response function from Green's functions
or the evaluation of the self-energy.
Cubically scaling parts are the construction of the Green's functions or diagonalisation of the 
density matrix.
In practice, one observes approximately an increase of computational time by a factor of 1.7 
when the basis-set cut-off is increased by 100~eV (recall that we typically use basis-set cut-offs
between 500 and 1000~eV, increased in steps of 100~eV).
The memory requirements increase by a similar factor.

\begin{table}
\caption{The CPU hours required to obtain the lattice energies for the RPA and RPA+GWSE methods.
The calculations for molecules are performed for different values of basis set cut-off and 
cell volumes, all these contributions were summed to obtain the presented value.
Moreover, the data include the time required to obtain the corrections 
with `hard' PAW potentials.
The data for solids collect contributions from calculations at different k-points and basis
set cut-offs, as well as the `hard' correction.
The calculations were run at the Salomon supercomputer equipped with Intel Xeon E5-2680v3
CPUs.}
\label{tab:rpa_CPU}
\begin{center}
\begin{tabular}{   l   c c c  }
\hline \hline
System & Part	&	RPA	&	+GWSE	\\
\hline
CO$_2$ & molecule &  2790 & 20810 \\
CO$_2$ & solid &  978 & 5806 \\
\hline
NH$_3$ & molecule & 2484 & 9793\\
NH$_3$ & solid & 312 & 2039\\
\hline
C$_6$H$_6$ & molecule & 1114 & 9043\\
C$_6$H$_6$ & solid & 3229 & 21066\\
\hline
C$_{10}$H$_8$ & molecule & 2695 & 21014\\
C$_{10}$H$_8$ & solid & 7501 &13451 \\
\hline
C$_{14}$H$_{10}$ & molecule & 9845 & 25849\\
C$_{14}$H$_{10}$ & solid & 5462 & 24794\\
\hline
water & molecule & 1899 & 29359 \\
ice Ih & solid & 2139 & 45732\\
ice II & solid & 1546 & 29285\\
ice VIII & solid & 647 & 3872\\
\hline \hline
\end{tabular}
\end{center}
\end{table}

\newpage

\section{ Comparison with other computational methods }\label{sec:othermethods}

Table~\ref{tab:crysSI} is an extended version of Table~I of the main paper, reporting several values published in the literature, as obtained from different groups with different setups. Values used for Table~I are highlighted.
More details about the DMC setup are given in the following sections. Concerning the other methods, see the corresponding references.

\begin{table*}[hb!]
\caption{
Extended version of Table~I of the main paper, reporting the lattice energy [kJ/mol] for the molecular crystals under consideration in this work,
as obtained with different computational approaches. 
Values used in Fig.~2 of the main paper are in boldface.
The RPA, RPA+rSE, and RPA+GWSE values for the ice polymorphs have been calculated in this work, with the same setup used in Ref.~\cite{Jiri:2016}.
}\label{tab:crysSI}
{\footnotesize
\begin{center}
\begin{tabular}{   l   c c c c c c c c }
\hline \hline
	&	Ice Ih			&	Ice II			&	Ice VIII			&	Carbon dioxide			&	Ammonia			&	Benzene			&	Naphthalene			&	Anthracene			\\
	&				&				&				&	CO$_2$			&	NH$_3$			&	C$_6$H$_6$			&	C$_{10}$H$_8$			&	C$_{14}$H$_{10}$			\\
\hline \hline																																	

\textbf{DMC}																																	\\
\textbf{DMC(large-cell)}	&	\textbf{-59.3(5)}			&	\textbf{-59.1(6)}		&	\textbf{-57.3(6)}		&	\textbf{-28.2(13)}			&	\textbf{-37.1(4)}		&	\textbf{-52.1(4)}		&	\textbf{-78.8(8)}		&	\textbf{-105.5(1.7)}			\\
\textbf{DMC(small-cell)}	&	\textbf{-59.2(2)}		&	\textbf{-59.0(3)}		&	\textbf{-57.4(1)}			&	\textbf{-28.5(4)}		&	\textbf{-37.5(1)}		&	\textbf{-51.2(2)}		&	\textbf{-78.0(6)}		&	\textbf{-103.9(1.0)}			\\
																																	
\hline
\textbf{MP2}	\\ 
DB-RI-MP2/CBS \cite{Wen:2011gm}	&	-59.9			&	n.a.			&	n.a.			&	\textbf{-29.1}			&	\textbf{-39.3}			&	\textbf{-61.6}			&	n.a.			&	n.a.			\\
LMP2/p-aug-6-31G(d,p) \cite{Cutini:2016fh}	&	n.a.			&	n.a.			&	n.a.			&	-22.7			&	-34.1			&	-57.7			&	\textbf{-91.5}			&	\textbf{-127}			\\
HF+$\Delta$MP2(EMBE-2) \cite{Gillan:enebench:jcp2013}	&	\textbf{-58.7}			&	\textbf{-58.4}			&	\textbf{-56.3}			&	n.a.			&	n.a.			&	n.a.			&	n.a.			&	n.a.			\\
																																	
\hline
\textbf{CCSD(T)}	\\
$\Delta$CCSD(T) \cite{Wen:2011gm}	&	-60.4			&	n.a.			&	n.a.			&	\textbf{-29.5}			&	\textbf{-40.2}			&	-51.2			&	n.a.			&	n.a.			\\
$\Delta$CCSD(T)  \cite{Gillan:enebench:jcp2013}	&	\textbf{-58.0}			&	\textbf{-58.0}			&	\textbf{-55.4}			&	n.a.			&	n.a.			&	n.a.			&	n.a.			&	n.a.			\\
OSV-LCCSD(T0)-F12 \cite{GarnetChan:Sci2014}	&	n.a.			&	n.a.			&	n.a.			&	n.a.			&	n.a.			&	\textbf{-54.6(8)}			&	n.a.			&	n.a.			\\
																																	
\hline
\textbf{RPA}	\\
EXX+RPA@PBE \cite{Jiri:2016}	&	\textbf{-52.0}			&	\textbf{-51.7}			&	\textbf{-49.5}			&	\textbf{-24.1}			&	\textbf{-31.5}			&	\textbf{-45.2}			&	\textbf{-68.4}			&	\textbf{-92.6}			\\
EXX+RPA@PBE \cite{RPAice:2014}	&	-52.5			&	-51.7			&	-49.3			&	n.a.			&	n.a.			&	n.a.			&	n.a.			&	n.a.			\\
HF+RPA@PBE \cite{RPAice:2014}	&	-65.6			&	-64.8			&	-63.6			&	n.a.			&	n.a.			&	n.a.			&	n.a.			&	n.a.			\\
\hline
\textbf{RPA+singles} \\
{RPA+rSE}  \cite{Jiri:2016}	&	-61.1			&	-61.1			&	-59.0			&	-26.9			&	-37.9			&	-49.1			&	-73.7			&	-98.9			\\
\textbf{RPA+GWSE}  \cite{Jiri:2016}	&	\textbf{-60.2}			&	\textbf{-60.1}			&	\textbf{-57.9}			&	\textbf{-27.3}			&	\textbf{-37.6}			&	\textbf{-51.5}			&	\textbf{-77.6}			&	\textbf{-103.5}			\\
\hline \hline
\end{tabular}
\end{center}
}
\end{table*}

It can be observed that there are big differences among values obtained with the same method but by different groups, see for instance the case of MP2 for carbon dioxide, ammonia and benzene, or CCSD(T) for the benzene case.
A  fraction of this energy difference can be due to differences in the underlying configurations.
For instance, the reported CCSD(T) value for benzene, -54.6$\pm$0.8~kJ/mol, 
obtained in Ref.~\citenum{GarnetChan:Sci2014} is from a configuration obtained from neutron diffraction at a temperature of 138~K, but they have estimated that the 0~K structure should be $\sim$1.3~kJ/mol lower in energy.
However, differences in structure cannot always explain these energy differences; other sources of disagreement can be more technical, such as the size of the basis set, the procedure for the fragment decomposition, etc.

\section{Lattice energy from experiments}\label{sec:ElattEXP}

The lattice energy $\Elatt$ is not directly measurable in experiments, but it can be obtained from measures of the sublimation enthalpy $\DHsub(T)$, 
by subtracting the contribution due to thermal and zero point motion $\DHcor(T)$, as given in Eq.~2 of the manuscript.
We report in Table~\ref{tab:DH} the values $\DHsub$, $\DHcor$ and $\Elatt$ for the molecular crystal studied in this work, excluded water ice polymorphs, which will be discussed separately in Sec.~\ref{sec:ice}.
\begin{table*}[bp!]
\caption{
We report the evaluations of the 
sublimation enthalpy $\DHsub(T)$ corrected to room temperature $T=298.15$~K, the fraction of energy due to thermal and zero point effects $\DHcor(T)$ (see Sec.~\ref{sec:DHcor}), and lattice energy $\Elatt$, for the molecular crystals studied here.
All values are in kJ/mol.
Values used for Table~I and Fig.~2 of the main paper are highlighted in boldface.
For $\DHcor(T)$, we show values obtained using difference approaches, for instance the harmonic approach ({\em Har.}, Eqs.~\ref{eq:DHcor}-\ref{Evib}), or the thermodynamic integration ({\em T.Int.}, Eqs.~\ref{HfromCp}-\ref{eq:DHcor2}) of experimental $C_p$s, to which the ZPE contribution has to be added.
We indicate also the exchange correlation DFT functional used to calculate the thermal and/or zero point energy. 
}\label{tab:DH}
\begin{center}
\begin{tabular}{   l  l  c c c c c }
	&& Carbon dioxide			&	Ammonia			&	Benzene			&	Naphthalene			&	Anthracene			\\
	&& CO$_2$			&	NH$_3$			&	C$_6$H$_6$			&	C$_{10}$H$_8$			&	C$_{14}$H$_{10}$			\\
\hline \hline
\multicolumn{7}{l}{\textbf{$\Delta_\textrm{sub}H(298.15~K)$ : sublimation enthalpy at room temperature }}	\\
\hline
\multicolumn{2}{l}{  Ref.~\onlinecite{Roux:2008dr}	 } &&&	44.7	&	72.6	&	101.9	\\
\multicolumn{2}{l}{ Refs.~\onlinecite{Elatt_ExpOJ, AcreeJr:2010ev}	 }&	24.6	&	29.8	&	45.1	&	71.3	&	98.2	\\
\multicolumn{2}{l}{ Ref.~\onlinecite{Ruzicka:2014} }	&&&	44.0	\\
Used	&& {\bf 24.6}$^a$	&	{\bf 29.8}$^a$	&	{\bf 44.0}$^b$	&	{\bf 71.3}$^a$	&	{\bf 98.2}$^a$	\\
\hline \hline
\multicolumn{7}{l}{\textbf{$\DHcor(298.15~K)$ : thermal and ZPE contribution to the sublimation enthalpy  }}	\\
\hline
$-2 R T$ &  	& -5.0 & -5.0 & -5.0 & -5.0 & -5.0 \\ 
Ref.~\onlinecite{Elatt_ExpOJ} 	&	{\em Har.}, PBE		& -3.2	&	-7.8	&	-5.3	&	-5.0	&	-2.4 \\
Ref.~\onlinecite{Elatt_ExpRT}	&	{\em Har.}, PBE+TS & {\bf -3.8}	&{\bf -7.4}	&{\bf -6.6}	&{\bf -7.9} &{\bf -7.6}	\\
Ref.~\onlinecite{Elatt_ExpRT}	&	{\em T.Int.}+ZPE, PBE+TS & & & & -10.5	&	-10.9 \\
\hline \hline
\multicolumn{7}{l}{\textbf{$\Elatt = -\DHsub(T) + \DHcor(T)$ : lattice energy }}	\\
\hline
Ref.~\onlinecite{Elatt_ExpOJ} 	&&	-27.8			&	-37.6			&	-50.4			&	-76.3			&	-100.6			\\
Ref.~\onlinecite{Elatt_ExpRT}	&&	-28.4			&	-37.2			&	-51.7			&	-81.7			&	-112.7			\\
Used && \textbf{-28.4}	&	\textbf{-37.2}			&	\textbf{-50.6}			&	\textbf{-79.2}			&	\textbf{-105.8}		\\
\hline \hline
\multicolumn{7}{p{6.5in}}{\footnotesize
$^a$~From Ref.~\cite{Elatt_ExpOJ}, corresponding to an 
average over experimental sublimation enthalpy measurements corrected to room temperature T=298.15~K via $C_p$ calculated
by group additivity as described in \citet{AcreeJr:2010ev}. 
$^b$~From \citet{Ruzicka:2014}. Notice that the triple point temperature of benzene is 278.674~K, thus the room temperature sublimation enthalpy is extrapolated. 
}
\end{tabular}
\end{center}
\end{table*}
We will discuss the accuracy and precision of the measured $\DHsub(T)$ for the molecular crystals under considerations in this work in Sec.~\ref{sec:DHsub}, and how to evaluate the term $\DHcor(T)$ in Sec.~\ref{sec:DHcor}.

\subsection{Experimental evaluations of the sublimation enthalpy}\label{sec:DHsub}

Whilst it is easy to find in the literature experimental values for $\DHsub$,
the accuracy and precision associated to the value is not always easy to assess.
%
%
Indeed, \citet{Chickos2003_DHerror} examined a big dataset ({\em i.e.} 80 compounds, 451 measurements) of published sublimation enthalpies by different laboratories and by different techniques, and found significant discrepancies among, with the agreement between laboratories is often not within the precision cited.
More specifically, the standard deviation between the mean for each compound and each measurement is $\pm 6.7$~kJ/mol, going down to $\pm 4.9$~kJ/mol when outliers are removed (outliers are considered measures more than $3\sigma$ far away). Moreover, compounds with larger sublimation enthalpy have larger errors: standard deviation of $\pm 4.6$~kJ/mol for sublimation enthalpies smaller than 100~kJ/mol; $\pm 5.6$~kJ/mol for the remaining.

In Fig.~\ref{fig:ExpDHsub} we show the available measures of $\DHsub$ as a function is the temperature $T$. 
It is clear that Chickos' conclusions remains correct also for the molecular crystals under consideration here.
In particular, the uncertainty associated to any sublimation enthalpy considered here is $>$~1~kJ/mol, and quite larger for naphthalene and anthracene.  
Moreover, \citet{Ruzicka:2014}
have recently discussed the measures of the sublimation enthalpy of benzene,
and it has observed that the values from the sources reported in the review paper by \citet{Roux:2008dr} differ from originally published ones as they were published in form of Antoine equation but recalculated in \citet{Roux:2008dr} to a constant value of sublimation enthalpy.

\begin{figure}[p!]
\includegraphics[width=2.8in]{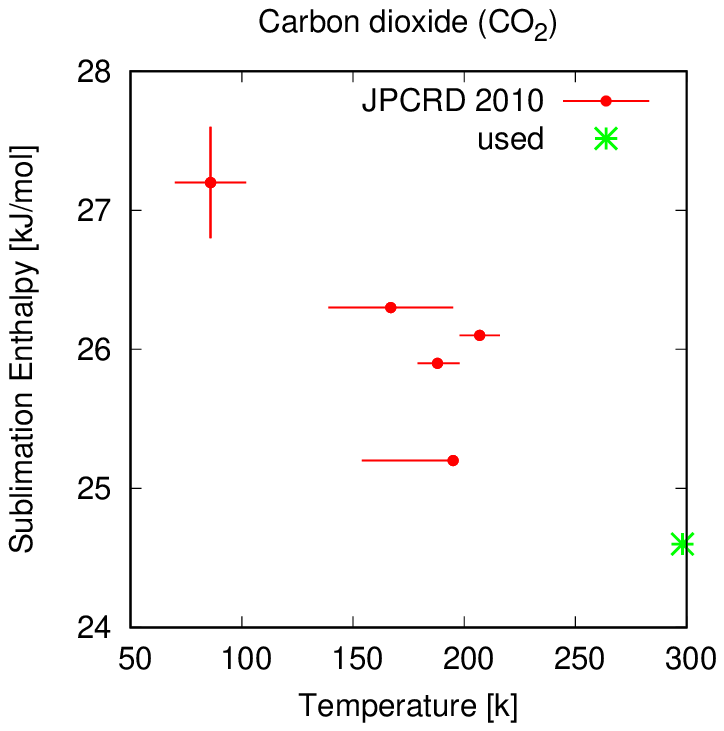}
\includegraphics[width=2.8in]{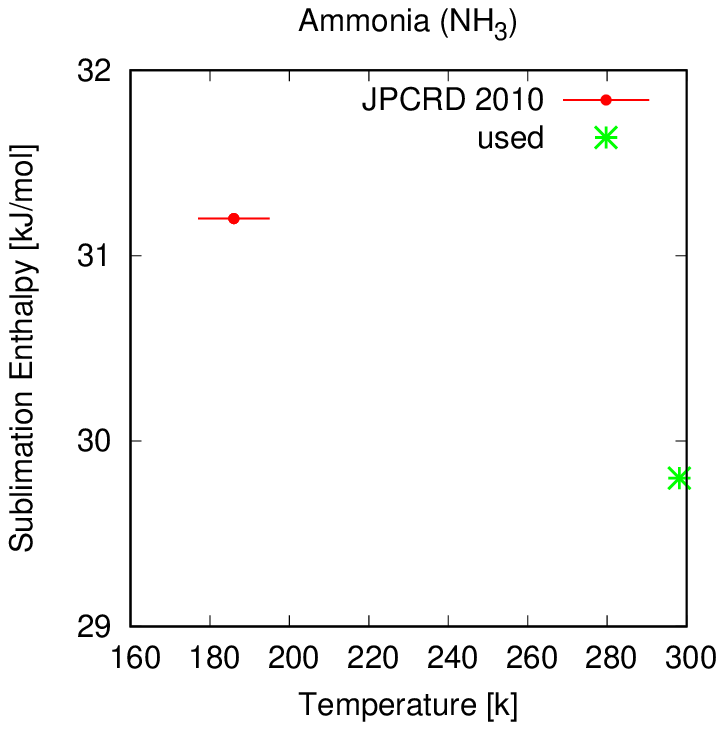} \\
\includegraphics[width=2.8in]{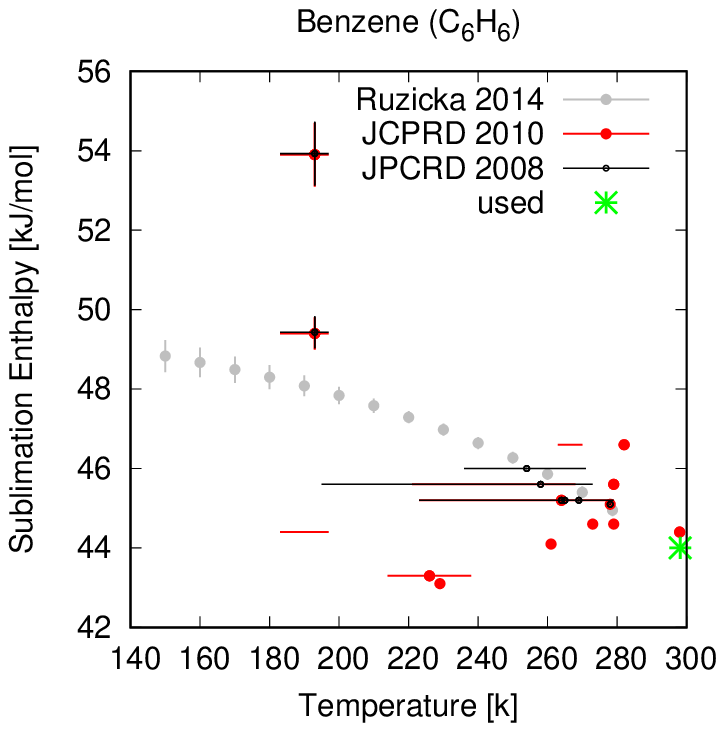}
\includegraphics[width=2.8in]{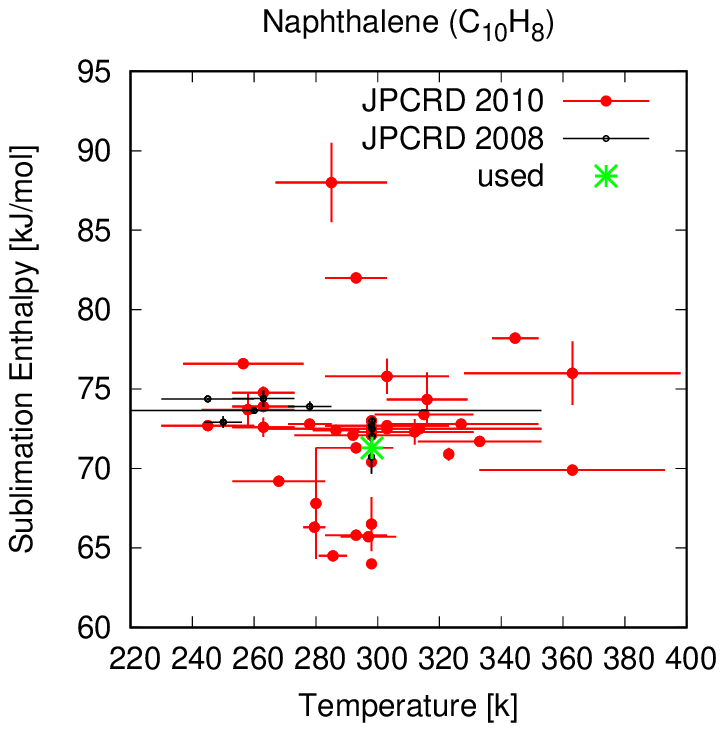}
\includegraphics[width=2.8in]{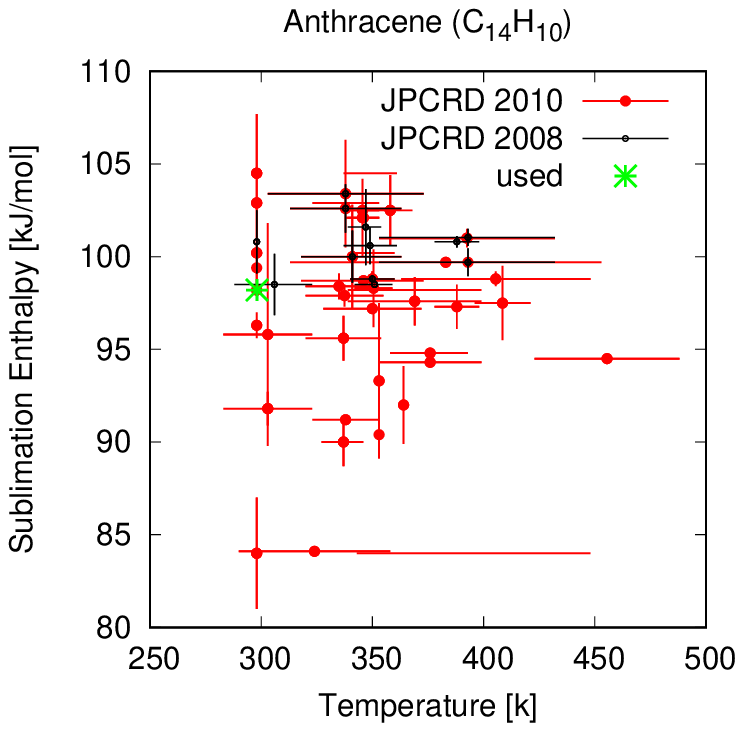}
\caption{\label{fig:ExpDHsub}
Experimental sublimation enthalpy $\DHsub(T)$ for carbon dioxide, ammonia, benzene, naphthalene and anthracene, as reported on the review papers by ~\citet{Roux:2008dr} (JPCRD 2008, black circles) and \citet{AcreeJr:2010ev} (JPCRD 2010, red dots). 
The $x$-bar represent the temperature range of the experiments, the $y$-bar correspond to the estimated error (one standard deviation).
The green asterisks correspond to the values of sublimation enthalpy at 298.15~K used for the evaluation of $\Elatt$, see Table~\ref{tab:DH}. 
For benzene, we also report the recent measures by \citet{Ruzicka:2014} (Ruzicka 2014, grey dots).
Notice that the triple point of benzene is $T_{tr}=$278.674~K, thus there are no experimental $\DHsub(T)$ for $T>T_{tr}$ and the value at 298.15~K is extrapolated.
}
\end{figure}

\newpage

\subsection{ Lattice energy from experimental measures of sublimation enthalpy }\label{sec:DHcor}

In this section we will show how to evaluate the term $\DHcor$.
It accounts for both thermal and quantum nuclear effects.
In order to derive $\DHcor$, we need to start from the definition of the sublimation enthalpy:
%
$\DHsub(T)$ is the difference between the enthalpy $H^g(T)$ of the gas and $H^s(T)$ of the crystal solid, both at temperature $T$. By separating the electronic ($el$), translational ($trans$), rotational ($rot$) and vibrational ($vib$) contributions, and noticing that in the crystal there are no trans-rotational contributions and the pressure times volume term, $pV$, is negligible, we have that:
\begin{equation}\label{eq:DH1}
\DHsub = E_{el}^g + E_{trans}^g + E_{rot}^g + E_{vib}^g + pV - (E_{el}^s + E_{vib}^s)
\end{equation}
where the superscript stands either for gas ($g$) or solid ($s$), and the time dependance has been dropped for the seek of brevity.
By assuming that the rigid rotor and ideal gas approximations are reliable (that is typically the case in these molecular systems), we have that 
$E_{trans}^g = 3/2 RT$, 
$E_{rot}^g = 3/2 RT$ if the molecule is non-linear, 
$E_{rot}^g = RT$ otherwise,
and
$pV = RT$.
Thus, Eq.~\ref{eq:DH1} simplifies into:
\begin{eqnarray}\label{eq:DH2}
\DHsub(T) &=& \Delta E_{el} + \Delta E_{vib}(T) + 4 RT \quad \textrm{for non-linear molecules} , \\
\DHsub(T) &=& \Delta E_{el} + \Delta E_{vib}(T) + 7/2 RT \quad \textrm{for linear molecules} , \nonumber
\end{eqnarray}
where the term $\Delta E_{vib}(T)$ contains both the thermal and the quantum nuclear contributions.
Notice that 
$\Delta E_{el}  = E_{el}^g - E_{el}^s$ 
is precisely the inverse of the lattice energy $\Elatt$, thus:
\begin{eqnarray}\label{eq:DHcor}
\DHcor(T) &=& \Delta E_{vib}(T) + 4 RT \quad \textrm{for non-linear molecules} , \\
\DHcor(T) &=& \Delta E_{vib}(T) + 7/2 RT \quad \textrm{for non-linear molecules} . \nonumber
\end{eqnarray}
Vibrations in the solid molecular crystals can usually be separated into intra-molecular and inter-molecular vibrations, $E_{vib}^{s} = E_{vib}^{s,intra} + E_{vib}^{s,inter}$, and the stiffest intra-molecular modes are decoupled from the intermolecular modes. Intra-molecular vibrations have similar modes and frequencies than the gas-phase molecule, thus we can conveniently write:
\begin{equation}\label{eq:Evib}
\Delta E_{vib} = \Delta E_{vib}^{relax} - E_{vib}^{s,inter}, \qquad 
\textrm{where} \quad \Delta E_{vib}^{relax} = E_{vib}^{g} - E_{vib}^{s,intra}
\end{equation}
is the change in (intra-molecular) vibrational energy given when molecules are packed in the crystal form.

At this point, 
a first approach can be to do a 
a drastic approximation (which is anyway often found in the literature), that is to assume that intra-molecular frequencies in the solid are exactly the same as in the gas phase ({\em i.e.}, $\Delta E_{vib}^{relax} \sim 0$), then to take the high temperature limit for the inter-molecular vibrations ({\em i.e.}, $E_{vib}^{s,intra} \sim 6 RT$) and to neglect any zero-point motion, yielding $\Delta E_{vib}(T) \sim -6 RT$ (Dulong-Petit law). 
In non-linear molecules that would imply that $\DHcor \sim -2 RT$, that is 4.96~kJ/mol at room temperature T=298.15~K and zero at T=0~K. 
This is a poor approximation, as shown by \citet{Elatt_ExpOJ} in a set of 21 molecular crystals.
This approximation is particularly bad for water ice:
according to \citet{ICE:Whalley1984} the hexagonal ice I$_h$ has a zero-point energy (ZPE) change with respect to gas-phase that is 
$\Delta \textrm{ZPE} = +11.5$~kJ/mol, see Sec.~\ref{sec:ice}.

A more reliable approach is to calculate the vibrational energies for the solid and gas phase in the harmonic limit, considering for each frequency $\omega$ a contribution
\begin{equation}\label{epsilonomega}
\epsilon(\omega,T) = 
{\hbar \omega \over 2} + {\hbar \omega \over \exp\left( {\hbar \omega \over k_B T} \right) -1 }
\, ,
\end{equation}
where the first term in the right hand size accounts for the ZPE contribution and the second for the thermal one.
This yield:
\begin{equation}\label{Evib}
E^g_{vib}(T) = \sum_i \epsilon(\omega_i,T) \, ,
\qquad
E^s_{vib}(T) = \int \! \epsilon(\omega,T) g(\omega) \, \textrm{d}\omega \, ,
\end{equation}
where $\omega_i$s are the frequencies of the isolated molecule, which are $3M-6$ ($M$ is the number of atoms in the molecule) for a non-linear molecule and $3M-5$ for a linear one;
$g(\omega)$ is the phonon density of states in the solid.
For a more detailed description about how to evaluate these quantities, see \citet{Elatt_ExpOJ} and \citet{Elatt_ExpRT}.

Notice that whenever we employ Eq.~\ref{Evib} to evaluate $\DHcor$, Eq.~\ref{eq:DHcor}, we are subject to errors not only coming from the harmonic approximation, but also from the limitations of the computational approaches (typically DFT) used for the evaluations of the frequencies and the phonon spectrum.
Different choices of the exchange-correlation functional in DFT can lead to differences in terms of $\DHcor$ quite larger than 1~kJ/mol.
In particular, inaccuracies on the evaluation of high frequency modes mostly affects the ZPE contribution, while low frequency modes affect mostly the thermal contribution.

There is a complementary approach, which can help to highlight shortfalls of the previous approach.
It employs the fact that the constant pressure heat capacity 
$C_p \equiv \left( {\partial H \over \partial T} \right)_p$,
which yields:
\begin{equation}\label{HfromCp}
H(T) = H(0\,\textrm{K}) + \int_0^T \! C_p(\tilde T) \, \textrm{d}\tilde T \,,
\end{equation}
unless there are phase transitions taking place between 0~K and T, which imply that in the right hand side we should add also the enthalpy of transformation. (However, none of the molecular crystals considered in this work has any phase transition occurring between 0~K and the temperature at which the sublimation enthalpy is measured.)
Eq.~\ref{HfromCp} can be used both for the gas and the solid state (but clearly the gas state at 0~K is fictitious as at some non-zero temperature there would be a condensation or a deposition).  
We have that $H(0\,\textrm{K}) = U + p V$, where $U$ is the internal energy, which includes both the electronic and the zero-point energy, and 
the $pV$ term at 0~K is $\ll$~1~kJ/mol both for the solid and the gas, and it can be neglected.
Thus, we have that:
\begin{equation}\label{eq:DHcor2}
\DHcor(T) = \Delta E_\textrm{ZPE} + \int_0^T \! \Delta C_p(\tilde T) \, \textrm{d}\tilde T \, , 
\end{equation}
where
$\Delta E_\textrm{ZPE} =  E_\textrm{ZPE}^g -  E_\textrm{ZPE}^s$,
and 
$\Delta C_p(\tilde T) = C_p^g(\tilde T) - C_p^s(\tilde T)$.
The $\Delta E_\textrm{ZPE}$ can to be calculated, in harmonic approximation, by taking only the first term in the right hand side of Eq.~\ref{epsilonomega}.
Also the $C_p(T)$ could be computed in harmonic approximation:
the temperature dependance of the electronic energy is negligible in our systems, so 
we can calculate the constant volume head capacity $C_V$ by taking the temperature derivative of $E_{vib}(T)$, 
$C_V(T) = {\partial E_{vib}(T) \over \partial T}$,
with $E_{vib}$ given by Eq.~\ref{Evib} and \ref{epsilonomega}. 
For the gas phase, under the ideal gas approximation, $C^g_p(T) = C^g_V(T) + R$, and $3RT$ has to be added to the integrated heat capacity in order to account for the roto-translational heat capacity (under rigid rotor hypothesis).
For the solid,
under harmonic approximation there is no thermal change of volume, so 
$C^s_p(T) = C^s_V(T)$.
All this leads to the identical result obtained by using Eq.~\ref{eq:DHcor} and Eq.~\ref{Evib}.
In fact, Eq.~\ref{eq:DHcor2} is useful if we instead use experimental values of the heat capacity at different temperatures, which,
at least for some of the systems here considered, are enough to interpolate $C_p(T)$ in the range of temperature for 0~K to the temperature $T$ at which the sublimation enthalpy is given.
See for instance the case of the benzene crystal, reported in Fig.~\ref{fig:cpintegrate}.
\begin{figure}[h]
\begin{center}
\includegraphics[width=0.7\textwidth]{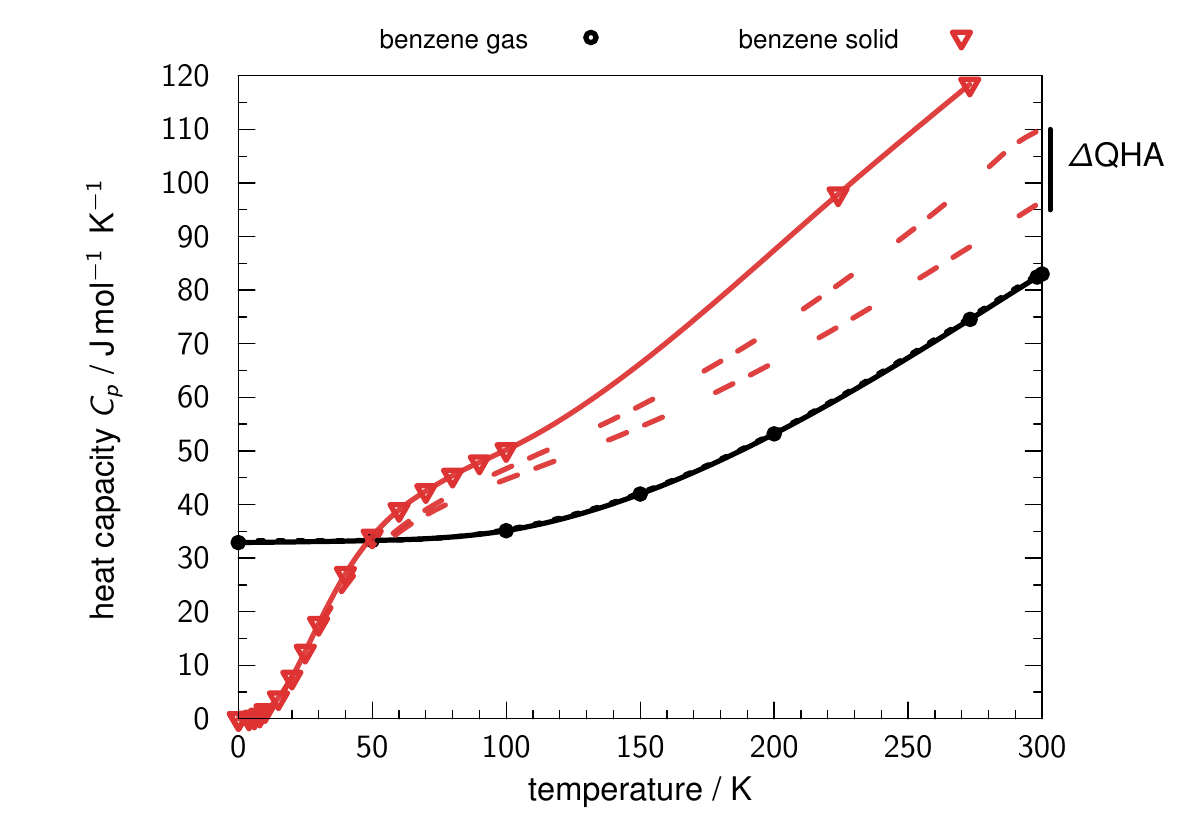}
\end{center}
\caption{\label{fig:cpintegrate}
Isobaric heat capacity $C_p$ of benzene in the solid and gas phase. Experimental points\cite{cpbenzene_gas,cpbenzene_solid} have been interpolated via cubic splines (straight lines) and the low temperature gas phase values are extrapolated to 0~K.
Heat capacities computed via harmonic HSE-3c frequencies are shown as dashed lines.
The impact of a quasi-harmonic treatment in the solid state is indicated.
}
\end{figure}
The advantage of this approach is that the thermal part of the evaluated $\DHcor$ is not affected by many of the shortfalls of Eq.~\ref{eq:DHcor} and Eq.~\ref{Evib}, such as the lack of anharmonic correction and the inaccuracy of the employed computational approach, whilst the ZPE part still is because it has to be computed. 
\citet{Elatt_ExpRT} have tested this second approach (Eq.~\ref{eq:DHcor2}), taking experimental values for $C_p$ for the solid but using computed values for the gas, and compared it with the harmonic approach (Eq.~\ref{eq:DHcor}), observing in a set of seven molecular crystals differences in $\DHcor$ of around 2 or 3~kJ/mol.
However, the non-symmetric choice to have measured $C_p$s only for the solid and computed (harmonic) ones for the gas could be not the best choice, as it cannot benefit from favorable error cancellations.
For this reason, in Table~I of the manuscript the reported the experimental valued of $\Elatt$ are obtained by employing the harmonic evaluation of $\DHcor$. 
In Table~\ref{tab:DH} we report the values for the $\DHcor$ obtained by following several approaches, and in Table~\ref{tab:DHnew} we report new evaluations at different temperatures and splitting the contributions due to thermal and zero point effects.
The difference of the evaluations obtained via different methods reflects the difficulty to have an accurate and precise evaluation of this quantity.

\begin{table*}[tp!]
\caption{
We report new evaluations of the $\DHcor(T)$ term to relate measured sublimation enthalpy $\DHsub(T)$ with the lattice energy $\Elatt$, see Eq.~1 in the main paper. We split the contribution due to thermal  ($\Delta_{\text{th}}$) and zero point ($\Delta E_{\text{ZPE}}$) effects, as discussed in Sec.~\ref{sec:DHcor}.
All energies are in kJ/mol and temperatures in K.
We show values obtained using difference approaches, for instance the harmonic approach ({\em Har.}, Eqs.~\ref{eq:DHcor}-\ref{Evib}), or the thermodynamic integration ({\em T.Int.}, Eqs.~\ref{HfromCp}-\ref{eq:DHcor2}) of experimental $C_p$s, to which the ZPE contribution has to be added.
The harmonic frequencies are obtained with the HSE-3c functional. 
}\label{tab:DHnew}
\begin{center}
\begin{tabular}{l r r r r}
\multicolumn{5}{c}{\bf Carbon dioxide}\\
                  & $\Delta E_{\text{ZPE}}$ & $\Delta_{\text{th}}$(70K) & $\Delta_{\text{th}}$(167K) & $\Delta_{\text{th}}$(207K)\\
harm. frequencies & -6.1 & +1.5 & +1.7 & +1.6\\
explt. $C_{p}$\cite{cpco2_gas,cpco2_solid}    & n.a.  & +0.9 &  -0.3 &  -1.2\\[0.2cm]
\multicolumn{5}{c}{\bf Ammonia}\\
                  & $\Delta E_{\text{ZPE}}$ & $\Delta_{\text{th}}$(177K) & $\Delta_{\text{th}}$(186K) & $\Delta_{\text{th}}$(195K)\\
harm. frequencies &-10.1 & +3.8 & +3.9 & +3.9\\
explt. $C_{p}$\cite{cpammonia}    & n.a.  & +2.0 & +1.9 & +1.9\\[0.2cm]
\multicolumn{5}{c}{\bf Benzene}\\
                  & $\Delta E_{\text{ZPE}}$ & $\Delta_{\text{th}}$(150K) & $\Delta_{\text{th}}$(270K) & $\Delta_{\text{th}}$(279K)\\
harm. frequencies & -6.4 & +2.0 &  -1.4 &  -1.5\\
explt. $C_{p}$\cite{cpbenzene_gas,cpbenzene_solid}    & n.a.  &  -0.5 &  -4.7 &  -5.0\\[0.2cm]
&\multicolumn{2}{c}{\bf Naphtalene}\\
                  & $\Delta E_{\text{ZPE}}$ & $\Delta_{\text{th}}$(293K) \\
harm. frequencies & -5.5 & +1.8 \\
explt. $C_{p}$\cite{cpnaphtalene,cpanthracene}    & n.a.  &  -3.7 \\[0.2cm]
&\multicolumn{2}{c}{\bf Anthracene}\\
                  & $\Delta E_{\text{ZPE}}$ & $\Delta_{\text{th}}$(293K)\\
harm. frequencies &  -6.7 & +2.3 \\
explt. $C_{p}$\cite{cpnaphtalene,cpanthracene}    & n.a. & -3.0 \\[0.2cm]
%
\end{tabular}
\end{center}
\end{table*}

\newpage

\subsection{Lattice energy for the ice polymorphs}\label{sec:ice}

For ice I$_h$, II and VIII, \citet{ICE:Whalley1984} provides the internal energy difference $\Delta U$ between ice and gas phase at 0~K. Thus, we need to account for the ZPE contribution:
$
\Delta U = \Elatt + \Delta  \textrm{ZPE} \,.
$
The $\Delta \textrm{ZPE}$ can be conveniently split into the intra-molecular contribution $\Delta \textrm{ZPE}_\textrm{intra}$ and the inter-molecular one $\textrm{ZPE}_\textrm{inter}$:
$
\Delta \textrm{ZPE} = \Delta \textrm{ZPE}_\textrm{intra} + \textrm{ZPE}_\textrm{inter}
\, .
$
The former is related with the different frequency of the two O--H stretching and the H--O--H bending modes for the molecule in gas and in the crystal phase.
The latter is proper only of the crystal phase.

The values for $\Elatt$, $\Delta U$ and ZPE obtained by \citet{ICE:Whalley1984} are reported in Table~\ref{tab:ice}.
Notice that the ZPE contribution is very important: 
in the hexagonal ice I$_h$   
$\Delta \textrm{ZPE} = +11.5$~kJ/mol, where the contribution of the intermolecular ZPE is 
$\Delta \textrm{ZPE}_\textrm{inter}=+16.5$~kJ/mol, and 
$\Delta \textrm{ZPE}_\textrm{intra}=-4.98$~kJ/mol.
Indeed, the lattice energy of I$_h$ ice is -58.82~kJ/mol, 
but the internal energy difference at zero Kelvin is 
$\Delta U = -47.341 \pm 0.015$~kJ/mol.

\begin{table*}[hbp!]
\caption{
Energies reported by \citet{ICE:Whalley1984} for Ice I$_h$, II and VIII.
All of them are in kJ/mol, and at 0~K.
$ \Delta U = \Elatt + \Delta \textrm{ZPE} $ is the internal energy difference between the ice and the gas phases, thus it includes both the electronic contribution $\Elatt$ and the zero point contribution $\Delta \textrm{ZPE}$ (which is split into the intra-molecular and inter-molecular contributions). Values in italic are not explicitly reported, but can be deduced by straightforward relations.
}\label{tab:ice}
\begin{center}
\begin{tabular}{   l  c c c }
	& Ice I$_h$		&		Ice II		&		Ice VIII		\\
\hline \hline
$\Delta U $	& -47.341$\pm$0.015 &  \textit{-47.3} & \textit{-44.2} \\
$\Delta U - \Delta U^{\textrm{I}_h}$	& 0	& 0.057 & 3.16 \\
\hline
$\textrm{ZPE}_\textrm{inter}$	&	16.5	&	\textit{16.5}	&	\textit{18.2}	\\
$\Delta \textrm{ZPE}_\textrm{intra}$	&	-4.98	& n.a. & n.a.	\\
$\Delta \textrm{ZPE}$	&	11.5 & n.a. & n.a. \\
\hline
$\Elatt+\Delta \textrm{ZPE}_\textrm{intra}$	&	-63.86 & -63.80 & -62.40 \\
\hline
$\Elatt$	&	-58.82 & n.a. & n.a. \\
\hline \hline
\end{tabular}
\end{center}
\end{table*}

Unfortunately, \citet{ICE:Whalley1984} provides $\Elatt$ only for ice I$_h$; 
for ice II and VIII only the $\Delta U$ and the $\Elatt+\Delta \textrm{ZPE}_\textrm{intra}$ 
are available.
It appears that the $\textrm{ZPE}_\textrm{inter}$ is different in the three ice polymorphs.
We can expect that $\Delta \textrm{ZPE}_\textrm{intra}$ is less affected by the ice polymorph, such that the difference of $\Delta \textrm{ZPE}$ among the polymorphs is the same of the difference of $\textrm{ZPE}_\textrm{inter}$. 
This assumption is supported by {\em ab-initio} computational simulations \cite{santra_on_2013,Ice:prl2011}, where the overall $\Delta \textrm{ZPE}$ has been computed for each polymorphs. 
This leads to the relations: 
$\Elatt^\textrm{II} \sim \Elatt^{\textrm{I}_h} = -58.8$~kJ/mol and
$
\Elatt^\textrm{VIII} \sim \Delta U^\textrm{VIII} 
- (\Delta \textrm{ZPE}_\textrm{intra}^{\textrm{I}_h} + \textrm{ZPE}_\textrm{inter}^\textrm{VIII}) 
= -57.4
$~kJ/mol.
These are the values used in Table~I of the main paper.
However, it must be clear that there is an assumption of the $\Delta \textrm{ZPE}^\textrm{VIII}$, thus the difference $\Elatt^\textrm{VIII} - \Elatt^{\textrm{I}_h} = 1.4$~kJ/mol has an uncertainty that we estimate $\ge$~1~kJ/mol.

\newpage

\newpage
\section{ Geometries of the molecular crystals used for the DMC, RPA and RPA-GWSE calculations }\label{sec:geoCrys}
The following configurations are given in cif format.
{\scriptsize
\begin{verbatim}
CARBON DIOXIDE
_cell_length_a       5.624
_cell_length_b       5.624
_cell_length_c       5.624
_cell_angle_alpha    90
_cell_angle_beta     90
_cell_angle_gamma    90

_symmetry_space_group_name_H-M    "P 1"
_symmetry_int_tables_number       1

loop_
  _symmetry_equiv_pos_as_xyz
  'x, y, z'

loop_
  _atom_site_label
  _atom_site_occupancy
  _atom_site_fract_x
  _atom_site_fract_y
  _atom_site_fract_z
  _atom_site_thermal_displace_type
  _atom_site_B_iso_or_equiv
  _atom_site_type_symbol
  C1       1.0000 0.00000  0.00000  0.00000  Biso   1.000  C
  C2       1.0000 0.50000  0.00000  0.50000  Biso   1.000  C
  C3       1.0000 0.00000  0.50000  0.50000  Biso   1.000  C
  C4       1.0000 0.50000  0.50000  0.00000  Biso   1.000  C
  O1       1.0000 0.11998  0.11998  0.11998  Biso   1.000  O
  O2       1.0000 0.88002  0.88002  0.88002  Biso   1.000  O
  O3       1.0000 0.38002  0.88002  0.61998  Biso   1.000  O
  O4       1.0000 0.61998  0.11998  0.38002  Biso   1.000  O
  O5       1.0000 0.88002  0.61998  0.38002  Biso   1.000  O
  O6       1.0000 0.11998  0.38002  0.61998  Biso   1.000  O
  O7       1.0000 0.61998  0.38002  0.88002  Biso   1.000  O
  O8       1.0000 0.38002  0.61998  0.11998  Biso   1.000  O

++++++++++++++++++++++++++++++++++++++++
AMMONIA
_cell_length_a       5.1305
_cell_length_b       5.1305
_cell_length_c       5.1305
_cell_angle_alpha    90
_cell_angle_beta     90
_cell_angle_gamma    90

_symmetry_space_group_name_H-M    "P 1"
_symmetry_int_tables_number       1

loop_
  _symmetry_equiv_pos_as_xyz
  'x, y, z'

loop_
  _atom_site_label
  _atom_site_occupancy
  _atom_site_fract_x
  _atom_site_fract_y
  _atom_site_fract_z
  _atom_site_thermal_displace_type
  _atom_site_B_iso_or_equiv
  _atom_site_type_symbol
  N1       1.0000 0.20363  0.20362  0.20363  Biso   1.000  N
  N2       1.0000 0.70363  0.29638  0.79637  Biso   1.000  N
  N3       1.0000 0.79638  0.70362  0.29637  Biso   1.000  N
  N4       1.0000 0.29638  0.79638  0.70363  Biso   1.000  N
  H1       1.0000 0.36292  0.27216  0.10382  Biso   1.000  H
  H2       1.0000 0.27217  0.10381  0.36294  Biso   1.000  H
  H3       1.0000 0.10380  0.36291  0.27216  Biso   1.000  H
  H4       1.0000 0.86292  0.22784  0.89618  Biso   1.000  H
  H5       1.0000 0.77217  0.39619  0.63706  Biso   1.000  H
  H6       1.0000 0.60380  0.13709  0.72784  Biso   1.000  H
  H7       1.0000 0.72783  0.60381  0.13706  Biso   1.000  H
  H8       1.0000 0.89620  0.86291  0.22784  Biso   1.000  H
  H9       1.0000 0.39620  0.63709  0.77216  Biso   1.000  H
  H10      1.0000 0.13708  0.72784  0.60382  Biso   1.000  H
  H11      1.0000 0.22783  0.89619  0.86294  Biso   1.000  H
  H12      1.0000 0.63707  0.77216  0.39618  Biso   1.000  H

++++++++++++++++++++++++++++++++++++++++
BENZENE 
_cell_length_a       7.39
_cell_length_b       9.42
_cell_length_c       6.81
_cell_angle_alpha    90
_cell_angle_beta     90
_cell_angle_gamma    90

_symmetry_space_group_name_H-M    "P 1"
_symmetry_int_tables_number       1

loop_
  _symmetry_equiv_pos_as_xyz
  'x, y, z'

loop_
  _atom_site_label
  _atom_site_occupancy
  _atom_site_fract_x
  _atom_site_fract_y
  _atom_site_fract_z
  _atom_site_thermal_displace_type
  _atom_site_B_iso_or_equiv
  _atom_site_type_symbol
  C1       1.0000 0.93997  0.14064  0.99361  Biso   1.000  C
  C2       1.0000 0.86203  0.04525  0.12613  Biso   1.000  C
  C3       1.0000 0.07749  0.09525  0.86753  Biso   1.000  C
  C4       1.0000 0.92251  0.90475  0.13247  Biso   1.000  C
  C5       1.0000 0.13797  0.95475  0.87387  Biso   1.000  C
  C6       1.0000 0.06003  0.85936  0.00639  Biso   1.000  C
  C7       1.0000 0.56003  0.64064  0.99361  Biso   1.000  C
  C8       1.0000 0.93997  0.35936  0.49361  Biso   1.000  C
  C9       1.0000 0.56003  0.85936  0.49361  Biso   1.000  C
  C10      1.0000 0.43997  0.35936  0.00639  Biso   1.000  C
  C11      1.0000 0.06003  0.64064  0.50639  Biso   1.000  C
  C12      1.0000 0.43997  0.14064  0.50639  Biso   1.000  C
  C13      1.0000 0.63797  0.54525  0.12613  Biso   1.000  C
  C14      1.0000 0.86203  0.45475  0.62613  Biso   1.000  C
  C15      1.0000 0.63797  0.95475  0.62613  Biso   1.000  C
  C16      1.0000 0.36203  0.45475  0.87387  Biso   1.000  C
  C17      1.0000 0.13797  0.54525  0.37387  Biso   1.000  C
  C18      1.0000 0.36203  0.04525  0.37387  Biso   1.000  C
  C19      1.0000 0.42251  0.59525  0.86753  Biso   1.000  C
  C20      1.0000 0.07749  0.40475  0.36753  Biso   1.000  C
  C21      1.0000 0.42251  0.90475  0.36753  Biso   1.000  C
  C22      1.0000 0.57749  0.40475  0.13247  Biso   1.000  C
  C23      1.0000 0.92251  0.59525  0.63247  Biso   1.000  C
  C24      1.0000 0.57749  0.09525  0.63247  Biso   1.000  C
  H1       1.0000 0.89362  0.25049  0.98833  Biso   1.000  H
  H2       1.0000 0.75368  0.07929  0.22407  Biso   1.000  H
  H3       1.0000 0.13726  0.16915  0.76303  Biso   1.000  H
  H4       1.0000 0.86274  0.83085  0.23697  Biso   1.000  H
  H5       1.0000 0.24632  0.92071  0.77593  Biso   1.000  H
  H6       1.0000 0.10638  0.74951  0.01167  Biso   1.000  H
  H7       1.0000 0.60638  0.75049  0.98833  Biso   1.000  H
  H8       1.0000 0.89362  0.24951  0.48833  Biso   1.000  H
  H9       1.0000 0.60638  0.74951  0.48833  Biso   1.000  H
  H10      1.0000 0.39362  0.24951  0.01167  Biso   1.000  H
  H11      1.0000 0.10638  0.75049  0.51167  Biso   1.000  H
  H12      1.0000 0.39362  0.25049  0.51167  Biso   1.000  H
  H13      1.0000 0.74632  0.57929  0.22407  Biso   1.000  H
  H14      1.0000 0.75368  0.42071  0.72407  Biso   1.000  H
  H15      1.0000 0.74632  0.92071  0.72407  Biso   1.000  H
  H16      1.0000 0.25368  0.42071  0.77593  Biso   1.000  H
  H17      1.0000 0.24632  0.57929  0.27593  Biso   1.000  H
  H18      1.0000 0.25368  0.07929  0.27593  Biso   1.000  H
  H19      1.0000 0.36274  0.66915  0.76303  Biso   1.000  H
  H20      1.0000 0.13726  0.33085  0.26303  Biso   1.000  H
  H21      1.0000 0.36274  0.83085  0.26303  Biso   1.000  H
  H22      1.0000 0.63726  0.33085  0.23697  Biso   1.000  H
  H23      1.0000 0.86274  0.66915  0.73697  Biso   1.000  H
  H24      1.0000 0.63726  0.16915  0.73697  Biso   1.000  H

++++++++++++++++++++++++++++++++++++++++
NAPHTHALENE
_cell_length_a       8.0846
_cell_length_b       5.9375
_cell_length_c       8.6335
_cell_angle_alpha    90
_cell_angle_beta     124.673
_cell_angle_gamma    90

_symmetry_space_group_name_H-M    "P 1"
_symmetry_int_tables_number       1

loop_
  _symmetry_equiv_pos_as_xyz
  'x, y, z'

loop_
  _atom_site_label
  _atom_site_occupancy
  _atom_site_fract_x
  _atom_site_fract_y
  _atom_site_fract_z
  _atom_site_thermal_displace_type
  _atom_site_B_iso_or_equiv
  _atom_site_type_symbol
  C1       1.0000 0.08311  0.01978  0.33017  Biso   1.000  C
  C2       1.0000 0.91689  0.98022  0.66983  Biso   1.000  C
  C3       1.0000 0.41689  0.51978  0.66983  Biso   1.000  C
  C4       1.0000 0.58311  0.48022  0.33017  Biso   1.000  C
  C5       1.0000 0.11355  0.16450  0.22369  Biso   1.000  C
  C6       1.0000 0.88645  0.83550  0.77631  Biso   1.000  C
  C7       1.0000 0.38645  0.66450  0.77631  Biso   1.000  C
  C8       1.0000 0.61355  0.33550  0.22369  Biso   1.000  C
  C9       1.0000 0.04847  0.10652  0.03765  Biso   1.000  C
  C10      1.0000 0.95153  0.89348  0.96235  Biso   1.000  C
  C11      1.0000 0.45153  0.60652  0.96235  Biso   1.000  C
  C12      1.0000 0.54847  0.39348  0.03765  Biso   1.000  C
  C13      1.0000 0.07642  0.25246  0.92415  Biso   1.000  C
  C14      1.0000 0.92358  0.74754  0.07585  Biso   1.000  C
  C15      1.0000 0.42358  0.75246  0.07585  Biso   1.000  C
  C16      1.0000 0.57642  0.24754  0.92415  Biso   1.000  C
  C17      1.0000 0.01285  0.19055  0.74428  Biso   1.000  C
  C18      1.0000 0.98715  0.80945  0.25572  Biso   1.000  C
  C19      1.0000 0.48715  0.69055  0.25572  Biso   1.000  C
  C20      1.0000 0.51285  0.30945  0.74428  Biso   1.000  C
  H1       1.0000 0.13301  0.06836  0.47227  Biso   1.000  H
  H2       1.0000 0.86699  0.93164  0.52773  Biso   1.000  H
  H3       1.0000 0.36699  0.56836  0.52773  Biso   1.000  H
  H4       1.0000 0.63301  0.43164  0.47227  Biso   1.000  H
  H5       1.0000 0.18671  0.32677  0.28011  Biso   1.000  H
  H6       1.0000 0.81329  0.67323  0.71989  Biso   1.000  H
  H7       1.0000 0.31329  0.82677  0.71989  Biso   1.000  H
  H8       1.0000 0.68671  0.17323  0.28011  Biso   1.000  H
  H9       1.0000 0.14995  0.41420  0.98198  Biso   1.000  H
  H10      1.0000 0.85005  0.58580  0.01802  Biso   1.000  H
  H11      1.0000 0.35005  0.91420  0.01802  Biso   1.000  H
  H12      1.0000 0.64995  0.08580  0.98198  Biso   1.000  H
  H13      1.0000 0.03729  0.30392  0.65957  Biso   1.000  H
  H14      1.0000 0.96271  0.69608  0.34043  Biso   1.000  H
  H15      1.0000 0.46271  0.80392  0.34043  Biso   1.000  H
  H16      1.0000 0.53729  0.19608  0.65957  Biso   1.000  H
++++++++++++++++++++++++++++++++++++++++
ANTHRACENE
_cell_length_a       8.4144
_cell_length_b       5.9903
_cell_length_c       11.0953
_cell_angle_alpha    90
_cell_angle_beta     125.293
_cell_angle_gamma    90

_symmetry_space_group_name_H-M    "P 1"
_symmetry_int_tables_number       1

loop_
  _symmetry_equiv_pos_as_xyz
  'x, y, z'

loop_
  _atom_site_label
  _atom_site_occupancy
  _atom_site_fract_x
  _atom_site_fract_y
  _atom_site_fract_z
  _atom_site_thermal_displace_type
  _atom_site_B_iso_or_equiv
  _atom_site_type_symbol
  C1       1.0000 0.08599  0.02583  0.36914  Biso   1.000  C
  C2       1.0000 0.11698  0.15621  0.28268  Biso   1.000  C
  C3       1.0000 0.05874  0.08212  0.14033  Biso   1.000  C
  C4       1.0000 0.08682  0.21225  0.04912  Biso   1.000  C
  C5       1.0000 0.96932  0.86452  0.08971  Biso   1.000  C
  C6       1.0000 0.94008  0.73438  0.18295  Biso   1.000  C
  C7       1.0000 0.99557  0.81283  0.31837  Biso   1.000  C
  C8       1.0000 0.03068  0.13548  0.91029  Biso   1.000  C
  C9       1.0000 0.91318  0.78775  0.95088  Biso   1.000  C
  C10      1.0000 0.94126  0.91788  0.85967  Biso   1.000  C
  C11      1.0000 0.05992  0.26562  0.81705  Biso   1.000  C
  C12      1.0000 0.88302  0.84379  0.71732  Biso   1.000  C
  C13      1.0000 0.00443  0.18717  0.68163  Biso   1.000  C
  C14      1.0000 0.91401  0.97417  0.63086  Biso   1.000  C
  C15      1.0000 0.41401  0.52583  0.63086  Biso   1.000  C
  C16      1.0000 0.58599  0.47417  0.36914  Biso   1.000  C
  C17      1.0000 0.38302  0.65621  0.71732  Biso   1.000  C
  C18      1.0000 0.61698  0.34379  0.28268  Biso   1.000  C
  C19      1.0000 0.44126  0.58212  0.85967  Biso   1.000  C
  C20      1.0000 0.55874  0.41788  0.14033  Biso   1.000  C
  C21      1.0000 0.41318  0.71225  0.95088  Biso   1.000  C
  C22      1.0000 0.58682  0.28775  0.04912  Biso   1.000  C
  C23      1.0000 0.53068  0.36452  0.91029  Biso   1.000  C
  C24      1.0000 0.46932  0.63548  0.08971  Biso   1.000  C
  C25      1.0000 0.55992  0.23438  0.81705  Biso   1.000  C
  C26      1.0000 0.44008  0.76562  0.18295  Biso   1.000  C
  C27      1.0000 0.50443  0.31283  0.68163  Biso   1.000  C
  C28      1.0000 0.49557  0.68717  0.31837  Biso   1.000  C
  H1       1.0000 0.13135  0.08582  0.47742  Biso   1.000  H
  H2       1.0000 0.18569  0.31962  0.32086  Biso   1.000  H
  H3       1.0000 0.15521  0.37608  0.08741  Biso   1.000  H
  H4       1.0000 0.87233  0.57083  0.14441  Biso   1.000  H
  H5       1.0000 0.97095  0.71082  0.38763  Biso   1.000  H
  H6       1.0000 0.84479  0.62392  0.91259  Biso   1.000  H
  H7       1.0000 0.12767  0.42917  0.85559  Biso   1.000  H
  H8       1.0000 0.81431  0.68038  0.67914  Biso   1.000  H
  H9       1.0000 0.02905  0.28918  0.61237  Biso   1.000  H
  H10      1.0000 0.86865  0.91418  0.52258  Biso   1.000  H
  H11      1.0000 0.36865  0.58582  0.52258  Biso   1.000  H
  H12      1.0000 0.63135  0.41418  0.47742  Biso   1.000  H
  H13      1.0000 0.31431  0.81962  0.67914  Biso   1.000  H
  H14      1.0000 0.68569  0.18038  0.32086  Biso   1.000  H
  H15      1.0000 0.34479  0.87608  0.91259  Biso   1.000  H
  H16      1.0000 0.65521  0.12392  0.08741  Biso   1.000  H
  H17      1.0000 0.62767  0.07083  0.85559  Biso   1.000  H
  H18      1.0000 0.37233  0.92917  0.14441  Biso   1.000  H
  H19      1.0000 0.52905  0.21082  0.61237  Biso   1.000  H
  H20      1.0000 0.47095  0.78918  0.38763  Biso   1.000  H
++++++++++++++++++++++++++++++++++++++++
ICE Ih
_cell_length_a                         7.76992
_cell_length_b                         7.76992
_cell_length_c                         7.32109
_cell_angle_alpha                      90
_cell_angle_beta                       90
_cell_angle_gamma                      60.00000
_symmetry_space_group_name_H-M         'P 1'
_symmetry_Int_Tables_number            1

loop_
_symmetry_equiv_pos_as_xyz
   'x, y, z'

loop_
   _atom_site_label
   _atom_site_occupancy
   _atom_site_fract_x
   _atom_site_fract_y
   _atom_site_fract_z
   _atom_site_adp_type
   _atom_site_B_iso_or_equiv
   _atom_site_type_symbol
   H1         1.0     0.334627      0.665351      0.198240     Biso  1.000000 H
   H2         1.0     0.334666      0.999979      0.698235     Biso  1.000000 H
   H3         1.0     0.453477      0.546508      0.020688     Biso  1.000000 H
   H4         1.0     0.453512      0.999982      0.520679     Biso  1.000000 H
   H5         1.0     0.332362      0.881324      0.984291     Biso  1.000000 H
   H6         1.0     0.213687      0.667621      0.484302     Biso  1.000000 H
   H7         1.0     0.332362      0.786290      0.484294     Biso  1.000000 H
   H8         1.0     0.213694      1.118665      0.984286     Biso  1.000000 H
   H9         1.0     0.665339      0.000009      0.198229     Biso  1.000000 H
   H10        1.0     0.665331      0.334660      0.698249     Biso  1.000000 H
   H11        1.0     0.546488      0.000005      0.020679     Biso  1.000000 H
   H12        1.0     0.546491      0.453493      0.520688     Biso  1.000000 H
   H13        1.0     0.786284      0.332375      0.984305     Biso  1.000000 H
   H14        1.0     0.667608      0.213708      0.984294     Biso  1.000000 H
   H15        1.0     0.667639      0.118663      0.484293     Biso  1.000000 H
   H16        1.0     0.786307     -0.118682      0.484284     Biso  1.000000 H
   H17        1.0     0.000006      0.334630      0.198247     Biso  1.000000 H
   H18        1.0    -0.000012      0.453492      0.020699     Biso  1.000000 H
   H19        1.0     0.118683      0.213668      0.484305     Biso  1.000000 H
   H20        1.0    -0.118661      0.332344      0.484305     Biso  1.000000 H
   H21        1.0     0.118645      0.667630      0.984298     Biso  1.000000 H
   H22        1.0    -0.000019      0.665331      0.698249     Biso  1.000000 H
   H23        1.0    -0.000001      0.546476      0.520695     Biso  1.000000 H
   H24        1.0    -0.118698      0.786302      0.984297     Biso  1.000000 H
   O1         1.0     0.331034      0.668942      0.062031     Biso  1.000000 O
   O2         1.0     0.335972      0.664006      0.437179     Biso  1.000000 O
   O3         1.0     0.331080      0.999973      0.562025     Biso  1.000000 O
   O4         1.0     0.335979      0.999992      0.937168     Biso  1.000000 O
   O5         1.0     0.668926      0.000011      0.062018     Biso  1.000000 O
   O6         1.0     0.668932      0.331063      0.562040     Biso  1.000000 O
   O7         1.0     0.663995      0.335997      0.937188     Biso  1.000000 O
   O8         1.0     0.664029     -0.000006      0.437162     Biso  1.000000 O
   O9         1.0     0.000001      0.331045      0.062038     Biso  1.000000 O
   O10        1.0     0.000012      0.335956      0.437190     Biso  1.000000 O
   O11        1.0    -0.000013      0.668920      0.562040     Biso  1.000000 O
   O12        1.0    -0.000026      0.664009      0.937191     Biso  1.000000 O
++++++++++++++++++++++++++++++++++++++++
ICE II
_cell_length_a                         7.74536
_cell_length_b                         7.74536
_cell_length_c                         7.74536
_cell_angle_alpha                      113.11900
_cell_angle_beta                       113.11900
_cell_angle_gamma                      113.11900
_symmetry_space_group_name_H-M         'P 1'
_symmetry_Int_Tables_number            1

loop_
_symmetry_equiv_pos_as_xyz
   'x, y, z'

loop_
   _atom_site_label
   _atom_site_occupancy
   _atom_site_fract_x
   _atom_site_fract_y
   _atom_site_fract_z
   _atom_site_adp_type
   _atom_site_B_iso_or_equiv
   _atom_site_type_symbol
   H1         1.0     0.037584      0.793687      0.142440     Biso  1.000000 H
   H2         1.0     0.192014      0.987960      0.423094     Biso  1.000000 H
   H3         1.0     0.590713      0.270286      0.587612     Biso  1.000000 H
   H4         1.0     0.626077      0.254891      0.794669     Biso  1.000000 H
   H5         1.0     0.142444      0.037570      0.793677     Biso  1.000000 H
   H6         1.0     0.423105      0.192016      0.987948     Biso  1.000000 H
   H7         1.0     0.587605      0.590691      0.270265     Biso  1.000000 H
   H8         1.0     0.794681      0.626087      0.254891     Biso  1.000000 H
   H9         1.0     0.793642      0.142413      0.037547     Biso  1.000000 H
   H10        1.0     0.987908      0.423070      0.191984     Biso  1.000000 H
   H11        1.0     0.270229      0.587582      0.590676     Biso  1.000000 H
   H12        1.0     0.254865      0.794660      0.626061     Biso  1.000000 H
   H13        1.0     0.962403      0.206300      0.857548     Biso  1.000000 H
   H14        1.0     0.807973      0.012027      0.576893     Biso  1.000000 H
   H15        1.0     0.409274      0.729701      0.412375     Biso  1.000000 H
   H16        1.0     0.373910      0.745096      0.205318     Biso  1.000000 H
   H17        1.0     0.857543      0.962417      0.206310     Biso  1.000000 H
   H18        1.0     0.576882      0.807971      0.012039     Biso  1.000000 H
   H19        1.0     0.412382      0.409296      0.729722     Biso  1.000000 H
   H20        1.0     0.205306      0.373900      0.745096     Biso  1.000000 H
   H21        1.0     0.206345      0.857574      0.962440     Biso  1.000000 H
   H22        1.0     0.012079      0.576917      0.808003     Biso  1.000000 H
   H23        1.0     0.729758      0.412405      0.409311     Biso  1.000000 H
   H24        1.0     0.745122      0.205327      0.373926     Biso  1.000000 H
   O1         1.0     0.018860      0.855482      0.267715     Biso  1.000000 O
   O2         1.0     0.517885      0.249302      0.666049     Biso  1.000000 O
   O3         1.0     0.267730      0.018859      0.855473     Biso  1.000000 O
   O4         1.0     0.666052      0.517874      0.249279     Biso  1.000000 O
   O5         1.0     0.855434      0.267693      0.018827     Biso  1.000000 O
   O6         1.0     0.249229      0.666011      0.517842     Biso  1.000000 O
   O7         1.0     0.981127      0.144505      0.732272     Biso  1.000000 O
   O8         1.0     0.482102      0.750685      0.333938     Biso  1.000000 O
   O9         1.0     0.732257      0.981128      0.144514     Biso  1.000000 O
   O10        1.0     0.333935      0.482113      0.750709     Biso  1.000000 O
   O11        1.0     0.144553      0.732294      0.981160     Biso  1.000000 O
   O12        1.0     0.750758      0.333976      0.482145     Biso  1.000000 O
++++++++++++++++++++++++++++++++++++++++
ICE VIII
_cell_length_a                         4.84947
_cell_length_b                         4.84947
_cell_length_c                         7.05651
_cell_angle_alpha                      90
_cell_angle_beta                       90
_cell_angle_gamma                      90
_symmetry_space_group_name_H-M         'P 1'
_symmetry_Int_Tables_number            1

loop_
_symmetry_equiv_pos_as_xyz
   'x, y, z'

loop_
   _atom_site_label
   _atom_site_occupancy
   _atom_site_fract_x
   _atom_site_fract_y
   _atom_site_fract_z
   _atom_site_adp_type
   _atom_site_B_iso_or_equiv
   _atom_site_type_symbol
   H1         1.0     0.000000      0.411155      0.193491     Biso  1.000000 H
   H2         1.0     0.500000      0.911155      0.693491     Biso  1.000000 H
   H3         1.0    -0.161155      0.750000      0.443491     Biso  1.000000 H
   H4         1.0     0.661155      0.250000      0.943491     Biso  1.000000 H
   H5         1.0     0.500000      0.411155      0.306509     Biso  1.000000 H
   H6         1.0     0.500000      0.088845      0.306509     Biso  1.000000 H
   H7         1.0     0.661155      0.750000      0.056509     Biso  1.000000 H
   H8         1.0    -0.161155      0.250000      0.556509     Biso  1.000000 H
   H9         1.0     0.338845      0.750000      0.056509     Biso  1.000000 H
   H10        1.0     0.338845      0.250000      0.943491     Biso  1.000000 H
   H11        1.0     0.000000      0.088845      0.193491     Biso  1.000000 H
   H12        1.0     0.161155      0.750000      0.443491     Biso  1.000000 H
   H13        1.0     0.161155      0.250000      0.556509     Biso  1.000000 H
   H14        1.0     0.500000      0.588845      0.693491     Biso  1.000000 H
   H15        1.0     0.000000      0.911155      0.806509     Biso  1.000000 H
   H16        1.0     0.000000      0.588845      0.806509     Biso  1.000000 H
   O1         1.0     0.000000      0.250000      0.108659     Biso  1.000000 O
   O2         1.0     0.000000      0.750000      0.358659     Biso  1.000000 O
   O3         1.0     0.500000      0.250000      0.858659     Biso  1.000000 O
   O4         1.0     0.500000      0.250000      0.391341     Biso  1.000000 O
   O5         1.0     0.500000      0.750000      0.141341     Biso  1.000000 O
   O6         1.0     0.000000      0.250000      0.641341     Biso  1.000000 O
   O7         1.0     0.500000      0.750000      0.608659     Biso  1.000000 O
   O8         1.0     0.000000      0.750000      0.891341     Biso  1.000000 O
\end{verbatim}
}

\newpage
\section{ Geometries of the reference molecules used for the DMC, RPA and RPA-GWSE calculations }\label{sec:geoMol}
The following configurations are given in xyz format.
{\scriptsize
\begin{verbatim}
3
CARBON DIOXIDE
C      4.81200000       4.81200000       5.62400000 
O      5.49105000       4.13279000       4.95756000 
O      4.13172000       5.49276000       6.28717000 
++++++++++++++++++++++++++++++++++++++++
4
AMMONIA
N    -0.0395476   -0.0395339   -0.0395339 
H     0.8032824    0.2694560   -0.5233139 
H     0.2695324   -0.5233139    0.8032260 
H    -0.5232475    0.8032360    0.2694660 
++++++++++++++++++++++++++++++++++++++++
12
BENZENE 
C     0.4153343   -1.3316828   -0.0204638 
C    -0.4153456    1.3317169    0.0204761 
C     0.9990143   -0.4326029    0.8725061 
C    -0.9990155    0.4326070   -0.8724938 
C     0.5836843    0.8990970    0.8929561 
C    -0.5836656   -0.8991329   -0.8929838 
H     0.7393443   -2.3704527   -0.0363838 
H    -0.7393556    2.3704368    0.0364061 
H     1.7782042   -0.7701029    1.5530460 
H    -1.7782355    0.7701070   -1.5530337 
H     1.0388543    1.6004269    1.5895060 
H    -1.0388855   -1.6004428   -1.5895037 
++++++++++++++++++++++++++++++++++++++++
18
NAPHTHALENE
C    -1.7405183   -1.8383406   -0.0000900 
C    -0.3638385   -1.8372006   -0.0002800 
C     0.3624115   -0.6189207   -0.0002100 
C    -0.3625085    0.6189791    0.0001000 
C    -1.7802083    0.5817892    0.0004300 
C    -2.4549083   -0.6181907    0.0002300 
C     1.7802713   -0.5818707   -0.0004100 
C     0.3637815    1.8371790    0.0002700 
C     1.7405513    1.8383290    0.0001500 
C     2.4548713    0.6182291   -0.0002400 
H     2.3293013   -1.5226707   -0.0005100 
H    -2.2852083   -2.7801805   -0.0000400 
H     0.1884015   -2.7761405   -0.0005000 
H    -2.3288983    1.5227591    0.0006100 
H    -3.5427082   -0.6322607    0.0005400 
H    -0.1881385    2.7762490    0.0005800 
H     2.2855713    2.7799990    0.0002000 
H     3.5428112    0.6324591   -0.0002800 
++++++++++++++++++++++++++++++++++++++++
24
ANTHRACENE
C    -2.7918100   -2.4660516   -0.0030577 
C    -1.4210601   -2.4658816    0.0018823 
C    -0.6878402   -1.2428117    0.0011423 
C    -1.4203801    0.0033881   -0.0048777 
C    -2.8455300   -0.0423519   -0.0097677 
C    -3.5123299   -1.2400917   -0.0089877 
C     0.7113697   -1.2102217    0.0058423 
C    -0.7114402    1.2101980   -0.0056577 
C     0.6878897    1.2427880   -0.0009877 
C     1.4203097   -0.0033819    0.0048923 
C     2.8455095    0.0424381    0.0096423 
C     3.5125095    1.2401380    0.0087423 
C     2.7918095    2.4660079    0.0030023 
C     1.4210097    2.4658179   -0.0018277 
H     3.3964095   -0.8971818    0.0142023 
H     1.2639097   -2.1502517    0.0102323 
H    -3.3379999   -3.4068815   -0.0026477 
H    -0.8678701   -3.4041315    0.0062723 
H    -3.3963799    0.8972481   -0.0142577 
H    -4.6001198   -1.2595417   -0.0126777 
H    -1.2640801    2.1502679   -0.0100577 
H     4.6002094    1.2596380    0.0125823 
H     3.3379095    3.4069878    0.0025523 
H     0.8678097    3.4040378   -0.0059877 
++++++++++++++++++++++++++++++++++++++++
3
WATER
H   6.95720000   6.00000000   6.00000000  
H   5.76001279   6.92662721   6.00000000  
O   6.00000000   6.00000000   6.00000000  
\end{verbatim}
}

\newpage

\end{widetext}

\bibliography{ref}

\begin{thebibliography}{91}%
\makeatletter
\providecommand \@ifxundefined [1]{%
 \@ifx{#1\undefined}
}%
\providecommand \@ifnum [1]{%
 \ifnum #1\expandafter \@firstoftwo
 \else \expandafter \@secondoftwo
 \fi
}%
\providecommand \@ifx [1]{%
 \ifx #1\expandafter \@firstoftwo
 \else \expandafter \@secondoftwo
 \fi
}%
\providecommand \natexlab [1]{#1}%
\providecommand \enquote  [1]{``#1''}%
\providecommand \bibnamefont  [1]{#1}%
\providecommand \bibfnamefont [1]{#1}%
\providecommand \citenamefont [1]{#1}%
\providecommand \href@noop [0]{\@secondoftwo}%
\providecommand \href [0]{\begingroup \@sanitize@url \@href}%
\providecommand \@href[1]{\@@startlink{#1}\@@href}%
\providecommand \@@href[1]{\endgroup#1\@@endlink}%
\providecommand \@sanitize@url [0]{\catcode `\\12\catcode `\$12\catcode
  `\&12\catcode `\#12\catcode `\^12\catcode `\_12\catcode `\%12\relax}%
\providecommand \@@startlink[1]{}%
\providecommand \@@endlink[0]{}%
\providecommand \url  [0]{\begingroup\@sanitize@url \@url }%
\providecommand \@url [1]{\endgroup\@href {#1}{\urlprefix }}%
\providecommand \urlprefix  [0]{URL }%
\providecommand \Eprint [0]{\href }%
\providecommand \doibase [0]{http://dx.doi.org/}%
\providecommand \selectlanguage [0]{\@gobble}%
\providecommand \bibinfo  [0]{\@secondoftwo}%
\providecommand \bibfield  [0]{\@secondoftwo}%
\providecommand \translation [1]{[#1]}%
\providecommand \BibitemOpen [0]{}%
\providecommand \bibitemStop [0]{}%
\providecommand \bibitemNoStop [0]{.\EOS\space}%
\providecommand \EOS [0]{\spacefactor3000\relax}%
\providecommand \BibitemShut  [1]{\csname bibitem#1\endcsname}%
\let\auto@bib@innerbib\@empty
\bibitem [{\citenamefont {Burke}(2012)}]{Burke2012:jcp_rev}%
  \BibitemOpen
  \bibfield  {author} {\bibinfo {author} {\bibfnamefont {K.}~\bibnamefont
  {Burke}},\ }\href@noop {} {\bibfield  {journal} {\bibinfo  {journal} {J.
  Chem. Phys.}\ }\textbf {\bibinfo {volume} {136}},\ \bibinfo {pages} {150901}
  (\bibinfo {year} {2012})}\BibitemShut {NoStop}%
\bibitem [{\citenamefont {Curtarolo}\ \emph {et~al.}(2013)\citenamefont
  {Curtarolo}, \citenamefont {Hart}, \citenamefont {Nardelli}, \citenamefont
  {Mingo}, \citenamefont {Sanvito},\ and\ \citenamefont
  {Levy}}]{Curtarolo:natmat2013}%
  \BibitemOpen
  \bibfield  {author} {\bibinfo {author} {\bibfnamefont {S.}~\bibnamefont
  {Curtarolo}}, \bibinfo {author} {\bibfnamefont {G.~L.~W.}\ \bibnamefont
  {Hart}}, \bibinfo {author} {\bibfnamefont {M.~B.}\ \bibnamefont {Nardelli}},
  \bibinfo {author} {\bibfnamefont {N.}~\bibnamefont {Mingo}}, \bibinfo
  {author} {\bibfnamefont {S.}~\bibnamefont {Sanvito}}, \ and\ \bibinfo
  {author} {\bibfnamefont {O.}~\bibnamefont {Levy}},\ }\href@noop {} {\bibfield
   {journal} {\bibinfo  {journal} {Nat. Mat.}\ }\textbf {\bibinfo {volume}
  {12}},\ \bibinfo {pages} {191} (\bibinfo {year} {2013})}\BibitemShut
  {NoStop}%
\bibitem [{\citenamefont {Marzari}(2016)}]{Marzari:natmat2016}%
  \BibitemOpen
  \bibfield  {author} {\bibinfo {author} {\bibfnamefont {N.}~\bibnamefont
  {Marzari}},\ }\href@noop {} {\bibfield  {journal} {\bibinfo  {journal} {Nat.
  Mat.}\ }\textbf {\bibinfo {volume} {15}},\ \bibinfo {pages} {381} (\bibinfo
  {year} {2016})}\BibitemShut {NoStop}%
\bibitem [{\citenamefont {Sun}\ \emph {et~al.}(2016)\citenamefont {Sun},
  \citenamefont {Remsing}, \citenamefont {Zhang}, \citenamefont {Sun},
  \citenamefont {Ruzsinszky}, \citenamefont {Peng}, \citenamefont {Yang},
  \citenamefont {Paul}, \citenamefont {Waghmare}, \citenamefont {Wu},
  \citenamefont {Klein},\ and\ \citenamefont {Perdew}}]{scan_natchem}%
  \BibitemOpen
  \bibfield  {author} {\bibinfo {author} {\bibfnamefont {J.}~\bibnamefont
  {Sun}}, \bibinfo {author} {\bibfnamefont {R.~C.}\ \bibnamefont {Remsing}},
  \bibinfo {author} {\bibfnamefont {Y.}~\bibnamefont {Zhang}}, \bibinfo
  {author} {\bibfnamefont {Z.}~\bibnamefont {Sun}}, \bibinfo {author}
  {\bibfnamefont {A.}~\bibnamefont {Ruzsinszky}}, \bibinfo {author}
  {\bibfnamefont {H.}~\bibnamefont {Peng}}, \bibinfo {author} {\bibfnamefont
  {Z.}~\bibnamefont {Yang}}, \bibinfo {author} {\bibfnamefont {A.}~\bibnamefont
  {Paul}}, \bibinfo {author} {\bibfnamefont {U.}~\bibnamefont {Waghmare}},
  \bibinfo {author} {\bibfnamefont {X.}~\bibnamefont {Wu}}, \bibinfo {author}
  {\bibfnamefont {M.~L.}\ \bibnamefont {Klein}}, \ and\ \bibinfo {author}
  {\bibfnamefont {J.~P.}\ \bibnamefont {Perdew}},\ }\href
  {http://dx.doi.org/10.1038/nchem.2535} {\bibfield  {journal} {\bibinfo
  {journal} {Nat. Chem.}\ }\textbf {\bibinfo {volume} {8}},\ \bibinfo {pages}
  {831} (\bibinfo {year} {2016})}\BibitemShut {NoStop}%
\bibitem [{\citenamefont {Cohen}\ \emph {et~al.}(2008)\citenamefont {Cohen},
  \citenamefont {Mori-Sanchez},\ and\ \citenamefont {Yang}}]{Cohen:2008fg}%
  \BibitemOpen
  \bibfield  {author} {\bibinfo {author} {\bibfnamefont {A.~J.}\ \bibnamefont
  {Cohen}}, \bibinfo {author} {\bibfnamefont {P.}~\bibnamefont {Mori-Sanchez}},
  \ and\ \bibinfo {author} {\bibfnamefont {W.}~\bibnamefont {Yang}},\
  }\href@noop {} {\bibfield  {journal} {\bibinfo  {journal} {Science (New York,
  NY)}\ }\textbf {\bibinfo {volume} {321}},\ \bibinfo {pages} {792} (\bibinfo
  {year} {2008})}\BibitemShut {NoStop}%
\bibitem [{\citenamefont {Peverati}\ and\ \citenamefont
  {Truhlar}(2014)}]{peverati2014}%
  \BibitemOpen
  \bibfield  {author} {\bibinfo {author} {\bibfnamefont {R.}~\bibnamefont
  {Peverati}}\ and\ \bibinfo {author} {\bibfnamefont {D.~G.}\ \bibnamefont
  {Truhlar}},\ }\href@noop {} {\bibfield  {journal} {\bibinfo  {journal} {Phil.
  Trans. R. Soc. A}\ }\textbf {\bibinfo {volume} {372}},\ \bibinfo {pages}
  {20120476} (\bibinfo {year} {2014})}\BibitemShut {NoStop}%
\bibitem [{\citenamefont {Medvedev}\ \emph {et~al.}(2017)\citenamefont
  {Medvedev}, \citenamefont {Bushmarinov}, \citenamefont {Sun}, \citenamefont
  {Perdew},\ and\ \citenamefont {Lyssenko}}]{dft_dens}%
  \BibitemOpen
  \bibfield  {author} {\bibinfo {author} {\bibfnamefont {M.~G.}\ \bibnamefont
  {Medvedev}}, \bibinfo {author} {\bibfnamefont {I.~S.}\ \bibnamefont
  {Bushmarinov}}, \bibinfo {author} {\bibfnamefont {J.}~\bibnamefont {Sun}},
  \bibinfo {author} {\bibfnamefont {J.~P.}\ \bibnamefont {Perdew}}, \ and\
  \bibinfo {author} {\bibfnamefont {K.~A.}\ \bibnamefont {Lyssenko}},\ }\href
  {\doibase 10.1126/science.aah5975} {\bibfield  {journal} {\bibinfo  {journal}
  {Science}\ }\textbf {\bibinfo {volume} {355}},\ \bibinfo {pages} {49}
  (\bibinfo {year} {2017})}\BibitemShut {NoStop}%
\bibitem [{\citenamefont {Hammes-Schiffer}(2017)}]{dft_dens_comment}%
  \BibitemOpen
  \bibfield  {author} {\bibinfo {author} {\bibfnamefont {S.}~\bibnamefont
  {Hammes-Schiffer}},\ }\href@noop {} {\bibfield  {journal} {\bibinfo
  {journal} {Science (New York, NY)}\ }\textbf {\bibinfo {volume} {355}},\
  \bibinfo {pages} {28} (\bibinfo {year} {2017})}\BibitemShut {NoStop}%
\bibitem [{\citenamefont {Grimme}\ \emph {et~al.}(2016)\citenamefont {Grimme},
  \citenamefont {Hansen}, \citenamefont {Brandenburg},\ and\ \citenamefont
  {Bannwarth}}]{grimme_chemrev}%
  \BibitemOpen
  \bibfield  {author} {\bibinfo {author} {\bibfnamefont {S.}~\bibnamefont
  {Grimme}}, \bibinfo {author} {\bibfnamefont {A.}~\bibnamefont {Hansen}},
  \bibinfo {author} {\bibfnamefont {J.~G.}\ \bibnamefont {Brandenburg}}, \ and\
  \bibinfo {author} {\bibfnamefont {C.}~\bibnamefont {Bannwarth}},\ }\href
  {\doibase 10.1021/acs.chemrev.5b00533} {\bibfield  {journal} {\bibinfo
  {journal} {{Chem. Rev.}}\ }\textbf {\bibinfo {volume} {116}},\ \bibinfo
  {pages} {5105} (\bibinfo {year} {2016})}\BibitemShut {NoStop}%
\bibitem [{\citenamefont {Klime\v{s}}\ and\ \citenamefont
  {Michaelides}(2012)}]{vdw_perspective}%
  \BibitemOpen
  \bibfield  {author} {\bibinfo {author} {\bibfnamefont {J.}~\bibnamefont
  {Klime\v{s}}}\ and\ \bibinfo {author} {\bibfnamefont {A.}~\bibnamefont
  {Michaelides}},\ }\href {\doibase 10.1063/1.4754130} {\bibfield  {journal}
  {\bibinfo  {journal} {J. Chem. Phys.}\ }\textbf {\bibinfo {volume} {137}},\
  \bibinfo {pages} {120901} (\bibinfo {year} {2012})}\BibitemShut {NoStop}%
\bibitem [{\citenamefont {Beran}(2016)}]{beran_chemrev}%
  \BibitemOpen
  \bibfield  {author} {\bibinfo {author} {\bibfnamefont {G.~J.~O.}\
  \bibnamefont {Beran}},\ }\href@noop {} {\bibfield  {journal} {\bibinfo
  {journal} {Chem. Rev.}\ }\textbf {\bibinfo {volume} {116}},\ \bibinfo {pages}
  {5567} (\bibinfo {year} {2016})}\BibitemShut {NoStop}%
\bibitem [{\citenamefont {Hermann}\ \emph {et~al.}(2017)\citenamefont
  {Hermann}, \citenamefont {DiStasio},\ and\ \citenamefont
  {Tkatchenko}}]{ChemRev2017:Tkatchenko}%
  \BibitemOpen
  \bibfield  {author} {\bibinfo {author} {\bibfnamefont {J.}~\bibnamefont
  {Hermann}}, \bibinfo {author} {\bibfnamefont {R.~A.}\ \bibnamefont
  {DiStasio}, \bibfnamefont {Jr.}}, \ and\ \bibinfo {author} {\bibfnamefont
  {A.}~\bibnamefont {Tkatchenko}},\ }\href@noop {} {\bibfield  {journal}
  {\bibinfo  {journal} {Chem. Rev.}\ }\textbf {\bibinfo {volume} {117}},\
  \bibinfo {pages} {4714} (\bibinfo {year} {2017})}\BibitemShut {NoStop}%
\bibitem [{\citenamefont {Sun}\ \emph {et~al.}(2015)\citenamefont {Sun},
  \citenamefont {Ruzsinszky},\ and\ \citenamefont {Perdew}}]{scan}%
  \BibitemOpen
  \bibfield  {author} {\bibinfo {author} {\bibfnamefont {J.}~\bibnamefont
  {Sun}}, \bibinfo {author} {\bibfnamefont {A.}~\bibnamefont {Ruzsinszky}}, \
  and\ \bibinfo {author} {\bibfnamefont {J.~P.}\ \bibnamefont {Perdew}},\
  }\href@noop {} {\bibfield  {journal} {\bibinfo  {journal} {Phys. Rev. Lett.}\
  }\textbf {\bibinfo {volume} {115}},\ \bibinfo {pages} {036402} (\bibinfo
  {year} {2015})}\BibitemShut {NoStop}%
\bibitem [{\citenamefont {Mardirossian}\ and\ \citenamefont
  {Head-Gordon}(2014)}]{wb97xv}%
  \BibitemOpen
  \bibfield  {author} {\bibinfo {author} {\bibfnamefont {N.}~\bibnamefont
  {Mardirossian}}\ and\ \bibinfo {author} {\bibfnamefont {M.}~\bibnamefont
  {Head-Gordon}},\ }\href@noop {} {\bibfield  {journal} {\bibinfo  {journal}
  {Phys. Chem. Chem. Phys.}\ }\textbf {\bibinfo {volume} {16}},\ \bibinfo
  {pages} {9904} (\bibinfo {year} {2014})}\BibitemShut {NoStop}%
\bibitem [{\citenamefont {Wang}\ \emph {et~al.}(2017)\citenamefont {Wang},
  \citenamefont {Jin}, \citenamefont {Yu}, \citenamefont {Truhlar},\ and\
  \citenamefont {He}}]{revm06l}%
  \BibitemOpen
  \bibfield  {author} {\bibinfo {author} {\bibfnamefont {Y.}~\bibnamefont
  {Wang}}, \bibinfo {author} {\bibfnamefont {X.}~\bibnamefont {Jin}}, \bibinfo
  {author} {\bibfnamefont {H.~S.}\ \bibnamefont {Yu}}, \bibinfo {author}
  {\bibfnamefont {D.~G.}\ \bibnamefont {Truhlar}}, \ and\ \bibinfo {author}
  {\bibfnamefont {X.}~\bibnamefont {He}},\ }\href@noop {} {\bibfield  {journal}
  {\bibinfo  {journal} {Proc Natl Acad Sci USA}\ }\textbf {\bibinfo {volume}
  {114}},\ \bibinfo {pages} {8487} (\bibinfo {year} {2017})}\BibitemShut
  {NoStop}%
\bibitem [{\citenamefont {Cruz-Cabeza}\ \emph {et~al.}(2015)\citenamefont
  {Cruz-Cabeza}, \citenamefont {Reutzel-Edens},\ and\ \citenamefont
  {Bernstein}}]{polymophism}%
  \BibitemOpen
  \bibfield  {author} {\bibinfo {author} {\bibfnamefont {A.~J.}\ \bibnamefont
  {Cruz-Cabeza}}, \bibinfo {author} {\bibfnamefont {S.~M.}\ \bibnamefont
  {Reutzel-Edens}}, \ and\ \bibinfo {author} {\bibfnamefont {J.}~\bibnamefont
  {Bernstein}},\ }\href {\doibase 10.1039/C5CS00227C} {\bibfield  {journal}
  {\bibinfo  {journal} {Chem. Soc. Rev.}\ }\textbf {\bibinfo {volume} {44}},\
  \bibinfo {pages} {8619} (\bibinfo {year} {2015})}\BibitemShut {NoStop}%
\bibitem [{\citenamefont {{A. M. Reilly {\em et al.}}}(2016)}]{CSP2016}%
  \BibitemOpen
  \bibfield  {author} {\bibinfo {author} {\bibnamefont {{A. M. Reilly {\em et
  al.}}}},\ }\href@noop {} {\bibfield  {journal} {\bibinfo  {journal} {Acta
  Cryst. B}\ }\textbf {\bibinfo {volume} {72}},\ \bibinfo {pages} {439}
  (\bibinfo {year} {2016})}\BibitemShut {NoStop}%
\bibitem [{\citenamefont {Booth}\ \emph {et~al.}(2013)\citenamefont {Booth},
  \citenamefont {Gr{\"u}neis}, \citenamefont {Kresse},\ and\ \citenamefont
  {Alavi}}]{FCIQMC:Nat2013}%
  \BibitemOpen
  \bibfield  {author} {\bibinfo {author} {\bibfnamefont {G.~H.}\ \bibnamefont
  {Booth}}, \bibinfo {author} {\bibfnamefont {A.}~\bibnamefont {Gr{\"u}neis}},
  \bibinfo {author} {\bibfnamefont {G.}~\bibnamefont {Kresse}}, \ and\ \bibinfo
  {author} {\bibfnamefont {A.}~\bibnamefont {Alavi}},\ }\href@noop {}
  {\bibfield  {journal} {\bibinfo  {journal} {Nature}\ }\textbf {\bibinfo
  {volume} {493}},\ \bibinfo {pages} {365} (\bibinfo {year}
  {2013})}\BibitemShut {NoStop}%
\bibitem [{\citenamefont {Wen}\ \emph {et~al.}(2012)\citenamefont {Wen},
  \citenamefont {Nanda}, \citenamefont {Huang},\ and\ \citenamefont
  {Beran}}]{Wen:2012cu}%
  \BibitemOpen
  \bibfield  {author} {\bibinfo {author} {\bibfnamefont {S.}~\bibnamefont
  {Wen}}, \bibinfo {author} {\bibfnamefont {K.}~\bibnamefont {Nanda}}, \bibinfo
  {author} {\bibfnamefont {Y.}~\bibnamefont {Huang}}, \ and\ \bibinfo {author}
  {\bibfnamefont {G.~J.~O.}\ \bibnamefont {Beran}},\ }\href@noop {} {\bibfield
  {journal} {\bibinfo  {journal} {Phys. Chem. Chem. Phys.}\ }\textbf {\bibinfo
  {volume} {14}},\ \bibinfo {pages} {7578} (\bibinfo {year}
  {2012})}\BibitemShut {NoStop}%
\bibitem [{\citenamefont {Bygrave}\ \emph {et~al.}(2012)\citenamefont
  {Bygrave}, \citenamefont {Allan},\ and\ \citenamefont
  {Manby}}]{Bygrave:2012eo}%
  \BibitemOpen
  \bibfield  {author} {\bibinfo {author} {\bibfnamefont {P.~J.}\ \bibnamefont
  {Bygrave}}, \bibinfo {author} {\bibfnamefont {N.~L.}\ \bibnamefont {Allan}},
  \ and\ \bibinfo {author} {\bibfnamefont {F.~R.}\ \bibnamefont {Manby}},\
  }\href@noop {} {\bibfield  {journal} {\bibinfo  {journal} {J. Chem. Phys.}\
  }\textbf {\bibinfo {volume} {137}},\ \bibinfo {pages} {164102} (\bibinfo
  {year} {2012})}\BibitemShut {NoStop}%
\bibitem [{\citenamefont {Yang}\ \emph {et~al.}(2014)\citenamefont {Yang},
  \citenamefont {Hu}, \citenamefont {Usvyat}, \citenamefont {Matthews},
  \citenamefont {Schutz},\ and\ \citenamefont {Chan}}]{GarnetChan:Sci2014}%
  \BibitemOpen
  \bibfield  {author} {\bibinfo {author} {\bibfnamefont {J.}~\bibnamefont
  {Yang}}, \bibinfo {author} {\bibfnamefont {W.}~\bibnamefont {Hu}}, \bibinfo
  {author} {\bibfnamefont {D.}~\bibnamefont {Usvyat}}, \bibinfo {author}
  {\bibfnamefont {D.}~\bibnamefont {Matthews}}, \bibinfo {author}
  {\bibfnamefont {M.}~\bibnamefont {Schutz}}, \ and\ \bibinfo {author}
  {\bibfnamefont {G.~K.~L.}\ \bibnamefont {Chan}},\ }\href@noop {} {\bibfield
  {journal} {\bibinfo  {journal} {Science (New York, NY)}\ }\textbf {\bibinfo
  {volume} {345}},\ \bibinfo {pages} {640} (\bibinfo {year}
  {2014})}\BibitemShut {NoStop}%
\bibitem [{\citenamefont {Werner}\ and\ \citenamefont
  {Sch{\"u}tz}(2011)}]{Werner:2011jv}%
  \BibitemOpen
  \bibfield  {author} {\bibinfo {author} {\bibfnamefont {H.-J.}\ \bibnamefont
  {Werner}}\ and\ \bibinfo {author} {\bibfnamefont {M.}~\bibnamefont
  {Sch{\"u}tz}},\ }\href@noop {} {\bibfield  {journal} {\bibinfo  {journal} {J.
  Chem. Phys.}\ }\textbf {\bibinfo {volume} {135}},\ \bibinfo {pages} {144116}
  (\bibinfo {year} {2011})}\BibitemShut {NoStop}%
\bibitem [{\citenamefont {Schimka}\ \emph {et~al.}(2010)\citenamefont
  {Schimka}, \citenamefont {Harl}, \citenamefont {Stroppa}, \citenamefont
  {Gr{\"u}neis}, \citenamefont {Marsman}, \citenamefont {Mittendorfer},\ and\
  \citenamefont {Kresse}}]{RPA_Schimka:NatMatt2010}%
  \BibitemOpen
  \bibfield  {author} {\bibinfo {author} {\bibfnamefont {L.}~\bibnamefont
  {Schimka}}, \bibinfo {author} {\bibfnamefont {J.}~\bibnamefont {Harl}},
  \bibinfo {author} {\bibfnamefont {A.}~\bibnamefont {Stroppa}}, \bibinfo
  {author} {\bibfnamefont {A.}~\bibnamefont {Gr{\"u}neis}}, \bibinfo {author}
  {\bibfnamefont {M.}~\bibnamefont {Marsman}}, \bibinfo {author} {\bibfnamefont
  {F.}~\bibnamefont {Mittendorfer}}, \ and\ \bibinfo {author} {\bibfnamefont
  {G.}~\bibnamefont {Kresse}},\ }\href@noop {} {\bibfield  {journal} {\bibinfo
  {journal} {Nat. Mater.}\ }\textbf {\bibinfo {volume} {9}},\ \bibinfo {pages}
  {741} (\bibinfo {year} {2010})}\BibitemShut {NoStop}%
\bibitem [{\citenamefont {Ren}\ \emph {et~al.}(2011)\citenamefont {Ren},
  \citenamefont {Tkatchenko}, \citenamefont {Rinke},\ and\ \citenamefont
  {Scheffler}}]{Ren:2011ht}%
  \BibitemOpen
  \bibfield  {author} {\bibinfo {author} {\bibfnamefont {X.}~\bibnamefont
  {Ren}}, \bibinfo {author} {\bibfnamefont {A.}~\bibnamefont {Tkatchenko}},
  \bibinfo {author} {\bibfnamefont {P.}~\bibnamefont {Rinke}}, \ and\ \bibinfo
  {author} {\bibfnamefont {M.}~\bibnamefont {Scheffler}},\ }\href@noop {}
  {\bibfield  {journal} {\bibinfo  {journal} {Phys. Rev. Lett.}\ }\textbf
  {\bibinfo {volume} {106}},\ \bibinfo {pages} {153003} (\bibinfo {year}
  {2011})}\BibitemShut {NoStop}%
\bibitem [{\citenamefont {Klime{\v s}}(2016)}]{Jiri:2016}%
  \BibitemOpen
  \bibfield  {author} {\bibinfo {author} {\bibfnamefont {J.}~\bibnamefont
  {Klime{\v s}}},\ }\href@noop {} {\bibfield  {journal} {\bibinfo  {journal}
  {J. Chem. Phys.}\ }\textbf {\bibinfo {volume} {145}},\ \bibinfo {pages}
  {094506} (\bibinfo {year} {2016})}\BibitemShut {NoStop}%
\bibitem [{\citenamefont {Foulkes}\ \emph {et~al.}(2001)\citenamefont
  {Foulkes}, \citenamefont {Mitas}, \citenamefont {Needs},\ and\ \citenamefont
  {Rajagopal}}]{foulkes01}%
  \BibitemOpen
  \bibfield  {author} {\bibinfo {author} {\bibfnamefont {W.~M.~C.}\
  \bibnamefont {Foulkes}}, \bibinfo {author} {\bibfnamefont {L.}~\bibnamefont
  {Mitas}}, \bibinfo {author} {\bibfnamefont {R.~J.}\ \bibnamefont {Needs}}, \
  and\ \bibinfo {author} {\bibfnamefont {G.}~\bibnamefont {Rajagopal}},\
  }\href@noop {} {\bibfield  {journal} {\bibinfo  {journal} {Rev. Mod. Phys.}\
  }\textbf {\bibinfo {volume} {73}},\ \bibinfo {pages} {33} (\bibinfo {year}
  {2001})}\BibitemShut {NoStop}%
\bibitem [{\citenamefont {Dubeck\'y}\ \emph {et~al.}(2016)\citenamefont
  {Dubeck\'y}, \citenamefont {Mitas},\ and\ \citenamefont
  {Jure\v{c}ka}}]{noncov:chemrev2016}%
  \BibitemOpen
  \bibfield  {author} {\bibinfo {author} {\bibfnamefont {M.}~\bibnamefont
  {Dubeck\'y}}, \bibinfo {author} {\bibfnamefont {L.}~\bibnamefont {Mitas}}, \
  and\ \bibinfo {author} {\bibfnamefont {P.}~\bibnamefont {Jure\v{c}ka}},\
  }\href@noop {} {\bibfield  {journal} {\bibinfo  {journal} {Chem. Rev.}\
  }\textbf {\bibinfo {volume} {116}},\ \bibinfo {pages} {5188} (\bibinfo {year}
  {2016})}\BibitemShut {NoStop}%
\bibitem [{\citenamefont {Zen}\ \emph {et~al.}(2016)\citenamefont {Zen},
  \citenamefont {Sorella}, \citenamefont {Gillan}, \citenamefont
  {Michaelides},\ and\ \citenamefont {Alf\`e}}]{sizeconsDMC}%
  \BibitemOpen
  \bibfield  {author} {\bibinfo {author} {\bibfnamefont {A.}~\bibnamefont
  {Zen}}, \bibinfo {author} {\bibfnamefont {S.}~\bibnamefont {Sorella}},
  \bibinfo {author} {\bibfnamefont {M.~J.}\ \bibnamefont {Gillan}}, \bibinfo
  {author} {\bibfnamefont {A.}~\bibnamefont {Michaelides}}, \ and\ \bibinfo
  {author} {\bibfnamefont {D.}~\bibnamefont {Alf\`e}},\ }\href@noop {}
  {\bibfield  {journal} {\bibinfo  {journal} {Phys. Rev. B}\ }\textbf {\bibinfo
  {volume} {93}},\ \bibinfo {pages} {241118(R)} (\bibinfo {year}
  {2016})}\BibitemShut {NoStop}%
\bibitem [{\citenamefont {Fraser}\ \emph {et~al.}(1996)\citenamefont {Fraser},
  \citenamefont {Foulkes}, \citenamefont {Rajagopal}, \citenamefont {Needs},
  \citenamefont {Kenny},\ and\ \citenamefont {Williamson}}]{MPC:Fraser1996}%
  \BibitemOpen
  \bibfield  {author} {\bibinfo {author} {\bibfnamefont {L.~M.}\ \bibnamefont
  {Fraser}}, \bibinfo {author} {\bibfnamefont {W.~M.~C.}\ \bibnamefont
  {Foulkes}}, \bibinfo {author} {\bibfnamefont {G.}~\bibnamefont {Rajagopal}},
  \bibinfo {author} {\bibfnamefont {R.~J.}\ \bibnamefont {Needs}}, \bibinfo
  {author} {\bibfnamefont {S.~D.}\ \bibnamefont {Kenny}}, \ and\ \bibinfo
  {author} {\bibfnamefont {A.~J.}\ \bibnamefont {Williamson}},\ }\href@noop {}
  {\bibfield  {journal} {\bibinfo  {journal} {Phys. Rev. B}\ }\textbf {\bibinfo
  {volume} {53}},\ \bibinfo {pages} {1814} (\bibinfo {year}
  {1996})}\BibitemShut {NoStop}%
\bibitem [{\citenamefont {Beran}\ \emph {et~al.}(2016)\citenamefont {Beran},
  \citenamefont {Hartman},\ and\ \citenamefont {Heit}}]{beran_accounts}%
  \BibitemOpen
  \bibfield  {author} {\bibinfo {author} {\bibfnamefont {G.~J.~O.}\
  \bibnamefont {Beran}}, \bibinfo {author} {\bibfnamefont {J.~D.}\ \bibnamefont
  {Hartman}}, \ and\ \bibinfo {author} {\bibfnamefont {Y.~N.}\ \bibnamefont
  {Heit}},\ }\href {\doibase 10.1021/acs.accounts.6b00404} {\bibfield
  {journal} {\bibinfo  {journal} {Acc. Chem. Res.}\ }\textbf {\bibinfo {volume}
  {49}},\ \bibinfo {pages} {2501} (\bibinfo {year} {2016})}\BibitemShut
  {NoStop}%
\bibitem [{\citenamefont {Reilly}\ and\ \citenamefont
  {Tkatchenko}(2013)}]{Elatt_ExpRT}%
  \BibitemOpen
  \bibfield  {author} {\bibinfo {author} {\bibfnamefont {A.~M.}\ \bibnamefont
  {Reilly}}\ and\ \bibinfo {author} {\bibfnamefont {A.}~\bibnamefont
  {Tkatchenko}},\ }\href@noop {} {\bibfield  {journal} {\bibinfo  {journal} {J.
  Chem. Phys.}\ }\textbf {\bibinfo {volume} {139}},\ \bibinfo {pages} {024705}
  (\bibinfo {year} {2013})}\BibitemShut {NoStop}%
\bibitem [{\citenamefont {Otero-de-la Roza}\ and\ \citenamefont
  {Johnson}(2012)}]{Elatt_ExpOJ}%
  \BibitemOpen
  \bibfield  {author} {\bibinfo {author} {\bibfnamefont {A.}~\bibnamefont
  {Otero-de-la Roza}}\ and\ \bibinfo {author} {\bibfnamefont {E.~R.}\
  \bibnamefont {Johnson}},\ }\href@noop {} {\bibfield  {journal} {\bibinfo
  {journal} {J. Chem. Phys.}\ }\textbf {\bibinfo {volume} {137}},\ \bibinfo
  {pages} {054103} (\bibinfo {year} {2012})}\BibitemShut {NoStop}%
\bibitem [{\citenamefont {Price}\ and\ \citenamefont
  {Reutzel-Edens}(2016)}]{csp_for_drugs}%
  \BibitemOpen
  \bibfield  {author} {\bibinfo {author} {\bibfnamefont {S.~L.}\ \bibnamefont
  {Price}}\ and\ \bibinfo {author} {\bibfnamefont {S.~M.}\ \bibnamefont
  {Reutzel-Edens}},\ }\href {\doibase 10.1016/j.drudis.2016.01.014} {\bibfield
  {journal} {\bibinfo  {journal} {Drug Discov. Today}\ }\textbf {\bibinfo
  {volume} {21}},\ \bibinfo {pages} {912} (\bibinfo {year} {2016})}\BibitemShut
  {NoStop}%
\bibitem [{\citenamefont {Pulido}\ \emph {et~al.}(2017)\citenamefont {Pulido},
  \citenamefont {Chen}, \citenamefont {Kaczorowski}, \citenamefont {Holden},
  \citenamefont {Little}, \citenamefont {Chong}, \citenamefont {Slater},
  \citenamefont {McMahon}, \citenamefont {Bonillo}, \citenamefont {Stackhouse},
  \citenamefont {Stephenson}, \citenamefont {Kane}, \citenamefont {Clowes},
  \citenamefont {Hasell}, \citenamefont {Cooper},\ and\ \citenamefont
  {Day}}]{csp_nature}%
  \BibitemOpen
  \bibfield  {author} {\bibinfo {author} {\bibfnamefont {A.}~\bibnamefont
  {Pulido}}, \bibinfo {author} {\bibfnamefont {L.}~\bibnamefont {Chen}},
  \bibinfo {author} {\bibfnamefont {T.}~\bibnamefont {Kaczorowski}}, \bibinfo
  {author} {\bibfnamefont {D.}~\bibnamefont {Holden}}, \bibinfo {author}
  {\bibfnamefont {M.~A.}\ \bibnamefont {Little}}, \bibinfo {author}
  {\bibfnamefont {S.~Y.}\ \bibnamefont {Chong}}, \bibinfo {author}
  {\bibfnamefont {B.~J.}\ \bibnamefont {Slater}}, \bibinfo {author}
  {\bibfnamefont {D.~P.}\ \bibnamefont {McMahon}}, \bibinfo {author}
  {\bibfnamefont {B.}~\bibnamefont {Bonillo}}, \bibinfo {author} {\bibfnamefont
  {C.~J.}\ \bibnamefont {Stackhouse}}, \bibinfo {author} {\bibfnamefont
  {A.}~\bibnamefont {Stephenson}}, \bibinfo {author} {\bibfnamefont {C.~M.}\
  \bibnamefont {Kane}}, \bibinfo {author} {\bibfnamefont {R.}~\bibnamefont
  {Clowes}}, \bibinfo {author} {\bibfnamefont {T.}~\bibnamefont {Hasell}},
  \bibinfo {author} {\bibfnamefont {A.~I.}\ \bibnamefont {Cooper}}, \ and\
  \bibinfo {author} {\bibfnamefont {G.~M.}\ \bibnamefont {Day}},\ }\href
  {\doibase 10.1038/nature21419} {\bibfield  {journal} {\bibinfo  {journal}
  {Nature}\ }\textbf {\bibinfo {volume} {543}},\ \bibinfo {pages} {657}
  (\bibinfo {year} {2017})}\BibitemShut {NoStop}%
\bibitem [{\citenamefont {Santra}\ \emph
  {et~al.}(2011{\natexlab{a}})\citenamefont {Santra}, \citenamefont
  {Klime\v{s}}, \citenamefont {Alf\`{e}}, \citenamefont {Tkatchenko},
  \citenamefont {Slater}, \citenamefont {Michaelides}, \citenamefont {Car},\
  and\ \citenamefont {Scheffler}}]{santra_hydrogen_2011}%
  \BibitemOpen
  \bibfield  {author} {\bibinfo {author} {\bibfnamefont {B.}~\bibnamefont
  {Santra}}, \bibinfo {author} {\bibfnamefont {J.}~\bibnamefont {Klime\v{s}}},
  \bibinfo {author} {\bibfnamefont {D.}~\bibnamefont {Alf\`{e}}}, \bibinfo
  {author} {\bibfnamefont {A.}~\bibnamefont {Tkatchenko}}, \bibinfo {author}
  {\bibfnamefont {B.}~\bibnamefont {Slater}}, \bibinfo {author} {\bibfnamefont
  {A.}~\bibnamefont {Michaelides}}, \bibinfo {author} {\bibfnamefont
  {R.}~\bibnamefont {Car}}, \ and\ \bibinfo {author} {\bibfnamefont
  {M.}~\bibnamefont {Scheffler}},\ }\href@noop {} {\bibfield  {journal}
  {\bibinfo  {journal} {Phys. Rev. Lett.}\ }\textbf {\bibinfo {volume} {107}},\
  \bibinfo {pages} {185701} (\bibinfo {year} {2011}{\natexlab{a}})}\BibitemShut
  {NoStop}%
\bibitem [{\citenamefont {Gillan}\ \emph {et~al.}(2013)\citenamefont {Gillan},
  \citenamefont {Alf\`{e}}, \citenamefont {Bygrave}, \citenamefont {Taylor},\
  and\ \citenamefont {Manby}}]{Gillan:enebench:jcp2013}%
  \BibitemOpen
  \bibfield  {author} {\bibinfo {author} {\bibfnamefont {M.~J.}\ \bibnamefont
  {Gillan}}, \bibinfo {author} {\bibfnamefont {D.}~\bibnamefont {Alf\`{e}}},
  \bibinfo {author} {\bibfnamefont {P.~J.}\ \bibnamefont {Bygrave}}, \bibinfo
  {author} {\bibfnamefont {C.~R.}\ \bibnamefont {Taylor}}, \ and\ \bibinfo
  {author} {\bibfnamefont {F.~R.}\ \bibnamefont {Manby}},\ }\href@noop {}
  {\bibfield  {journal} {\bibinfo  {journal} {J. Chem. Phys.}\ }\textbf
  {\bibinfo {volume} {139}},\ \bibinfo {pages} {114101} (\bibinfo {year}
  {2013})}\BibitemShut {NoStop}%
\bibitem [{\citenamefont {Cutini}\ \emph {et~al.}(2016)\citenamefont {Cutini},
  \citenamefont {Civalleri}, \citenamefont {Corno}, \citenamefont {Orlando},
  \citenamefont {Brandenburg}, \citenamefont {Maschio},\ and\ \citenamefont
  {Ugliengo}}]{Cutini:2016fh}%
  \BibitemOpen
  \bibfield  {author} {\bibinfo {author} {\bibfnamefont {M.}~\bibnamefont
  {Cutini}}, \bibinfo {author} {\bibfnamefont {B.}~\bibnamefont {Civalleri}},
  \bibinfo {author} {\bibfnamefont {M.}~\bibnamefont {Corno}}, \bibinfo
  {author} {\bibfnamefont {R.}~\bibnamefont {Orlando}}, \bibinfo {author}
  {\bibfnamefont {J.~G.}\ \bibnamefont {Brandenburg}}, \bibinfo {author}
  {\bibfnamefont {L.}~\bibnamefont {Maschio}}, \ and\ \bibinfo {author}
  {\bibfnamefont {P.}~\bibnamefont {Ugliengo}},\ }\href@noop {} {\bibfield
  {journal} {\bibinfo  {journal} {J. Chem. Theory Comput.}\ }\textbf {\bibinfo
  {volume} {12}},\ \bibinfo {pages} {3340} (\bibinfo {year}
  {2016})}\BibitemShut {NoStop}%
\bibitem [{\citenamefont {Wen}\ and\ \citenamefont {Beran}(2011)}]{Wen:2011gm}%
  \BibitemOpen
  \bibfield  {author} {\bibinfo {author} {\bibfnamefont {S.}~\bibnamefont
  {Wen}}\ and\ \bibinfo {author} {\bibfnamefont {G.~J.~O.}\ \bibnamefont
  {Beran}},\ }\href@noop {} {\bibfield  {journal} {\bibinfo  {journal} {J.
  Chem. Theory Comput.}\ }\textbf {\bibinfo {volume} {7}},\ \bibinfo {pages}
  {3733} (\bibinfo {year} {2011})}\BibitemShut {NoStop}%
\bibitem [{\citenamefont {Podeszwa}\ \emph {et~al.}(2008)\citenamefont
  {Podeszwa}, \citenamefont {Rice},\ and\ \citenamefont
  {Szalewicz}}]{Podeszwa:2008fq}%
  \BibitemOpen
  \bibfield  {author} {\bibinfo {author} {\bibfnamefont {R.}~\bibnamefont
  {Podeszwa}}, \bibinfo {author} {\bibfnamefont {B.~M.}\ \bibnamefont {Rice}},
  \ and\ \bibinfo {author} {\bibfnamefont {K.}~\bibnamefont {Szalewicz}},\
  }\href@noop {} {\bibfield  {journal} {\bibinfo  {journal} {Phys. Rev. Lett.}\
  }\textbf {\bibinfo {volume} {101}},\ \bibinfo {pages} {115503} (\bibinfo
  {year} {2008})}\BibitemShut {NoStop}%
\bibitem [{\citenamefont {Bludsk{\'y}}\ \emph {et~al.}(2008)\citenamefont
  {Bludsk{\'y}}, \citenamefont {Rube{\v s}},\ and\ \citenamefont
  {Sold{\'a}n}}]{Bludsky:2008bu}%
  \BibitemOpen
  \bibfield  {author} {\bibinfo {author} {\bibfnamefont {O.}~\bibnamefont
  {Bludsk{\'y}}}, \bibinfo {author} {\bibfnamefont {M.}~\bibnamefont {Rube{\v
  s}}}, \ and\ \bibinfo {author} {\bibfnamefont {P.}~\bibnamefont
  {Sold{\'a}n}},\ }\href@noop {} {\bibfield  {journal} {\bibinfo  {journal}
  {Phys. Rev. B}\ }\textbf {\bibinfo {volume} {77}},\ \bibinfo {pages} {092103}
  (\bibinfo {year} {2008})}\BibitemShut {NoStop}%
\bibitem [{\citenamefont {Ringer}\ and\ \citenamefont
  {Sherrill}(2008)}]{Ringer:2008kq}%
  \BibitemOpen
  \bibfield  {author} {\bibinfo {author} {\bibfnamefont {A.~L.}\ \bibnamefont
  {Ringer}}\ and\ \bibinfo {author} {\bibfnamefont {C.~D.}\ \bibnamefont
  {Sherrill}},\ }\href@noop {} {\bibfield  {journal} {\bibinfo  {journal}
  {Chem. Eur. J.}\ }\textbf {\bibinfo {volume} {14}},\ \bibinfo {pages} {2542}
  (\bibinfo {year} {2008})}\BibitemShut {NoStop}%
\bibitem [{\citenamefont {Klime{\v s}}\ \emph {et~al.}(2010)\citenamefont
  {Klime{\v s}}, \citenamefont {Bowler},\ and\ \citenamefont
  {Michaelides}}]{klimes-vdW-DF}%
  \BibitemOpen
  \bibfield  {author} {\bibinfo {author} {\bibfnamefont {J.}~\bibnamefont
  {Klime{\v s}}}, \bibinfo {author} {\bibfnamefont {D.~R.}\ \bibnamefont
  {Bowler}}, \ and\ \bibinfo {author} {\bibfnamefont {A.}~\bibnamefont
  {Michaelides}},\ }\href@noop {} {\bibfield  {journal} {\bibinfo  {journal}
  {J. Phys.: Cond. Mat.}\ }\textbf {\bibinfo {volume} {22}},\ \bibinfo {pages}
  {022201} (\bibinfo {year} {2010})}\BibitemShut {NoStop}%
\bibitem [{\citenamefont {Needs}\ \emph {et~al.}(2010)\citenamefont {Needs},
  \citenamefont {Towler}, \citenamefont {Drummond},\ and\ \citenamefont
  {Rios}}]{casino}%
  \BibitemOpen
  \bibfield  {author} {\bibinfo {author} {\bibfnamefont {R.~J.}\ \bibnamefont
  {Needs}}, \bibinfo {author} {\bibfnamefont {M.~D.}\ \bibnamefont {Towler}},
  \bibinfo {author} {\bibfnamefont {N.~D.}\ \bibnamefont {Drummond}}, \ and\
  \bibinfo {author} {\bibfnamefont {P.~L.}\ \bibnamefont {Rios}},\ }\href@noop
  {} {\bibfield  {journal} {\bibinfo  {journal} {J. Phys.: Condens. Matter}\
  }\textbf {\bibinfo {volume} {22}},\ \bibinfo {pages} {023201} (\bibinfo
  {year} {2010})}\BibitemShut {NoStop}%
\bibitem [{\citenamefont {Trail}\ and\ \citenamefont
  {Needs}(2005{\natexlab{a}})}]{trail05_NCHF}%
  \BibitemOpen
  \bibfield  {author} {\bibinfo {author} {\bibfnamefont {J.~R.}\ \bibnamefont
  {Trail}}\ and\ \bibinfo {author} {\bibfnamefont {R.~J.}\ \bibnamefont
  {Needs}},\ }\href@noop {} {\bibfield  {journal} {\bibinfo  {journal} {J.
  Chem. Phys.}\ }\textbf {\bibinfo {volume} {122}},\ \bibinfo {pages} {014112}
  (\bibinfo {year} {2005}{\natexlab{a}})}\BibitemShut {NoStop}%
\bibitem [{\citenamefont {Trail}\ and\ \citenamefont
  {Needs}(2005{\natexlab{b}})}]{trail05_SRHF}%
  \BibitemOpen
  \bibfield  {author} {\bibinfo {author} {\bibfnamefont {J.~R.}\ \bibnamefont
  {Trail}}\ and\ \bibinfo {author} {\bibfnamefont {R.~J.}\ \bibnamefont
  {Needs}},\ }\href@noop {} {\bibfield  {journal} {\bibinfo  {journal} {J.
  Chem. Phys.}\ }\textbf {\bibinfo {volume} {122}},\ \bibinfo {pages} {174109}
  (\bibinfo {year} {2005}{\natexlab{b}})}\BibitemShut {NoStop}%
\bibitem [{\citenamefont {Mitas}\ \emph {et~al.}(1991)\citenamefont {Mitas},
  \citenamefont {Shirley},\ and\ \citenamefont {Ceperley}}]{mitas91}%
  \BibitemOpen
  \bibfield  {author} {\bibinfo {author} {\bibfnamefont {L.}~\bibnamefont
  {Mitas}}, \bibinfo {author} {\bibfnamefont {E.~L.}\ \bibnamefont {Shirley}},
  \ and\ \bibinfo {author} {\bibfnamefont {D.~M.}\ \bibnamefont {Ceperley}},\
  }\href@noop {} {\bibfield  {journal} {\bibinfo  {journal} {J. Chem. Phys.}\
  }\textbf {\bibinfo {volume} {95}},\ \bibinfo {pages} {3467} (\bibinfo {year}
  {1991})}\BibitemShut {NoStop}%
\bibitem [{pws()}]{pwscf}%
  \BibitemOpen
  \href@noop {} {}\bibinfo {note} {S. Baroni, A. Dal Corso, S. de Gironcoli,
  and P. Giannozzi, http://www.pwscf.org}\BibitemShut {NoStop}%
\bibitem [{\citenamefont {Alf\`{e}}\ and\ \citenamefont
  {Gillan}(2004)}]{alfe04}%
  \BibitemOpen
  \bibfield  {author} {\bibinfo {author} {\bibfnamefont {D.}~\bibnamefont
  {Alf\`{e}}}\ and\ \bibinfo {author} {\bibfnamefont {M.~J.}\ \bibnamefont
  {Gillan}},\ }\href@noop {} {\bibfield  {journal} {\bibinfo  {journal} {Phys.
  Rev. B}\ }\textbf {\bibinfo {volume} {70}},\ \bibinfo {pages} {161101}
  (\bibinfo {year} {2004})}\BibitemShut {NoStop}%
\bibitem [{\citenamefont {Trail}\ and\ \citenamefont
  {Needs}(2013)}]{CEPP:JCP2013}%
  \BibitemOpen
  \bibfield  {author} {\bibinfo {author} {\bibfnamefont {J.~R.}\ \bibnamefont
  {Trail}}\ and\ \bibinfo {author} {\bibfnamefont {R.~J.}\ \bibnamefont
  {Needs}},\ }\href@noop {} {\bibfield  {journal} {\bibinfo  {journal} {J.
  Chem. Phys.}\ }\textbf {\bibinfo {volume} {139}},\ \bibinfo {pages} {014101}
  (\bibinfo {year} {2013})}\BibitemShut {NoStop}%
\bibitem [{\citenamefont {Lin}\ \emph {et~al.}(2001)\citenamefont {Lin},
  \citenamefont {Zong},\ and\ \citenamefont
  {Ceperley}}]{Lin:qmctwistavg:pre2001}%
  \BibitemOpen
  \bibfield  {author} {\bibinfo {author} {\bibfnamefont {C.}~\bibnamefont
  {Lin}}, \bibinfo {author} {\bibfnamefont {F.~H.}\ \bibnamefont {Zong}}, \
  and\ \bibinfo {author} {\bibfnamefont {D.~M.}\ \bibnamefont {Ceperley}},\
  }\href@noop {} {\bibfield  {journal} {\bibinfo  {journal} {Phys. Rev. E}\
  }\textbf {\bibinfo {volume} {64}},\ \bibinfo {pages} {016702} (\bibinfo
  {year} {2001})}\BibitemShut {NoStop}%
\bibitem [{\citenamefont {Williamson}\ \emph {et~al.}(1997)\citenamefont
  {Williamson}, \citenamefont {Rajagopal}, \citenamefont {Needs}, \citenamefont
  {Fraser}, \citenamefont {Foulkes}, \citenamefont {Wang},\ and\ \citenamefont
  {Chou}}]{MPC:Will1997}%
  \BibitemOpen
  \bibfield  {author} {\bibinfo {author} {\bibfnamefont {A.~J.}\ \bibnamefont
  {Williamson}}, \bibinfo {author} {\bibfnamefont {G.}~\bibnamefont
  {Rajagopal}}, \bibinfo {author} {\bibfnamefont {R.~J.}\ \bibnamefont
  {Needs}}, \bibinfo {author} {\bibfnamefont {L.~M.}\ \bibnamefont {Fraser}},
  \bibinfo {author} {\bibfnamefont {W.~M.~C.}\ \bibnamefont {Foulkes}},
  \bibinfo {author} {\bibfnamefont {Y.}~\bibnamefont {Wang}}, \ and\ \bibinfo
  {author} {\bibfnamefont {M.-Y.}\ \bibnamefont {Chou}},\ }\href@noop {}
  {\bibfield  {journal} {\bibinfo  {journal} {Phys. Rev. B}\ }\textbf {\bibinfo
  {volume} {55}},\ \bibinfo {pages} {R4851} (\bibinfo {year}
  {1997})}\BibitemShut {NoStop}%
\bibitem [{\citenamefont {Kent}\ \emph {et~al.}(1999)\citenamefont {Kent},
  \citenamefont {Hood}, \citenamefont {Williamson}, \citenamefont {Needs},
  \citenamefont {Foulkes},\ and\ \citenamefont {Rajagopal}}]{MPC:Kent1999}%
  \BibitemOpen
  \bibfield  {author} {\bibinfo {author} {\bibfnamefont {P.~R.~C.}\
  \bibnamefont {Kent}}, \bibinfo {author} {\bibfnamefont {R.~Q.}\ \bibnamefont
  {Hood}}, \bibinfo {author} {\bibfnamefont {A.~J.}\ \bibnamefont
  {Williamson}}, \bibinfo {author} {\bibfnamefont {R.~J.}\ \bibnamefont
  {Needs}}, \bibinfo {author} {\bibfnamefont {W.~M.~C.}\ \bibnamefont
  {Foulkes}}, \ and\ \bibinfo {author} {\bibfnamefont {G.}~\bibnamefont
  {Rajagopal}},\ }\href@noop {} {\bibfield  {journal} {\bibinfo  {journal}
  {Phys. Rev. B}\ }\textbf {\bibinfo {volume} {59}},\ \bibinfo {pages} {1917}
  (\bibinfo {year} {1999})}\BibitemShut {NoStop}%
\bibitem [{\citenamefont {Chiesa}\ \emph {et~al.}(2006)\citenamefont {Chiesa},
  \citenamefont {Ceperley}, \citenamefont {Martin},\ and\ \citenamefont
  {Holzmann}}]{Chiesa:size_effects:prl2006}%
  \BibitemOpen
  \bibfield  {author} {\bibinfo {author} {\bibfnamefont {S.}~\bibnamefont
  {Chiesa}}, \bibinfo {author} {\bibfnamefont {D.~M.}\ \bibnamefont
  {Ceperley}}, \bibinfo {author} {\bibfnamefont {R.~M.}\ \bibnamefont
  {Martin}}, \ and\ \bibinfo {author} {\bibfnamefont {M.}~\bibnamefont
  {Holzmann}},\ }\href@noop {} {\bibfield  {journal} {\bibinfo  {journal}
  {Phys. Rev. Lett.}\ }\textbf {\bibinfo {volume} {97}},\ \bibinfo {pages}
  {076404} (\bibinfo {year} {2006})}\BibitemShut {NoStop}%
\bibitem [{\citenamefont {Kwee}\ \emph {et~al.}(2008)\citenamefont {Kwee},
  \citenamefont {Zhang},\ and\ \citenamefont {Krakauer}}]{KZK:prl2008}%
  \BibitemOpen
  \bibfield  {author} {\bibinfo {author} {\bibfnamefont {H.}~\bibnamefont
  {Kwee}}, \bibinfo {author} {\bibfnamefont {S.}~\bibnamefont {Zhang}}, \ and\
  \bibinfo {author} {\bibfnamefont {H.}~\bibnamefont {Krakauer}},\ }\href@noop
  {} {\bibfield  {journal} {\bibinfo  {journal} {Phys. Rev. Lett.}\ }\textbf
  {\bibinfo {volume} {100}},\ \bibinfo {pages} {126404} (\bibinfo {year}
  {2008})}\BibitemShut {NoStop}%
\bibitem [{\citenamefont {Umrigar}\ \emph {et~al.}(1993)\citenamefont
  {Umrigar}, \citenamefont {Nightingale},\ and\ \citenamefont
  {Runge}}]{umrigar93}%
  \BibitemOpen
  \bibfield  {author} {\bibinfo {author} {\bibfnamefont {C.~J.}\ \bibnamefont
  {Umrigar}}, \bibinfo {author} {\bibfnamefont {M.~P.}\ \bibnamefont
  {Nightingale}}, \ and\ \bibinfo {author} {\bibfnamefont {K.~J.}\ \bibnamefont
  {Runge}},\ }\href@noop {} {\bibfield  {journal} {\bibinfo  {journal} {J.
  Chem. Phys.}\ }\textbf {\bibinfo {volume} {99}},\ \bibinfo {pages} {2865}
  (\bibinfo {year} {1993})}\BibitemShut {NoStop}%
\bibitem [{\citenamefont {Casula}(2006)}]{CasulaTmove}%
  \BibitemOpen
  \bibfield  {author} {\bibinfo {author} {\bibfnamefont {M.}~\bibnamefont
  {Casula}},\ }\href@noop {} {\bibfield  {journal} {\bibinfo  {journal}
  {Physical Review B}\ }\textbf {\bibinfo {volume} {74}},\ \bibinfo {pages}
  {161102} (\bibinfo {year} {2006})}\BibitemShut {NoStop}%
\bibitem [{\citenamefont {Casula}\ \emph {et~al.}(2010)\citenamefont {Casula},
  \citenamefont {Moroni}, \citenamefont {Sorella},\ and\ \citenamefont
  {Filippi}}]{casula10}%
  \BibitemOpen
  \bibfield  {author} {\bibinfo {author} {\bibfnamefont {M.}~\bibnamefont
  {Casula}}, \bibinfo {author} {\bibfnamefont {S.}~\bibnamefont {Moroni}},
  \bibinfo {author} {\bibfnamefont {S.}~\bibnamefont {Sorella}}, \ and\
  \bibinfo {author} {\bibfnamefont {C.}~\bibnamefont {Filippi}},\ }\href@noop
  {} {\bibfield  {journal} {\bibinfo  {journal} {J. Chem. Phys.}\ }\textbf
  {\bibinfo {volume} {132}},\ \bibinfo {pages} {154113} (\bibinfo {year}
  {2010})}\BibitemShut {NoStop}%
\bibitem [{\citenamefont {Drummond}\ \emph {et~al.}(2008)\citenamefont
  {Drummond}, \citenamefont {Needs}, \citenamefont {Sorouri},\ and\
  \citenamefont {Foulkes}}]{FSEqmc:PRB2008}%
  \BibitemOpen
  \bibfield  {author} {\bibinfo {author} {\bibfnamefont {N.~D.}\ \bibnamefont
  {Drummond}}, \bibinfo {author} {\bibfnamefont {R.~J.}\ \bibnamefont {Needs}},
  \bibinfo {author} {\bibfnamefont {A.}~\bibnamefont {Sorouri}}, \ and\
  \bibinfo {author} {\bibfnamefont {W.~M.~C.}\ \bibnamefont {Foulkes}},\ }\href
  {\doibase 10.1103/PhysRevB.78.125106} {\bibfield  {journal} {\bibinfo
  {journal} {Phys. Rev. B}\ }\textbf {\bibinfo {volume} {78}},\ \bibinfo
  {pages} {125106} (\bibinfo {year} {2008})}\BibitemShut {NoStop}%
\bibitem [{\citenamefont {Holzmann}\ \emph {et~al.}(2016)\citenamefont
  {Holzmann}, \citenamefont {Clay}, \citenamefont {Morales}, \citenamefont
  {Tubman}, \citenamefont {Ceperley},\ and\ \citenamefont
  {Pierleoni}}]{FSE:Ceperley2016}%
  \BibitemOpen
  \bibfield  {author} {\bibinfo {author} {\bibfnamefont {M.}~\bibnamefont
  {Holzmann}}, \bibinfo {author} {\bibfnamefont {R.~C.}\ \bibnamefont {Clay}},
  \bibinfo {author} {\bibfnamefont {M.~A.}\ \bibnamefont {Morales}}, \bibinfo
  {author} {\bibfnamefont {N.~M.}\ \bibnamefont {Tubman}}, \bibinfo {author}
  {\bibfnamefont {D.~M.}\ \bibnamefont {Ceperley}}, \ and\ \bibinfo {author}
  {\bibfnamefont {C.}~\bibnamefont {Pierleoni}},\ }\href@noop {} {\bibfield
  {journal} {\bibinfo  {journal} {Phys. Rev. B}\ }\textbf {\bibinfo {volume}
  {94}},\ \bibinfo {pages} {035126} (\bibinfo {year} {2016})}\BibitemShut
  {NoStop}%
\bibitem [{Note1()}]{Note1}%
  \BibitemOpen
  \bibinfo {note} {This process can become tricky for metals, but in this work
  we are only interested to insulators, for which there are no expected
  difficulties.}\BibitemShut {Stop}%
\bibitem [{\citenamefont {Rajagopal}\ \emph {et~al.}(1994)\citenamefont
  {Rajagopal}, \citenamefont {Needs}, \citenamefont {Kenny}, \citenamefont
  {Foulkes},\ and\ \citenamefont {James}}]{Rajagopal:1994cc}%
  \BibitemOpen
  \bibfield  {author} {\bibinfo {author} {\bibfnamefont {G.}~\bibnamefont
  {Rajagopal}}, \bibinfo {author} {\bibfnamefont {R.~J.}\ \bibnamefont
  {Needs}}, \bibinfo {author} {\bibfnamefont {S.}~\bibnamefont {Kenny}},
  \bibinfo {author} {\bibfnamefont {W.~M.~C.}\ \bibnamefont {Foulkes}}, \ and\
  \bibinfo {author} {\bibfnamefont {A.}~\bibnamefont {James}},\ }\href@noop {}
  {\bibfield  {journal} {\bibinfo  {journal} {Phys Rev Lett}\ }\textbf
  {\bibinfo {volume} {73}},\ \bibinfo {pages} {1959} (\bibinfo {year}
  {1994})}\BibitemShut {NoStop}%
\bibitem [{\citenamefont {Rajagopal}\ \emph {et~al.}(1995)\citenamefont
  {Rajagopal}, \citenamefont {Needs}, \citenamefont {James}, \citenamefont
  {Kenny},\ and\ \citenamefont {Foulkes}}]{Rajagopal:1995fk}%
  \BibitemOpen
  \bibfield  {author} {\bibinfo {author} {\bibfnamefont {G.}~\bibnamefont
  {Rajagopal}}, \bibinfo {author} {\bibfnamefont {R.~J.}\ \bibnamefont
  {Needs}}, \bibinfo {author} {\bibfnamefont {A.}~\bibnamefont {James}},
  \bibinfo {author} {\bibfnamefont {S.~D.}\ \bibnamefont {Kenny}}, \ and\
  \bibinfo {author} {\bibfnamefont {W.~M.~C.}\ \bibnamefont {Foulkes}},\
  }\href@noop {} {\bibfield  {journal} {\bibinfo  {journal} {Phys. Rev. B}\
  }\textbf {\bibinfo {volume} {51}},\ \bibinfo {pages} {10591} (\bibinfo {year}
  {1995})}\BibitemShut {NoStop}%
\bibitem [{\citenamefont {Dagrada}\ \emph {et~al.}(2016)\citenamefont
  {Dagrada}, \citenamefont {Karakuzu}, \citenamefont {Vildosola},\ and\
  \citenamefont {Casula}}]{Dagrada:2016}%
  \BibitemOpen
  \bibfield  {author} {\bibinfo {author} {\bibfnamefont {M.}~\bibnamefont
  {Dagrada}}, \bibinfo {author} {\bibfnamefont {S.}~\bibnamefont {Karakuzu}},
  \bibinfo {author} {\bibfnamefont {V.~L.}\ \bibnamefont {Vildosola}}, \ and\
  \bibinfo {author} {\bibfnamefont {M.}~\bibnamefont {Casula}},\ }\href@noop {}
  {\bibfield  {journal} {\bibinfo  {journal} {Phys. Rev. B}\ }\textbf {\bibinfo
  {volume} {94}},\ \bibinfo {pages} {245108} (\bibinfo {year}
  {2016})}\BibitemShut {NoStop}%
\bibitem [{\citenamefont {Baldereschi}(1973)}]{Baldereschi}%
  \BibitemOpen
  \bibfield  {author} {\bibinfo {author} {\bibfnamefont {A.}~\bibnamefont
  {Baldereschi}},\ }\href@noop {} {\bibfield  {journal} {\bibinfo  {journal}
  {Phys. Rev. B}\ }\textbf {\bibinfo {volume} {7}},\ \bibinfo {pages} {5212}
  (\bibinfo {year} {1973})}\BibitemShut {NoStop}%
\bibitem [{\citenamefont {Anderson}(1976)}]{FNApp:Anderson}%
  \BibitemOpen
  \bibfield  {author} {\bibinfo {author} {\bibfnamefont {J.~B.}\ \bibnamefont
  {Anderson}},\ }\href@noop {} {\bibfield  {journal} {\bibinfo  {journal} {J.
  Chem. Phys.}\ }\textbf {\bibinfo {volume} {65}},\ \bibinfo {pages} {4121}
  (\bibinfo {year} {1976})}\BibitemShut {NoStop}%
\bibitem [{\citenamefont {Reynolds}\ \emph {et~al.}(1982)\citenamefont
  {Reynolds}, \citenamefont {Ceperley}, \citenamefont {Alder},\ and\
  \citenamefont {Lester}}]{FNApp:Reynolds}%
  \BibitemOpen
  \bibfield  {author} {\bibinfo {author} {\bibfnamefont {P.~J.}\ \bibnamefont
  {Reynolds}}, \bibinfo {author} {\bibfnamefont {D.~M.}\ \bibnamefont
  {Ceperley}}, \bibinfo {author} {\bibfnamefont {B.~J.}\ \bibnamefont {Alder}},
  \ and\ \bibinfo {author} {\bibfnamefont {W.~A.}\ \bibnamefont {Lester}},\
  }\href@noop {} {\bibfield  {journal} {\bibinfo  {journal} {J. Chem. Phys.}\
  }\textbf {\bibinfo {volume} {77}},\ \bibinfo {pages} {5593} (\bibinfo {year}
  {1982})}\BibitemShut {NoStop}%
\bibitem [{\citenamefont {Ortiz}\ \emph {et~al.}(1993)\citenamefont {Ortiz},
  \citenamefont {Ceperley},\ and\ \citenamefont {Martin}}]{FPapp}%
  \BibitemOpen
  \bibfield  {author} {\bibinfo {author} {\bibfnamefont {G.}~\bibnamefont
  {Ortiz}}, \bibinfo {author} {\bibfnamefont {D.~M.}\ \bibnamefont {Ceperley}},
  \ and\ \bibinfo {author} {\bibfnamefont {R.~M.}\ \bibnamefont {Martin}},\
  }\href@noop {} {\bibfield  {journal} {\bibinfo  {journal} {Phys. Rev. Lett.}\
  }\textbf {\bibinfo {volume} {71}},\ \bibinfo {pages} {2777} (\bibinfo {year}
  {1993})}\BibitemShut {NoStop}%
\bibitem [{Note2()}]{Note2}%
  \BibitemOpen
  \bibinfo {note} {For instance, the primitive cell of the ammonia crystal has
  4 molecules, each having 8 electrons, for a total of 32, but if FSE is large
  we have to take the $2\times 2\times 2$ supercell, having $4\times 2^3 = 32$
  molecules, 256 electrons; if FSE are still non negligible we need the
  $3\times 3\times 3$ supercell, having 108 molecules, 864 electrons, and so
  on. The situation is even worse with large molecules, where already a
  $2\times 2\times 2$ supercell may have over a thousand
  electrons.}\BibitemShut {Stop}%
\bibitem [{\citenamefont {Kresse}\ and\ \citenamefont
  {Furthm\"{u}ller}(1996)}]{kresse_efficient_1996}%
  \BibitemOpen
  \bibfield  {author} {\bibinfo {author} {\bibfnamefont {G.}~\bibnamefont
  {Kresse}}\ and\ \bibinfo {author} {\bibfnamefont {J.}~\bibnamefont
  {Furthm\"{u}ller}},\ }\href@noop {} {\bibfield  {journal} {\bibinfo
  {journal} {Phys. Rev. B}\ }\textbf {\bibinfo {volume} {54}},\ \bibinfo
  {pages} {11169} (\bibinfo {year} {1996})}\BibitemShut {NoStop}%
\bibitem [{\citenamefont {Kresse}\ and\ \citenamefont
  {Joubert}(1999)}]{kresse_ultrasoft_1999}%
  \BibitemOpen
  \bibfield  {author} {\bibinfo {author} {\bibfnamefont {G.}~\bibnamefont
  {Kresse}}\ and\ \bibinfo {author} {\bibfnamefont {D.}~\bibnamefont
  {Joubert}},\ }\href@noop {} {\bibfield  {journal} {\bibinfo  {journal} {Phys.
  Rev. B}\ }\textbf {\bibinfo {volume} {59}},\ \bibinfo {pages} {1758}
  (\bibinfo {year} {1999})}\BibitemShut {NoStop}%
\bibitem [{\citenamefont {Kaltak}\ \emph
  {et~al.}(2014{\natexlab{a}})\citenamefont {Kaltak}, \citenamefont
  {Klime\v{s}},\ and\ \citenamefont {Kresse}}]{kaltak2014rpa2}%
  \BibitemOpen
  \bibfield  {author} {\bibinfo {author} {\bibfnamefont {M.}~\bibnamefont
  {Kaltak}}, \bibinfo {author} {\bibfnamefont {J.}~\bibnamefont {Klime\v{s}}},
  \ and\ \bibinfo {author} {\bibfnamefont {G.}~\bibnamefont {Kresse}},\
  }\href@noop {} {\bibfield  {journal} {\bibinfo  {journal} {Phys. Rev. B}\
  }\textbf {\bibinfo {volume} {90}},\ \bibinfo {pages} {{054115}} (\bibinfo
  {year} {2014}{\natexlab{a}})}\BibitemShut {NoStop}%
\bibitem [{\citenamefont {Klime\v{s}}\ \emph {et~al.}(2015)\citenamefont
  {Klime\v{s}}, \citenamefont {Kaltak}, \citenamefont {Maggio},\ and\
  \citenamefont {Kresse}}]{klimes2015}%
  \BibitemOpen
  \bibfield  {author} {\bibinfo {author} {\bibfnamefont {J.}~\bibnamefont
  {Klime\v{s}}}, \bibinfo {author} {\bibfnamefont {M.}~\bibnamefont {Kaltak}},
  \bibinfo {author} {\bibfnamefont {E.}~\bibnamefont {Maggio}}, \ and\ \bibinfo
  {author} {\bibfnamefont {G.}~\bibnamefont {Kresse}},\ }\href@noop {}
  {\bibfield  {journal} {\bibinfo  {journal} {J. Chem. Phys.}\ }\textbf
  {\bibinfo {volume} {143}},\ \bibinfo {pages} {{102816}} (\bibinfo {year}
  {2015})}\BibitemShut {NoStop}%
\bibitem [{\citenamefont {Perdew}\ \emph {et~al.}(1996)\citenamefont {Perdew},
  \citenamefont {Burke},\ and\ \citenamefont
  {Ernzerhof}}]{perdew_generalized_1996}%
  \BibitemOpen
  \bibfield  {author} {\bibinfo {author} {\bibfnamefont {J.~P.}\ \bibnamefont
  {Perdew}}, \bibinfo {author} {\bibfnamefont {K.}~\bibnamefont {Burke}}, \
  and\ \bibinfo {author} {\bibfnamefont {M.}~\bibnamefont {Ernzerhof}},\
  }\href@noop {} {\bibfield  {journal} {\bibinfo  {journal} {Phys. Rev. Lett.}\
  }\textbf {\bibinfo {volume} {77}},\ \bibinfo {pages} {3865} (\bibinfo {year}
  {1996})}\BibitemShut {NoStop}%
\bibitem [{\citenamefont {Kaltak}\ \emph
  {et~al.}(2014{\natexlab{b}})\citenamefont {Kaltak}, \citenamefont
  {Klime\v{s}},\ and\ \citenamefont {Kresse}}]{kaltak2014rpa1}%
  \BibitemOpen
  \bibfield  {author} {\bibinfo {author} {\bibfnamefont {M.}~\bibnamefont
  {Kaltak}}, \bibinfo {author} {\bibfnamefont {J.}~\bibnamefont {Klime\v{s}}},
  \ and\ \bibinfo {author} {\bibfnamefont {G.}~\bibnamefont {Kresse}},\
  }\href@noop {} {\bibfield  {journal} {\bibinfo  {journal} {J. Chem. Theo.
  Comput.}\ }\textbf {\bibinfo {volume} {10}},\ \bibinfo {pages} {{2498}}
  (\bibinfo {year} {2014}{\natexlab{b}})}\BibitemShut {NoStop}%
\bibitem [{\citenamefont {Rozzi}\ \emph {et~al.}(2006)\citenamefont {Rozzi},
  \citenamefont {Varsano}, \citenamefont {Marini}, \citenamefont {Gross},\ and\
  \citenamefont {Rubio}}]{rozzi2006}%
  \BibitemOpen
  \bibfield  {author} {\bibinfo {author} {\bibfnamefont {C.~A.}\ \bibnamefont
  {Rozzi}}, \bibinfo {author} {\bibfnamefont {D.}~\bibnamefont {Varsano}},
  \bibinfo {author} {\bibfnamefont {A.}~\bibnamefont {Marini}}, \bibinfo
  {author} {\bibfnamefont {E.~K.~U.}\ \bibnamefont {Gross}}, \ and\ \bibinfo
  {author} {\bibfnamefont {A.}~\bibnamefont {Rubio}},\ }\href@noop {}
  {\bibfield  {journal} {\bibinfo  {journal} {Phys. Rev. B}\ }\textbf {\bibinfo
  {volume} {73}},\ \bibinfo {pages} {205119} (\bibinfo {year}
  {2006})}\BibitemShut {NoStop}%
\bibitem [{\citenamefont {Harl}\ and\ \citenamefont
  {Kresse}(2008)}]{RPA:Kresse2008}%
  \BibitemOpen
  \bibfield  {author} {\bibinfo {author} {\bibfnamefont {J.}~\bibnamefont
  {Harl}}\ and\ \bibinfo {author} {\bibfnamefont {G.}~\bibnamefont {Kresse}},\
  }\href@noop {} {\bibfield  {journal} {\bibinfo  {journal} {Phys. Rev. B}\
  }\textbf {\bibinfo {volume} {77}},\ \bibinfo {pages} {045136} (\bibinfo
  {year} {2008})}\BibitemShut {NoStop}%
\bibitem [{\citenamefont {Macher}\ \emph {et~al.}(2014)\citenamefont {Macher},
  \citenamefont {Klime{\v s}}, \citenamefont {Franchini},\ and\ \citenamefont
  {Kresse}}]{RPAice:2014}%
  \BibitemOpen
  \bibfield  {author} {\bibinfo {author} {\bibfnamefont {M.}~\bibnamefont
  {Macher}}, \bibinfo {author} {\bibfnamefont {J.}~\bibnamefont {Klime{\v s}}},
  \bibinfo {author} {\bibfnamefont {C.}~\bibnamefont {Franchini}}, \ and\
  \bibinfo {author} {\bibfnamefont {G.}~\bibnamefont {Kresse}},\ }\href@noop {}
  {\bibfield  {journal} {\bibinfo  {journal} {J. Chem. Phys.}\ }\textbf
  {\bibinfo {volume} {140}},\ \bibinfo {pages} {084502} (\bibinfo {year}
  {2014})}\BibitemShut {NoStop}%
\bibitem [{\citenamefont {Roux}\ \emph {et~al.}(2008)\citenamefont {Roux},
  \citenamefont {Temprado}, \citenamefont {Chickos},\ and\ \citenamefont
  {Nagano}}]{Roux:2008dr}%
  \BibitemOpen
  \bibfield  {author} {\bibinfo {author} {\bibfnamefont {M.~V.}\ \bibnamefont
  {Roux}}, \bibinfo {author} {\bibfnamefont {M.}~\bibnamefont {Temprado}},
  \bibinfo {author} {\bibfnamefont {J.~S.}\ \bibnamefont {Chickos}}, \ and\
  \bibinfo {author} {\bibfnamefont {Y.}~\bibnamefont {Nagano}},\ }\href@noop {}
  {\bibfield  {journal} {\bibinfo  {journal} {J. Phys. Chem. Ref. Data}\
  }\textbf {\bibinfo {volume} {37}},\ \bibinfo {pages} {1855} (\bibinfo {year}
  {2008})}\BibitemShut {NoStop}%
\bibitem [{\citenamefont {Acree}\ and\ \citenamefont
  {Chickos}(2010)}]{AcreeJr:2010ev}%
  \BibitemOpen
  \bibfield  {author} {\bibinfo {author} {\bibfnamefont {W.}~\bibnamefont
  {Acree}, \bibfnamefont {Jr.}}\ and\ \bibinfo {author} {\bibfnamefont {J.~S.}\
  \bibnamefont {Chickos}},\ }\href@noop {} {\bibfield  {journal} {\bibinfo
  {journal} {J. Phys. Chem. Ref. Data}\ }\textbf {\bibinfo {volume} {39}},\
  \bibinfo {pages} {043101} (\bibinfo {year} {2010})}\BibitemShut {NoStop}%
\bibitem [{\citenamefont {R{\r u}{\v z}i{\v c}ka}\ \emph
  {et~al.}(2014)\citenamefont {R{\r u}{\v z}i{\v c}ka}, \citenamefont {Fulem},\
  and\ \citenamefont {{\v C}ervinka}}]{Ruzicka:2014}%
  \BibitemOpen
  \bibfield  {author} {\bibinfo {author} {\bibfnamefont {K.}~\bibnamefont {R{\r
  u}{\v z}i{\v c}ka}}, \bibinfo {author} {\bibfnamefont {M.}~\bibnamefont
  {Fulem}}, \ and\ \bibinfo {author} {\bibfnamefont {C.}~\bibnamefont {{\v
  C}ervinka}},\ }\href@noop {} {\bibfield  {journal} {\bibinfo  {journal} {J.
  Chem. Thermodyn.}\ }\textbf {\bibinfo {volume} {68}},\ \bibinfo {pages} {40}
  (\bibinfo {year} {2014})}\BibitemShut {NoStop}%
\bibitem [{\citenamefont {Chickos}(2003)}]{Chickos2003_DHerror}%
  \BibitemOpen
  \bibfield  {author} {\bibinfo {author} {\bibfnamefont {J.~S.}\ \bibnamefont
  {Chickos}},\ }\href@noop {} {\bibfield  {journal} {\bibinfo  {journal} {Netsu
  Sokutei}\ } (\bibinfo {year} {2003})}\BibitemShut {NoStop}%
\bibitem [{\citenamefont {Whalley}(1984)}]{ICE:Whalley1984}%
  \BibitemOpen
  \bibfield  {author} {\bibinfo {author} {\bibfnamefont {E.}~\bibnamefont
  {Whalley}},\ }\href@noop {} {\bibfield  {journal} {\bibinfo  {journal} {J.
  Chem. Phys.}\ }\textbf {\bibinfo {volume} {81}},\ \bibinfo {pages} {4087}
  (\bibinfo {year} {1984})}\BibitemShut {NoStop}%
\bibitem [{cpb()}]{cpbenzene_gas}%
  \BibitemOpen
  \href@noop {} {}\bibinfo {note} {Thermodynamics Research Center, Selected
  Values of Properties of Chemical Compounds., Thermodynamics Research Center,
  Texas A\&M University, College Station, Texas, 1997.}\BibitemShut {Stop}%
\bibitem [{\citenamefont {Ahlberg}\ \emph {et~al.}(1937)\citenamefont
  {Ahlberg}, \citenamefont {Blanchard},\ and\ \citenamefont
  {Lundberg}}]{cpbenzene_solid}%
  \BibitemOpen
  \bibfield  {author} {\bibinfo {author} {\bibfnamefont {J.~E.}\ \bibnamefont
  {Ahlberg}}, \bibinfo {author} {\bibfnamefont {E.~R.}\ \bibnamefont
  {Blanchard}}, \ and\ \bibinfo {author} {\bibfnamefont {W.~O.}\ \bibnamefont
  {Lundberg}},\ }\href {\doibase 10.1063/1.1750072} {\bibfield  {journal}
  {\bibinfo  {journal} {J. Chem. Phys.}\ }\textbf {\bibinfo {volume} {5}},\
  \bibinfo {pages} {539} (\bibinfo {year} {1937})}\BibitemShut {NoStop}%
\bibitem [{\citenamefont {Woolley}(1954)}]{cpco2_gas}%
  \BibitemOpen
  \bibfield  {author} {\bibinfo {author} {\bibfnamefont {H.~W.}\ \bibnamefont
  {Woolley}},\ }\href@noop {} {\bibfield  {journal} {\bibinfo  {journal} {J.
  Res. NBS}\ }\textbf {\bibinfo {volume} {52}},\ \bibinfo {pages} {289}
  (\bibinfo {year} {1954})}\BibitemShut {NoStop}%
\bibitem [{\citenamefont {Manzheli}(1971)}]{cpco2_solid}%
  \BibitemOpen
  \bibfield  {author} {\bibinfo {author} {\bibfnamefont {V.~G.}\ \bibnamefont
  {Manzheli}},\ }\href@noop {} {\bibfield  {journal} {\bibinfo  {journal}
  {Phys. Stat. Sol.}\ }\textbf {\bibinfo {volume} {44}},\ \bibinfo {pages} {39}
  (\bibinfo {year} {1971})}\BibitemShut {NoStop}%
\bibitem [{\citenamefont {Popov}\ \emph {et~al.}(1971)\citenamefont {Popov},
  \citenamefont {Manzhelii},\ and\ \citenamefont {Bagatskii}}]{cpammonia}%
  \BibitemOpen
  \bibfield  {author} {\bibinfo {author} {\bibfnamefont {V.~A.}\ \bibnamefont
  {Popov}}, \bibinfo {author} {\bibfnamefont {V.~G.}\ \bibnamefont
  {Manzhelii}}, \ and\ \bibinfo {author} {\bibfnamefont {M.~I.}\ \bibnamefont
  {Bagatskii}},\ }\href@noop {} {\bibfield  {journal} {\bibinfo  {journal} {J.
  Low Temp. Phys.}\ }\textbf {\bibinfo {volume} {5}},\ \bibinfo {pages} {4}
  (\bibinfo {year} {1971})}\BibitemShut {NoStop}%
\bibitem [{\citenamefont {Goursot}\ \emph {et~al.}(1970)\citenamefont
  {Goursot}, \citenamefont {Girdhar},\ and\ \citenamefont
  {Westrum}}]{cpnaphtalene}%
  \BibitemOpen
  \bibfield  {author} {\bibinfo {author} {\bibfnamefont {P.}~\bibnamefont
  {Goursot}}, \bibinfo {author} {\bibfnamefont {H.~L.}\ \bibnamefont
  {Girdhar}}, \ and\ \bibinfo {author} {\bibfnamefont {E.~F.}\ \bibnamefont
  {Westrum}},\ }\href@noop {} {\bibfield  {journal} {\bibinfo  {journal} {J.
  Phys. Chem.}\ }\textbf {\bibinfo {volume} {74}},\ \bibinfo {pages} {2538}
  (\bibinfo {year} {1970})}\BibitemShut {NoStop}%
\bibitem [{\citenamefont {McCullough}\ \emph {et~al.}(1957)\citenamefont
  {McCullough}, \citenamefont {Finke}, \citenamefont {Messerly}, \citenamefont
  {Todd}, \citenamefont {Kincheloe},\ and\ \citenamefont
  {Waddington}}]{cpanthracene}%
  \BibitemOpen
  \bibfield  {author} {\bibinfo {author} {\bibfnamefont {P.}~\bibnamefont
  {McCullough}}, \bibinfo {author} {\bibfnamefont {H.~L.}\ \bibnamefont
  {Finke}}, \bibinfo {author} {\bibfnamefont {J.~F.}\ \bibnamefont {Messerly}},
  \bibinfo {author} {\bibfnamefont {S.~S.}\ \bibnamefont {Todd}}, \bibinfo
  {author} {\bibfnamefont {T.~C.}\ \bibnamefont {Kincheloe}}, \ and\ \bibinfo
  {author} {\bibfnamefont {G.}~\bibnamefont {Waddington}},\ }\href@noop {}
  {\bibfield  {journal} {\bibinfo  {journal} {J. Phys. Chem.}\ }\textbf
  {\bibinfo {volume} {61}},\ \bibinfo {pages} {1105} (\bibinfo {year}
  {1957})}\BibitemShut {NoStop}%
\bibitem [{\citenamefont {Santra}\ \emph {et~al.}(2013)\citenamefont {Santra},
  \citenamefont {Klime\v{s}}, \citenamefont {Tkatchenko}, \citenamefont
  {Alf\`{e}}, \citenamefont {Slater}, \citenamefont {Michaelides},
  \citenamefont {Car},\ and\ \citenamefont {Scheffler}}]{santra_on_2013}%
  \BibitemOpen
  \bibfield  {author} {\bibinfo {author} {\bibfnamefont {B.}~\bibnamefont
  {Santra}}, \bibinfo {author} {\bibfnamefont {J.}~\bibnamefont {Klime\v{s}}},
  \bibinfo {author} {\bibfnamefont {A.}~\bibnamefont {Tkatchenko}}, \bibinfo
  {author} {\bibfnamefont {D.}~\bibnamefont {Alf\`{e}}}, \bibinfo {author}
  {\bibfnamefont {B.}~\bibnamefont {Slater}}, \bibinfo {author} {\bibfnamefont
  {A.}~\bibnamefont {Michaelides}}, \bibinfo {author} {\bibfnamefont
  {R.}~\bibnamefont {Car}}, \ and\ \bibinfo {author} {\bibfnamefont
  {M.}~\bibnamefont {Scheffler}},\ }\href@noop {} {\bibfield  {journal}
  {\bibinfo  {journal} {J. Chem. Phys.}\ }\textbf {\bibinfo {volume} {139}}
  (\bibinfo {year} {2013})}\BibitemShut {NoStop}%
\bibitem [{\citenamefont {Santra}\ \emph
  {et~al.}(2011{\natexlab{b}})\citenamefont {Santra}, \citenamefont {Klime{\v
  s}}, \citenamefont {Alf{\`e}}, \citenamefont {Tkatchenko}, \citenamefont
  {Slater}, \citenamefont {Michaelides}, \citenamefont {Car},\ and\
  \citenamefont {Scheffler}}]{Ice:prl2011}%
  \BibitemOpen
  \bibfield  {author} {\bibinfo {author} {\bibfnamefont {B.}~\bibnamefont
  {Santra}}, \bibinfo {author} {\bibfnamefont {J.}~\bibnamefont {Klime{\v s}}},
  \bibinfo {author} {\bibfnamefont {D.}~\bibnamefont {Alf{\`e}}}, \bibinfo
  {author} {\bibfnamefont {A.}~\bibnamefont {Tkatchenko}}, \bibinfo {author}
  {\bibfnamefont {B.}~\bibnamefont {Slater}}, \bibinfo {author} {\bibfnamefont
  {A.}~\bibnamefont {Michaelides}}, \bibinfo {author} {\bibfnamefont
  {R.}~\bibnamefont {Car}}, \ and\ \bibinfo {author} {\bibfnamefont
  {M.}~\bibnamefont {Scheffler}},\ }\href@noop {} {\bibfield  {journal}
  {\bibinfo  {journal} {Phys. Rev. Lett.}\ }\textbf {\bibinfo {volume} {107}},\
  \bibinfo {pages} {185701} (\bibinfo {year} {2011}{\natexlab{b}})}\BibitemShut
  {NoStop}%
\end{thebibliography}%

\end{document}